\documentclass[twocolumn,showpacs,prd,aps,superscriptaddress,tightenlines,floatfix]{revtex4-1} 

\usepackage{graphics,graphicx}
\usepackage{color}

\newcommand{\beq}{\begin{equation}}
\newcommand{\eeq}{\end{equation}}
\newcommand{\bea}{\begin{eqnarray}}
\newcommand{\eea}{\end{eqnarray}}

\begin{document}

\title{
Eccentric black hole mergers and zoom-whirl behavior
from elliptic inspirals to hyperbolic encounters 
}

%
\author{Roman Gold}
\affiliation{Theoretical Physics Institute, 
University of Jena, 07743 Jena, Germany}
\affiliation{Department of Physics, University of Illinois at Urbana-Champaign, Urbana, Illinois 61801}
%
%
\author{Bernd Br\"ugmann}
\affiliation{Theoretical Physics Institute, 
University of Jena, 07743 Jena, Germany}
\medskip
\date{September 12, 2012}
\begin{abstract}
We perform a parameter study of non-spinning, equal and unequal mass
black hole binaries on generic, eccentric orbits in numerical
relativity. The linear momentum considered ranges from that of a circular 
orbit to ten times that value.  
We discuss the different manifestations of zoom-whirl behavior in the 
hyperbolic and the elliptic regime. 
The hyperbolic data set applies to dynamical capture scenarios (e.g.
in globular clusters). Evolutions in the elliptic regime correspond 
to possible end states of supermassive black hole
binaries. 
We spot zoom-whirl behavior for eccentricities as low as $e\sim0.5$,
i.e.\ within the expected range of eccentricities in massive black hole
binaries from galaxy mergers and binaries near galactic centers.
The resulting gravitational waveforms reveal a rich structure, which will 
effectively break degeneracies in parameter space improving parameter 
estimation. 
\end{abstract}
\pacs{
04.25.D-, 
04.25.dg, 
04.70.Bw, 
95.30.Sf  
}
\maketitle
%

\section{Introduction}
\label{Introduction}

Zoom-whirl orbits arise as a general relativistic phenomenon of the
two-body problem. Such orbits do not exist in Newtonian gravity, where
the orbits are Kepler's conic sections, hence they represent
an important facet of one of the fundamental problems in general
relativity (GR). The term zoom-whirl was first used in
\cite{GlaKen02}\footnote{
  The term zoom-whirl is attributed to C. Cutler and E. Poisson (who
  in turn mention it was suggested by K. Thorne). 
}.
It refers to the orbits of eccentric binaries where tight and fast
revolutions (the whirls) are separated by phases in which the two
objects move out to larger distances and back in (the zooms).

The physics behind this effect is precession. For bound orbits, one
can define the precession per orbit as the angle between two
consecutive apocenters. In general, precession accumulates
continuously (with respect to some external reference frame) and
amounts to an excess angle beyond the Newtonian motion.
Precession is strongest for small separations and 
is therefore significant especially for 
eccentric orbits. In the solar
system, the precession of the orbit of Mercury due to general
relativity is $43$ arcseconds per century, or
$(2.9\cdot10^{-5})^\circ$ per orbit. For binary pulsars, precession
is not necessarily much larger.
For the Hulse-Taylor pulsar the precession is $4.2^\circ$ per year,
but due to its short orbital period this amounts to
$(3.7\cdot10^{-3})^\circ$ per orbit. Observed are on the order of
$(2\cdot10^{-2})^\circ$ per orbit in some
cases~\cite{HulTay75,Bur03,KraStaMan06}. 
A SMBH binary model \cite{Sillanpaa:1988zz,valtonen-2009} 
fitted to the optical light curve of the quasar 
OJ-287 predicts (model dependent) orbital parameters with precession 
as large as $\sim 40^\circ$ per orbit. 

In theory, general relativistic orbits with yet larger 
precession can be easily constructed by choosing appropriate orbital
parameters. This is possible for test particles following geodesics
around a black hole, but also for comparable mass compact
objects in the post-Newtonian (PN) approximation \cite{LevCon10}.
As long as the particle or compact object orbits well outside the
innermost stable circular orbit (ISCO), the classical picture of a
slowly precessing ellipse applies. If the orbital parameters are
chosen such that the object approaches distances close to or even
inside the ISCO, it may follow an unstable circular orbit for some 
time. After this it either plunges or escapes to larger distances
(infinity if the motion is unbound), which is the zoom-whirl behavior
we are interested in.
In the whirl regime orbits exhibit extreme precession with
precession angles comparable to or larger than $2\pi$, wrapping the
inner part of the orbit once or even several times around its center.

The basic features of zoom-whirl orbits were first discussed in the
context of geodesics in a stationary black hole spacetime
(e.g.~\cite{Cha83,Mar03,LevGiz08}), in extreme mass ratio
inspirals (e.g.~\cite{GlaKen02,DraHug05,Haas:2007kz}) and
PN evolutions \cite{LevGro08,GroLev08}.
For geodesics, it is a matter of fine-tuning the initial parameters of
the orbit to obtain a certain number of whirls. In fact, for geodesics the
number of whirls can be made arbitrarily large since there is no
gravitational radiation, see \cite{Mar03} for an example with 
$6$ orbits during a whirl.
Going beyond the test mass limit, including radiation loss is
a key task, e.g.\ \cite{GlaKen02,hughes-2005-94,PatWil02}.

The main question about zoom-whirl orbits in full GR is how the
classic, well known picture of zoom-whirl geodesics changes for
binaries with comparable masses in configurations where radiation
damping becomes significant.
Naively, we do not expect the binary to radiate away more than its
total mass, i.e.\ the number of orbits is finite since it is limited by the
energy and angular momentum radiated away during each whirl. 
In fact, for comparable masses one might have questioned whether it is
possible to obtain even a single (full) whirl.  
Since the whirls happen at high velocity and small separation (even
inside the innermost stable circular orbit), the PN approximation is
not directly applicable, e.g.\ \cite{LevCon10}.
However, recently some groups have performed numerical evolutions in
full general relativity of eccentric black hole binaries (BHBs). Zoom-whirl
orbits have indeed been found, although the number of whirls in these
experiments is less than three.

In~\cite{PreKhu07}, Pretorius and Khurana present the first example of
a whirl orbit for an equal mass binary.
In~\cite{SpeBerCar07,HinVaiHer07,WasHeaHer08,GolBru09,HeaLagMat09}, 
several
examples for the transition from inspiral to plunge, radiated energy,
angular momentum and the resulting final spin are investigated.
In~\cite{HeaLevSho09} longer evolutions of unequal masses and 
non-vanishing spin with up to three elliptic orbits which 
transition through the zoom-whirl regime prior to merger
are studied. 
The notion of marginally stable circular orbits in background spacetimes was
shown to be in close resemblance to whirl orbits in numerical evolutions of
finite mass ratio \cite{PhysRevD50-3816,PreKhu07}. 
The consequences for kicks are addressed in \cite{HeaHerHin08}. 
Implications for data analysis are studied in 
e.g.\ \cite{martel-1999-60,Mar99,BroZim09,Cokelaer:2009hj,VaiHinSho09,Mik12}.
In particular \cite{Cokelaer:2009hj,BroZim09} point out 
the potentially deteriorating effects in signal processing when 
eccentricity is ignored in the waveform models.

Eccentric neutron star and mixed binaries in dynamical spacetime have 
been studied in \cite{East:2012ww,EasPreSte11,GolBerThi11,SteEasPre11}, and in all  
cases zoom-whirl behavior has been identified.
The focus
in~\cite{ShiOkaYam08,Sperhake:2009jz,SpeCarPre08,
Berti:2010ce,Sperhake:2010dn}
is on high-energy collision.  Among the key results so far is that the
total energy radiated can easily exceed the $4\%$ of the total mass
radiated during the last stage of a quasi-circular inspiral. For
high-energy collision, up to $35\pm5\%$ have been
found~\cite{SpeCarPre08}.  
In \cite{GolBru09}, we found at low momentum multiple extrema in the
radiated energy as a function of the initial data, and that only a
modest amount of fine-tuning is required to spot these extrema. These
extrema should be compared to the variations in the mass and spin of
the merger remnant noted in~\cite{HeaLagMat09}.

Choosing different initial data and also different tuning strategies,
these investigations have been performed in different regions in
parameter space. In the present work we focus on an area that has received
relatively little attention so far, namely intermediate 
momenta and comparable but not necessarily equal masses. 
We extend the discussion of~\cite{GolBru09}, specifically we consider
mass ratios 1:1, 1:2, and 1:3, 
and linear momenta that are 1 to 6, and in one case 10 times
the value of a circular orbit, although not in all possible
combinations.

Zoom-whirl is sometimes thought to occur beyond a certain, rather
large eccentricity. Our results instead show that whirls can also be
found for modest eccentricities.  We give an analysis of the
gravitational waves (GWs) and how specific features in the radiated energy
are related to orbital characteristics.

A prerequisite for zoom-whirl orbits is eccentricity. 
\emph{Isolated} black hole binaries formed at typical
separations perform a sufficiently large number of orbits such that
the orbits become circularized long before entering the strong field
regime \cite{Pet64}.
However, it cannot be expected that all binary GW sources are sufficiently isolated and hence other effects have to be taken into account. 
In fact, SMBH binaries are expected to be formed in gas/star rich environments \cite{SesRoeRey11} with potentially large eccentricities \cite{Khan:2011gi,Preto:2011gu}. It is well understood that such binaries can gain eccentricity as a consequence of gravitational torques exchanged with the circumbinary disk \cite{PapNelMas01,Armitage:2005xq,Cuadra:2008xn,RoeSes11,SesRoeRey11}. 
Likewise, gravitational interaction with additional bodies (Kozai-oscillations, Hill-mechanism, mass segregation, gravitational focussing, etc.) generically induce eccentricity growth on a binary system.
Numerous studies of such effects 
\cite{zwart-1999,benacquista-2001-18,wen-2003-598,pooley-2003,miller-2002,LeeRamVen09,oleary-2008b, brown-2008ns, zwart-2005,
  blaes-2002-578, HopAle05, BroZim09, Mar99,
  martel-1999-60, ManBroGai07,iwasawa-2005,blaes-2002-578,volonteri-2003-582,Antonini:2012ad}
 suggest that the eccentricity of binaries, which emit significant gravitational radiation, cannot in general be ignored. 
Event rate estimates for eccentric compact object binaries 
\cite{oleary-2008b,LeeRamVen09,Kowalska:2010qg} suffer from large uncertainties 
and vary considerably. 
Some studies predict that advanced LIGO should 
detect such sources, but given the large uncertainties this should 
be taken with care. For third generation detectors the detection 
range will be larger. Eccentric binary mergers will therefore 
become more interesting sources in the future.
For supermassive BHBs, pulsar timing arrays will soon be able 
to resolve individual sources in a regime where many binaries are 
still expected to be eccentric \cite{Kocsis:2011ch}.

Given a population of eccentric binaries,
whether zoom-whirl orbits are of relevance to gravitational wave
astronomy depends on several factors. Even if the signals are
stronger, if excessive fine-tuning is required, then the population of
strong sources might amount to a very small corner of parameter space.
Conversely, if little tuning is involved, then zoom-whirl orbits can
be potential GW sources even for ground-based
detectors~\cite{Mar99,martel-1999-60,BroZim09,wen-2003-598}. 
With regard to GWs, for comparable mass
binaries with astrophysical momenta 
we loose the unlimited number of whirl orbits due to
gravitational radiation, but what is lost corresponds to a very small
part of parameter space anyway.
In any case, as a matter of principle we should be prepared to detect
and recognize GWs from all corners of parameter space including
zoom-whirl orbits.

The paper is organized as follows. In Sec.~\ref{Numerical} we describe
the basics of our numerical methods, our choice of initial
configurations, and give error estimates for typical runs.
We discuss orbital properties in Sec.~\ref{Orbits}, resulting
waveforms and radiated energy in \ref{Radiation}, and phase space
trajectories in \ref{sec:PhaseSpace}.
We conclude with Sec.~\ref{Conclusion}.

%
\section{Numerical methods and summary of simulations}
\label{Numerical}

\subsection{Method}

We performed a parameter study of the black hole binary problem using
3d numerical simulations obtained with the BAM code
\cite{BruGonHan06,BruTicJan03,HusGonHan07}.  Initial data for black
holes is computed by the puncture method \cite{BraBru97} using a
pseudo-spectral code \cite{AnsBruTic04}, and evolved with the
$\chi$-variant of the moving-puncture \cite{CamLouMar05,BakCenCho05}
version of the BSSN \cite{ShiNak95,BauSha98} formulation of the 3+1
Einstein evolution equations.  We use a $4^{th}$ order Runge-Kutta
method with $6^{th}$ order finite differencing.
The wave extraction and the calculation of the radiated energy is done
using a $4^{th}$ order accurate implementation of the Newman-Penrose
formalism. We extract $\Psi_4$ and thus also $E_{rad}$ at extraction
radii of $r_{GW}=60M,80M,100M$, 
where $M$ the total puncture mass $M$ (see below). 
Our grid is a box of typically $\gtrsim (640M)^3$ size, which is
sufficient to keep the boundaries causally disconnected from the GWs
for most of our runs. We employ bitant or quadrant symmetry when possible.
Usually, the grid consists of 9 levels of mesh refinement starting at the
coarsest level with resolution of $h=5M$ and increases by factors of
two, resulting in the resolution of $h\approx M/50$ at the finest
level. 
The inner, finer levels are evolved according to Berger-Oliger
timestepping while the outer levels do not follow the motion of the
punctures and are evolved at the fixed timestep given by the innermost
fixed level \cite{BruTicJan03,Sch10}. The fixed boxes have twice as many grid
points (for a more accurate wave propagation). Since some runs have
exceptional settings the parameters of our simulations are summarized
in Table~\ref{tab:accuracy}.
For our analysis we measured the GW emission and radiated energy
(normalized to the initial ADM-mass), studied the shape of the event
horizons, the coordinate distance over time and how much time the
binary spends at a separation $D$.
Moreover we investigated a new way of analyzing binary evolutions,
namely to look at the phase space with the coordinate
velocity and the separation of the punctures $(v,D)$ serving as
generalized coordinates. The velocity of the puncture is computed from
the shift as $v = \sqrt{\beta_i(x_p)\beta^i(x_p)}$, where $x_p$ is the
coordinate location of the puncture. $x_p$ is - as a diagnostic -
tracked by integrating $\partial_t x^i_p = -\beta^i(x^j_p)$ using the
ICN method as in \cite{CamLouMar05}.

\subsection{Black hole parameters}

The initial data for black hole binaries is characterized by a choice
of the following parameters.  In this work we set the spins to zero.
Input for the computation of the initial data are parameters $m_i$ for
the (bare) puncture masses, $\vec{P}_i$ for the momenta, and
$\vec{x}_i$ for the positions. The total puncture mass $M$ is defined
as $M=\sum_i m_i$.
Since the global mass scale in vacuum is arbitrary, the masses can be
characterized by one number, say the symmetric mass ratio denoted by $\nu =
m_1m_2/(m_1+m_2)^2$.

\begin{figure}[t]
\centerline{\resizebox{9cm}{!}{
\includegraphics{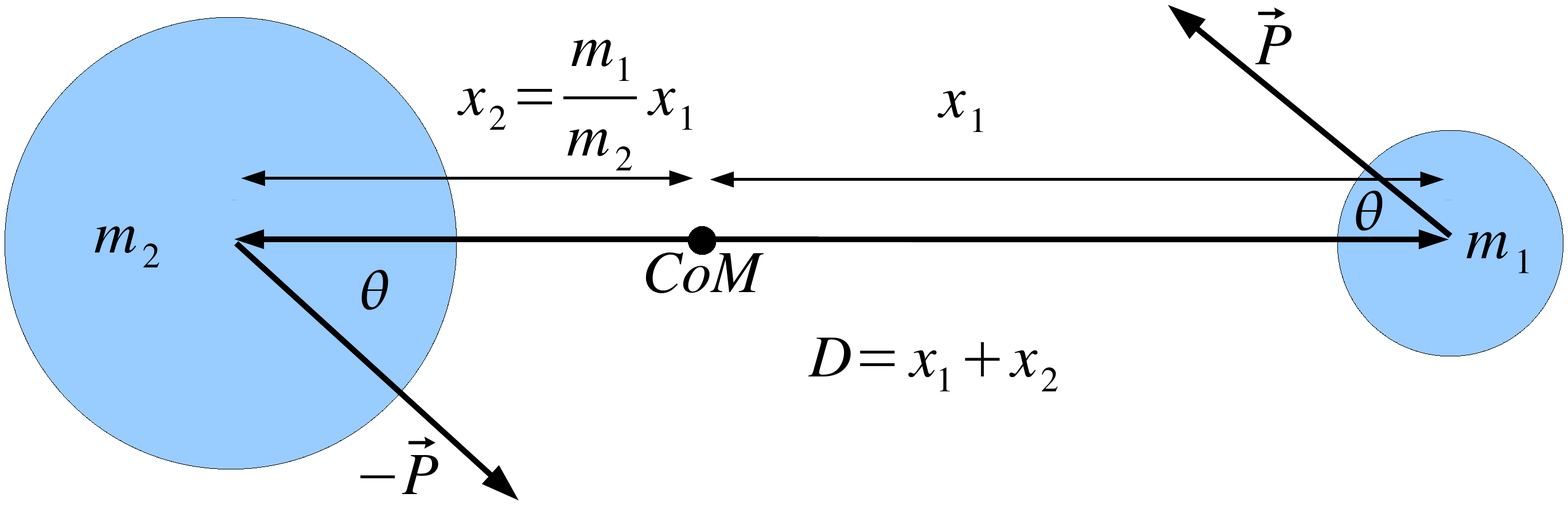}}}
\vspace{-3cm}
\caption{Configuration of initial data.}
\label{fig:Setup}
\end{figure}

We choose coordinates in which the punctures are initially located on
the $x$-axis, see Fig.~\ref{fig:Setup}. For equal masses we set
$x_{1,2}=\pm D/2$ for a coordinate separation $D$. For unequal masses
we leave $x_1$ unchanged but set $x_2=x_1m_1/m_2$. For the
momenta we choose $\vec{P}_{1,2} = \pm \vec{P}$. This implies together
with the choice of $x_{1,2}$ that initially the center of mass is at
rest and that mergers happen at the origin (except for a small merger
kick due to unequal masses). Concretely, we consider momenta in the
$x$-$y$-plane given by their magnitude $P$ and an angle $\Theta$ such
that
$\vec{P} = (-P\cos\Theta, P\sin\Theta, 0)$. 
Specifying the ``shooting angle'' $\Theta$ is equivalent to the choice of
an impact parameter. 
The magnitude $P$ of the momenta is chosen as a multiple of $P_{qc}$,
which denotes the magnitude of the momentum for a quasi-circular
inspiral at separation $D$.

Given the configuration in Fig.~\ref{fig:Setup}, 
numerical simulations are
parameterized by specific choices for $x_1$, $\nu$, $P$, and $\Theta$.
We set $x_1 = 10M$ for all runs, which implies $D=20M$ for
equal masses. For unequal masses, we position the larger
mass at $x_2$, i.e.\ $m_2>m_1$ and $|x_2|<x_1$.  Most of the
simulations we discuss are for equal masses ($\nu = 1/4$), but we also
consider a few examples for mass ratios 1:2 ($\nu=2/9$) and 1:3
($\nu=3/16$). 
%
Following \cite{WalBruMue09}, for equal masses at $D=20M$ the
magnitude of the momentum for quasi-circular inspiral is
$P_{qc}=0.061747M$. We consider $P/P_{qc} = 1$, $2$, $\ldots$, $6$,
and in one example as the most extreme case $P/P_{qc}=10$.
The direction of the momenta is given by $\Theta\in[0,90^\circ]$.
Here $\Theta=0$ corresponds to a head-on collision, while for
quasi-circular inspiraling orbits $\Theta$ is slightly smaller than
$90^\circ$ because the momentum has a small radial component. The case
$\Theta > 90^\circ$ with initially radially outgoing motion can be
ignored \footnote{Generally speaking, if such an orbit is unbound then the
black holes separate without close encounter. If bound, then after
reaching the first apocenter the orbit becomes inward-bound, which in
principle is already included for $\Theta\in[0,90^\circ]$.}.

The ADM mass at the $i$-th puncture and at infinity is
\bea
  M_{ADM}^{i}  &=& (1+u(\vec{x}_{i}))m_i + \frac{m_1m_2}{2D}, 
\\
  M_{ADM}^\infty &=& M_{ADM}^{1} + M_{ADM}^{2} + E_{bind}
\nonumber
\\
  \phantom{M_{ADM}^\infty} &=& m_1 + m_2 + \lim_{r \rightarrow \infty} (2ru),
\label{eqn:MADM}
\eea
respectively, where $u$ is the correction to the conformal factor in the
puncture framework and $E_{bind}$ the binding energy. Values for
$M_{ADM}^\infty$ range from $0.994$ for $1P_{qc}$ to $1.2$ 
for $6P_{qc}$.
Since the momenta are non-zero, we obtain larger physical masses
$M^i_{ADM}$ at the inner asymptotically flat ends of the punctures.
The difference between the masses $m_i$ and $M^i_{ADM}$ ranges from
$7\cdot10^{-3}M$ for $1P_{qc}$ to $3.5\cdot10^{-2}M$ for $6P_{qc}$, and
is essentially independent of $\Theta$.  

For the main part of this work, we first choose a specific mass ratio,
in particular we choose between equal and unequal masses. Second, we
choose one of several (low) momentum cases. Third, we vary the
shooting angle systematically, in particular searching for maxima and
minima in the total radiated energy, examining the number of whirls,
etc. There are some obvious alternatives to set up such parameter
scans, say fixing $\Theta$~\cite{WasHeaHer08}, using some measure
of eccentricity, the angular momentum~\cite{HeaLagMat09}, or the
binding energy~\cite{SpeBerCar07} as parameter.
Apart from having a simple interpretation as scattering
experiment with fixed momentum size, our setup also describes
simulations at roughly constant total energy, if in analogy to
classical point masses the total energy is defined as the sum of the
kinetic and potential energy (since $P$ and $D$ are constant while
varying $\Theta$).
Each run amounts to $500-30000$ CPUh (the latter one for
$1P_{qc},\Theta=60^\circ$), which is strongly dependent on how far and how many
times the orbits zoom out. 
We implemented Brent's method \cite{PreFlaTeu92}
to bracket local extrema in the efficiency of converting energy into
outgoing gravitational radiation, for which a small number of runs
sufficed.  This reduced the total number of runs to about a $130$
while still sampling the parameter space in an adaptive and accurate
way. In retrospect, we found that a golden section search
\cite{PreFlaTeu92} is, for the finite accuracy we
required, a better choice despite being only first order convergent.
The parabolic interpolation inside Brent's method chooses the new
guesses systematically towards the flatter part of the asymmetric
maxima.

\begin{table}[t]
\begin{ruledtabular}
\begin{tabular*}{0.5\textwidth}[b]{@{\extracolsep{\fill}} | c | c | c | c |c |}
  Model & $\frac{\Delta E_{rad}}{E_{rad}}$	& $\Delta A/A$ & Ord. & $t_m\,[M]$\\
  \hline
  $1P_{qc},\Theta=10^\circ$	& 	0.01	& 0.008 & 4 &  40.5 \\
  $1P_{qc},\Theta=40^\circ$	&	0.013	& 0.010	& 4 &  92.7\\
  $1P_{qc},\Theta=52^\circ$	&	0.015	& 0.012	& 4 & 766 \\
  $2P_{qc},\Theta=23.9^\circ$	&	0.017	& 0.016	& 4 & 79.9\\
  $2P_{qc},\Theta=26^\circ$	&	0.0013	& 0.002	& 4 & $\infty$ \\
  $3P_{qc},\Theta=18^\circ$	&	0.0095	& 0.003	& 5 & 69.4\\
  $4P_{qc},\Theta=15.6^\circ$	&	0.0095	& 0.012	& 4 & 39.0 \\
  $10P_{qc},\Theta^D_{50}=5.4^\circ$	&0.03	& 0.012	& 4 & 40.5
\end{tabular*}
\end{ruledtabular}
\caption{\label{tab:accuracy}List of selected runs with corresponding initial
data, order of convergence, estimated errors and merger times. A 3PN
estimate \cite{memmesheimer-2005-71,SpeBerCar07} of our lowest eccentric
run $1P_{qc},\Theta=60^\circ$ gives $e\approx 0.5-0.6$ 
(depending on which definition of $e$ is used).
}
\end{table}


\begin{figure}[h]
\centerline{\resizebox{9cm}{!}{
\includegraphics{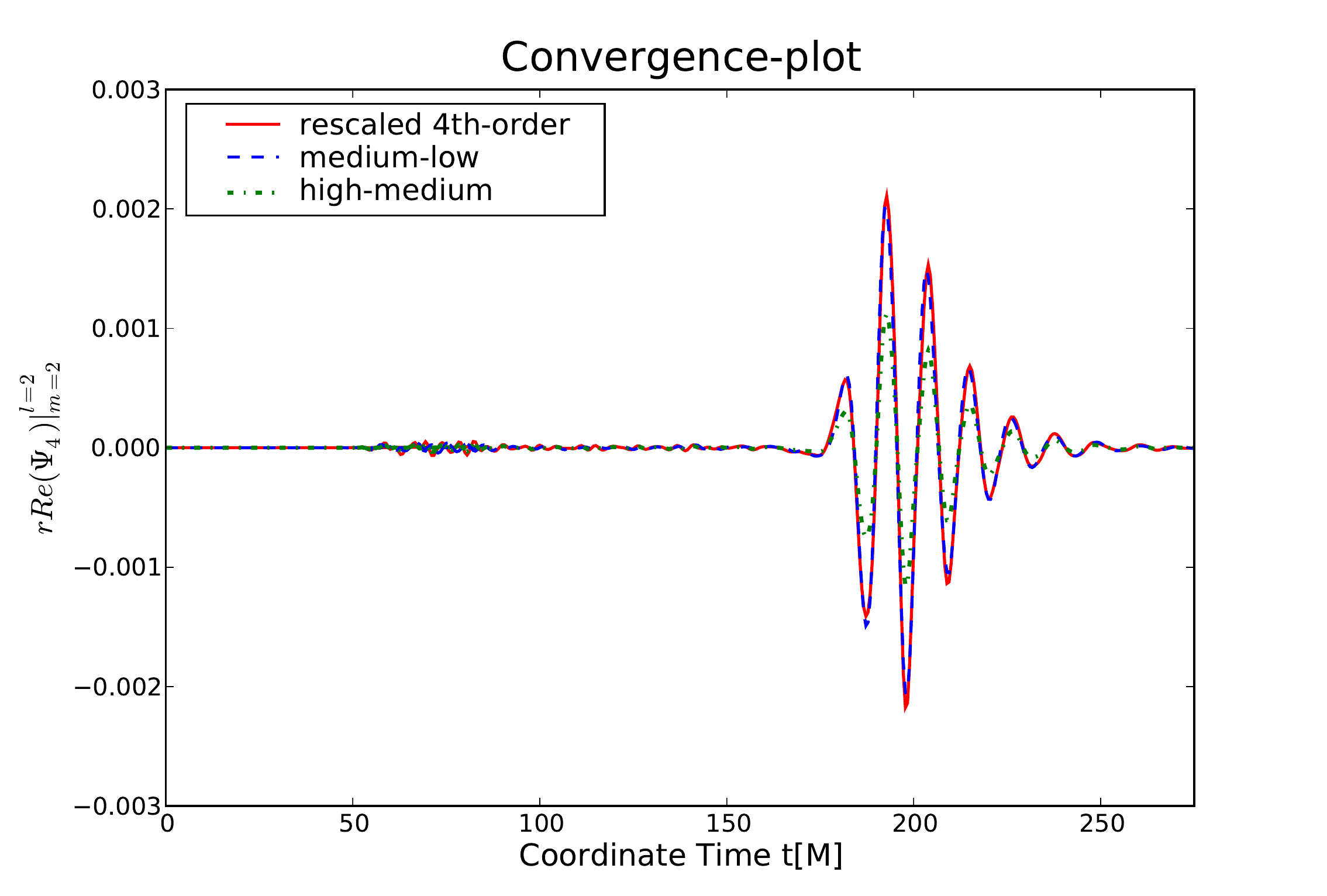}}}
\caption{
  Convergence plot of the
  22-mode of $r\Psi_4$ for $P=1P_{qc}$, $\Theta=40^\circ$. 
  The blue dashed line shows the difference of
  $r\mathcal{R}e(\Psi_4)|^{l=2}_{m=2}$ as computed from the medium and
  low resolution data, and similarly the green (dashed dotted) line for
  high and medium resolution. The solid line rescales the latter one
  assuming 4th order convergence. Both lines lie on top of each other
  demonstrating overall consistency of our results. The small noise at
  the beginning is due to the spurious junk radiation of
  conformally flat initial data.}
\label{fig:rPsi4Convergence}
\end{figure}

\begin{figure}[h!]
\centerline{\resizebox{9cm}{!}{
\includegraphics{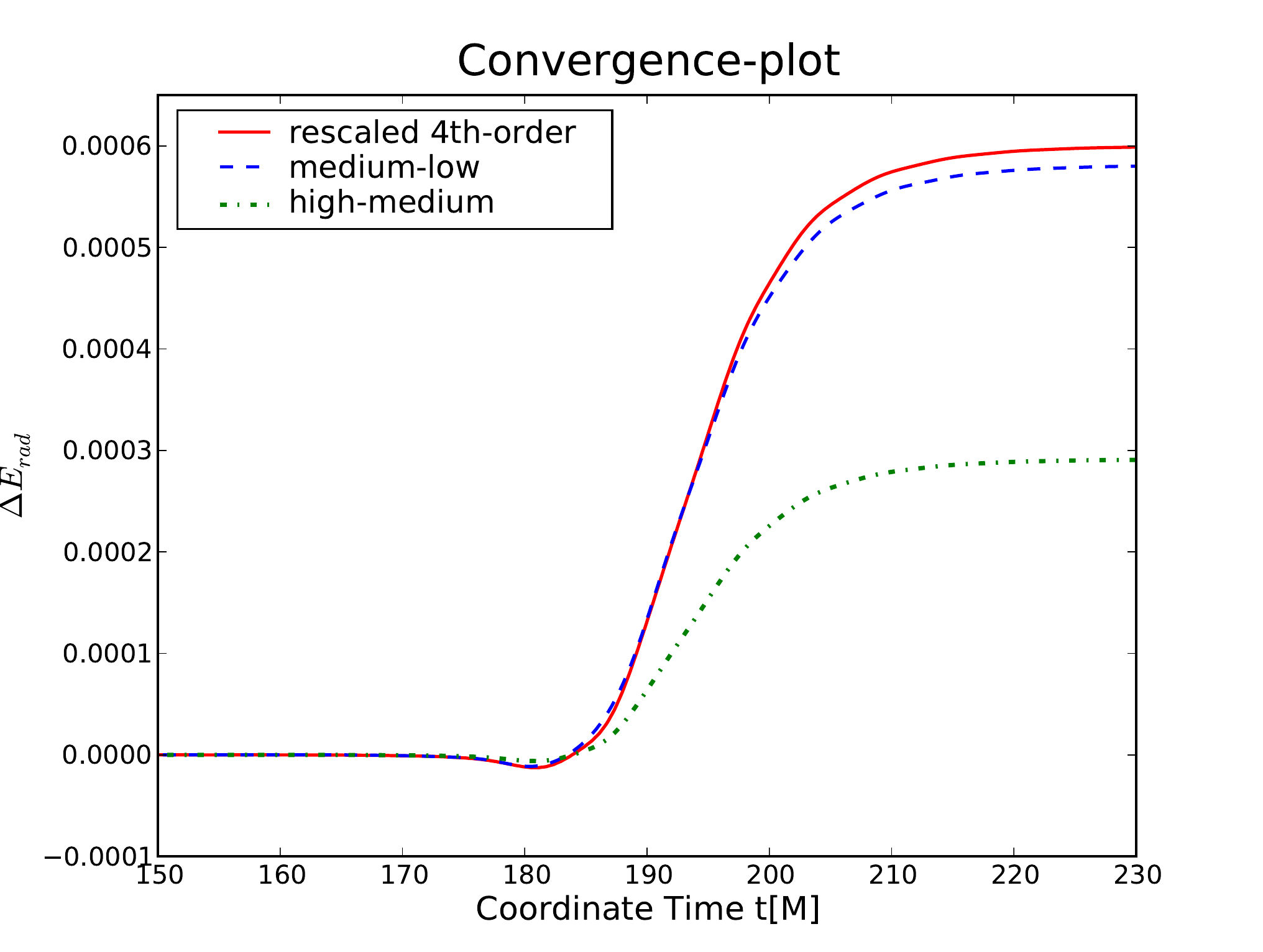}}}
\caption{Convergence plot of the radiated energy 
  for $P=1P_{qc}$, $\Theta=40^\circ$. The solid red
  line and the dashed blue one are almost on top of each other,
  demonstrating a convergence order very close to $4$ consistent with our
  numerical scheme.}
\label{fig:EConvergence}
\end{figure}


\subsection{Convergence and error estimates}
\label{Error}

We performed a convergence analysis for a representative subset of our
runs and in general found $4^{th}$ order convergence in the $22$-mode of
$r\Psi_4$ and in the radiated energy $E_{rad}$ demonstrating the
overall consistency of the code with respect to the order of the
Runge-Kutta integrator and the wave extraction routine. The errors due
to the finite radius of our wave extraction sphere are quantified by
the deviation from a $1/r$ fall-off as measured from the data taken at
three different extraction radii.

Error estimates based on this analysis are shown in
Tab.~\ref{tab:accuracy}, and selected convergence plots are shown in
Figs.~\ref{fig:rPsi4Convergence}, and \ref{fig:EConvergence}.
In general, highly eccentric orbits are
accurately treated by the BAM code, and also the presence of the
rather high momenta considered here can be dealt with consistently.
The relative errors in the 22-mode and the radiated energy both due to
finite resolution and extraction radius is around
$1\%$. 
For larger momenta the error from a finite extraction radius becomes
the dominating error ($\approx 2\%$). Increasing the initial momentum
leads to higher amounts of artificial (junk) radiation, enhances the ADM mass of
the initial time slice and reduces the BH horizons.  At some point
these effects contaminate the solution in the sense that its physical
relevance becomes questionable. However, we limited our data set to
those regimes where the artificial radiation is either entirely negligible or at
least small in comparison to the physical radiation. The $P=10P_{qc}$
sequence represents an exception, but we only used it in the context of
the Hawking limit and as an approximate extrapolation of our data set to
the ones obtained by other groups at larger momenta.

The error due to the junk radiation arising from the conformally
flat initial data can be reduced to some extent by choosing a
sufficiently large initial separation. In the very high momentum case
we needed both large separations and also much larger resolution until
the radiated energy results converged, but once the appropriate
resolution is used the accuracy compares favourably with the other
results. For yet higher momentum runs we refer to
\cite{SpeCarPre08,ShiOkaYam08}, who study momenta beyond the
rest-mass dominated regime.

Convergence for unequal masses can be shown only at the higher
resolutions. For the resolution we used the $l=2$, $m=2$ mode
converges at second order -- a common tendency when being at the edge
of the convergent regime. The somewhat lower accuracy for larger mass ratios
is a well-known effect of the gamma-driver condition we use
($\eta=const$). In \cite{MueBru09,MueGriBru10,Alic:2010wu} it is shown that a
generalization of this condition ($\eta$ dependent on
the local mass) leads to an improvement in accuracy.

Analyzing the dependence on resolution shows that the derived
errors in the energy are not behaving according to a Gaussian
distribution. There is a skewness in the actual (unknown)
distribution of our measurements such that higher resolutions
systematically produce higher energies. Hence, our errorbars should be
slightly more extended towards larger values of $E_{rad}$.

Summarizing, the simulations presented here 
do not pose new challenges to the numerical scheme, although there
are specific requirements for accuracy in the presence of whirls
together with long runtimes. In these cases there is a high
sensitivity to the parameters and during the long evolutions numerical
errors accumulate.  Nevertheless, those evolutions have similar
convergence behavior and error estimates, and only require a higher
resolution to obtain convergence.

%
\section{Results}
\label{Results}
%

%
%

\begin{figure}[t]
\centerline{\resizebox{8cm}{!}{
\includegraphics{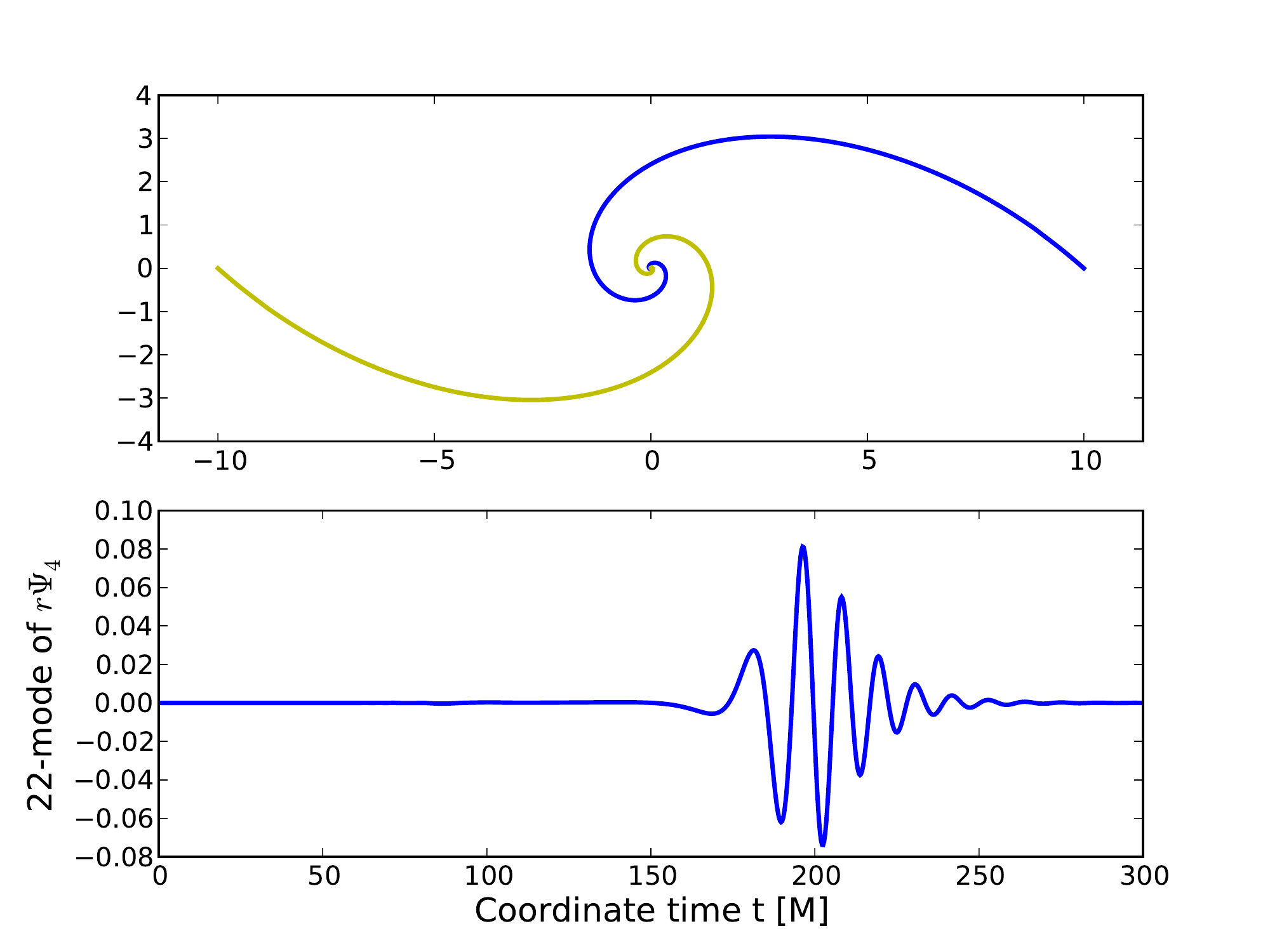}
}}
\caption{ 
  $P=P_{qc}$, $\Theta=42^\circ$.  
  From $\Theta=0^\circ$ to about $40^\circ$, the black holes collide
  within $200M$ of evolution time and perform less than one orbit 
  before merger.  The waves show a merger signal with a brief 
  ring-up and characteristic ring-down. In the limit of the head-on 
  collision, $\Theta\rightarrow0^\circ$, the $22$-mode vanishes and 
  the $20$-mode becomes the dominant mode.  }
\label{fig:1qcA}
\end{figure}

\begin{figure}[t!]
\centerline{\resizebox{8cm}{!}{
\includegraphics{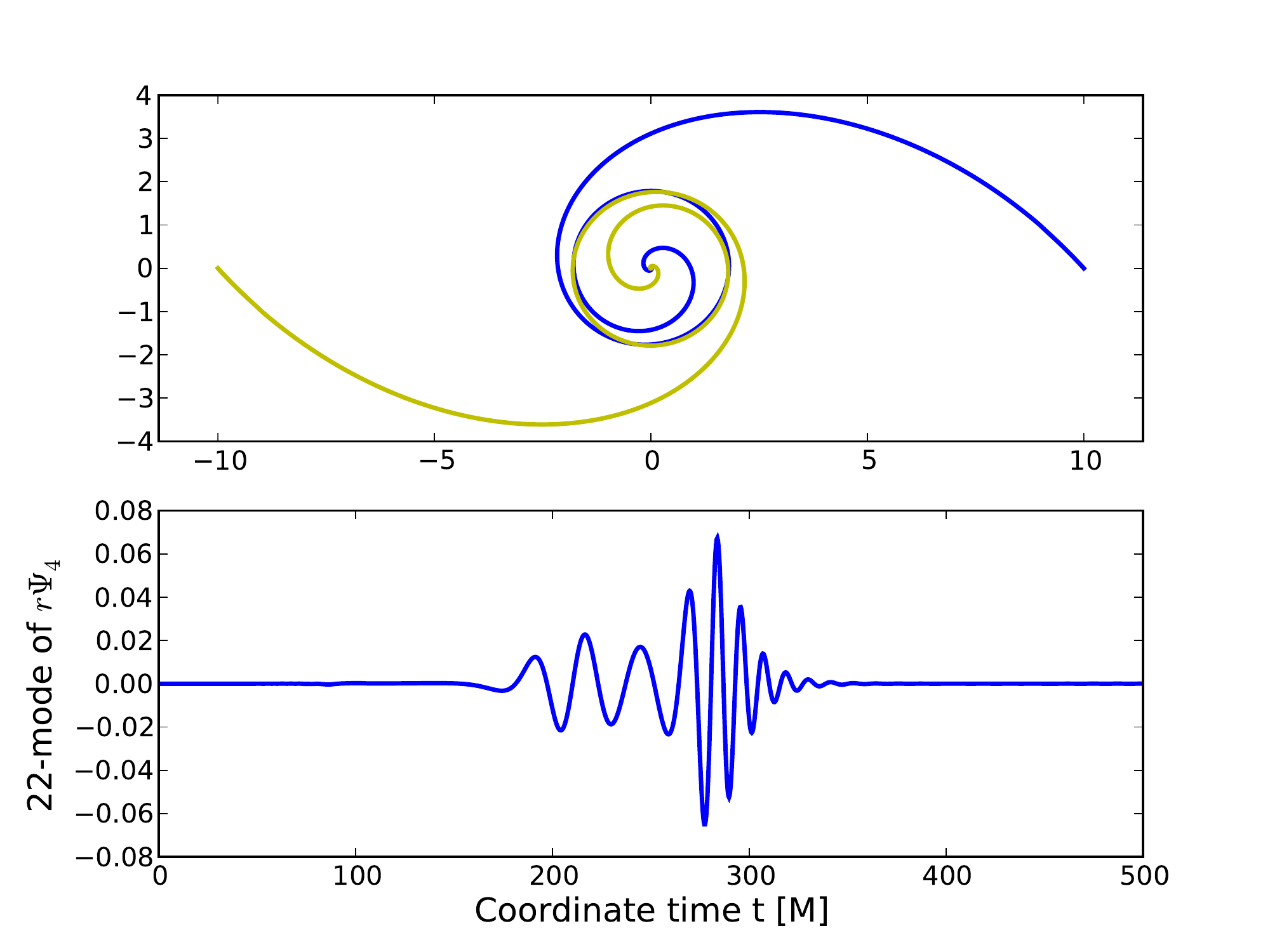}
}}
\caption{
  $P=P_{qc}$, $\Theta=48^\circ$.
  There is about one full whirl in a range of $\pm1^\circ$ around this
  shooting angle followed by the merger. The waveform clearly shows a
  wave associated with the whirl. Its amplitude is smaller than
  that of the ensuing merger signal. The diameter of the whirl is
  smaller than the innermost stable circular orbit of a 
  Schwarzschild BH with the same total ADM mass.}
\label{fig:1qcB}
\end{figure}

\begin{figure}[h!]
\centerline{\resizebox{8cm}{!}{
\includegraphics{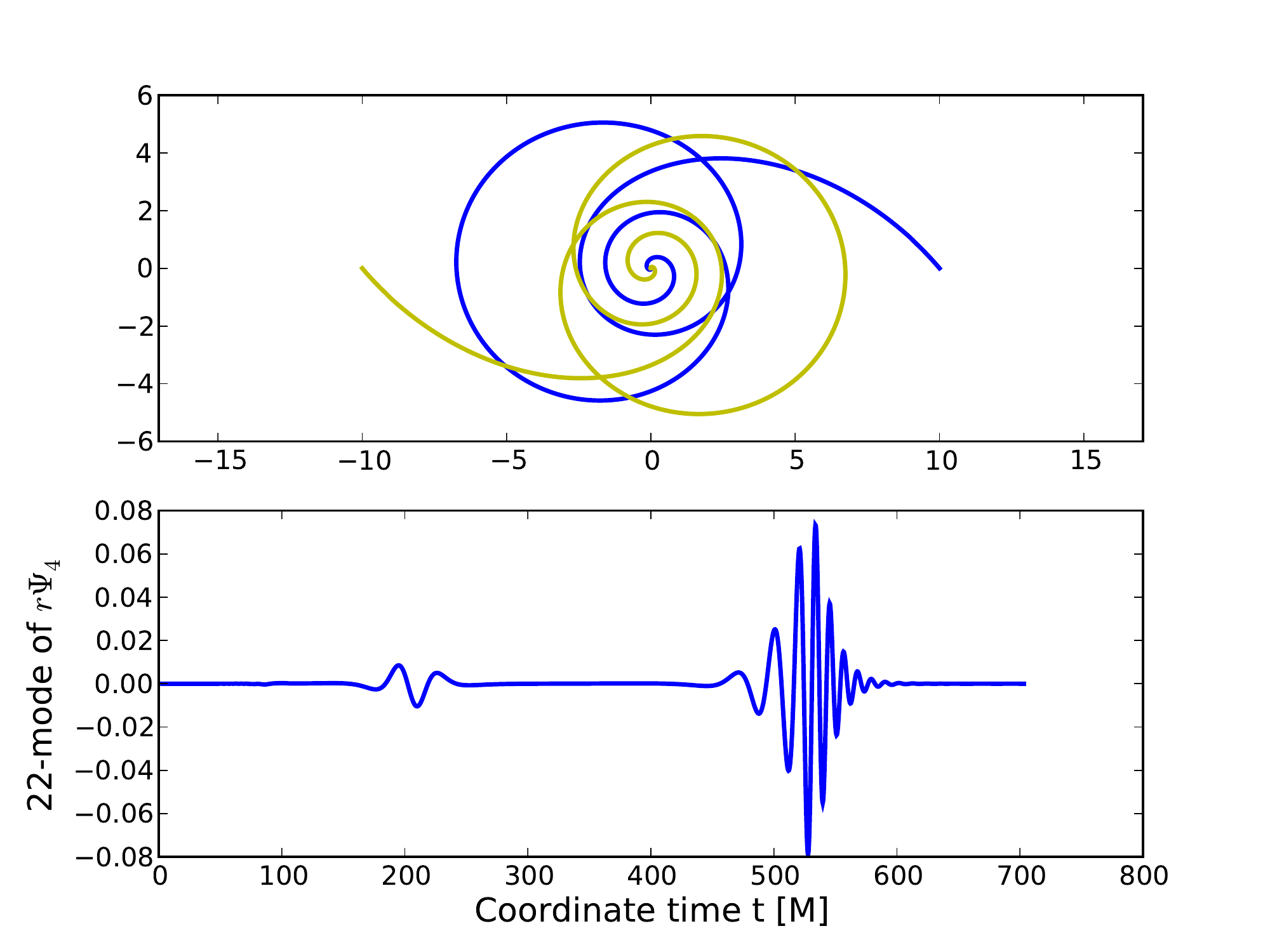}
}}
\caption{
  $P=P_{qc}$, $\Theta=50^\circ$.
  For a shooting angle two degrees larger than that leading to the
  {strong whirl, there is a close encounter with a precession 
  of about half an orbit}, followed by a zoom out
  to about three times the {radius at pericenter}, followed by a short
  inspiral and merger that starts with significantly reduced
  eccentricity. Note the comparatively small and short wave pulse
  associated with the {close encounter}, again at about $200M$ of evolution.  }
\label{fig:1qcC}
\end{figure}

\begin{figure}[h!]
\centerline{\resizebox{8cm}{!}{
\includegraphics{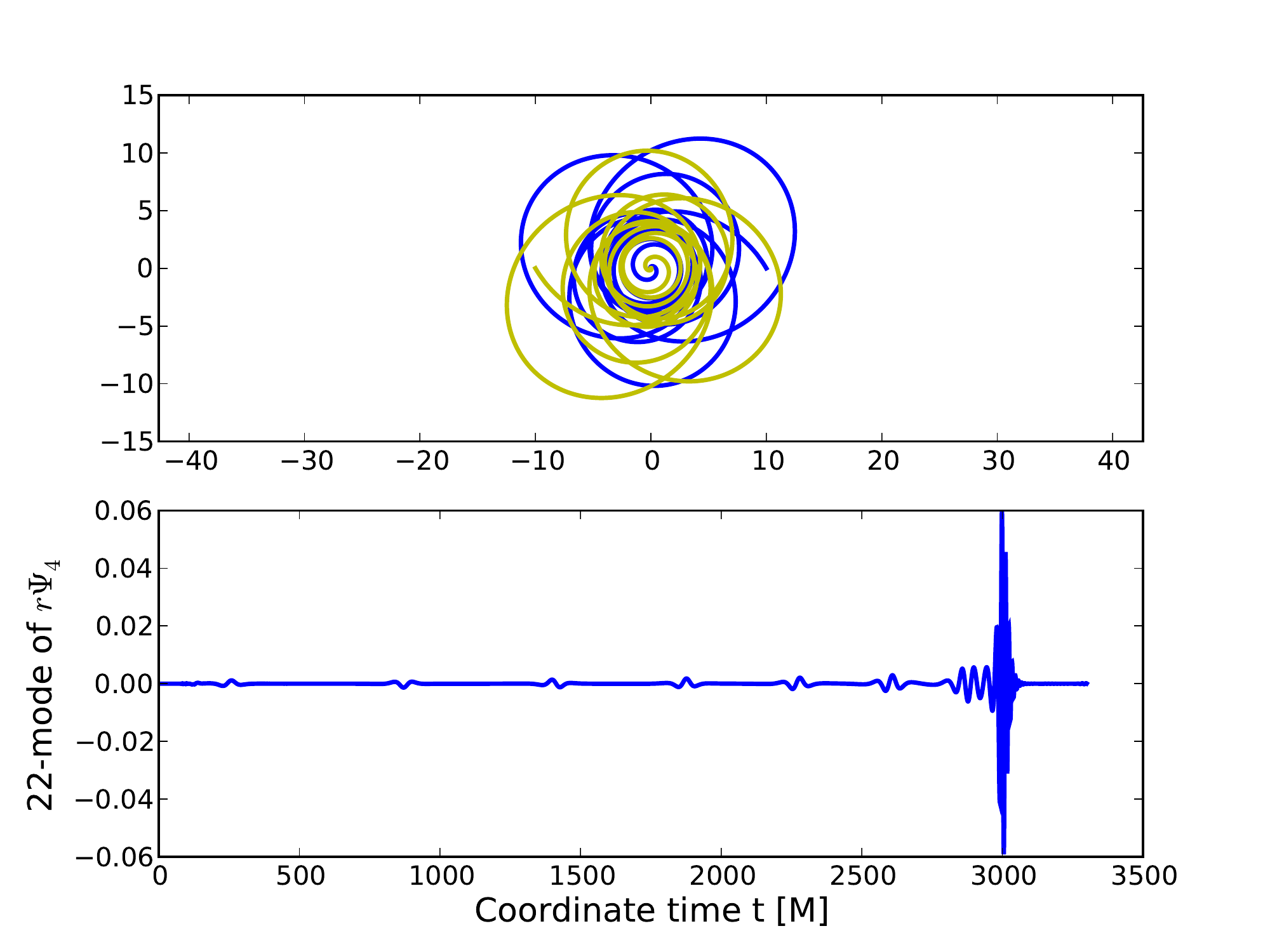}
}}
\caption{
  $P=P_{qc}$, $\Theta=60^\circ$.
  Increasing the shooting angle beyond $50^\circ$, one can find an
  increasing number of 
  elliptic orbits. Early on the orbit resembles the classical picture
  of a (strongly) precessing ellipse. The plot shows a transition 
  through plunge through a full whirl phase at the onset of merger 
  with a clear corresponding wave signal. }
\label{fig:1qcD}
\end{figure}

\begin{figure}[h!]
\centerline{\resizebox{8cm}{!}{
\includegraphics{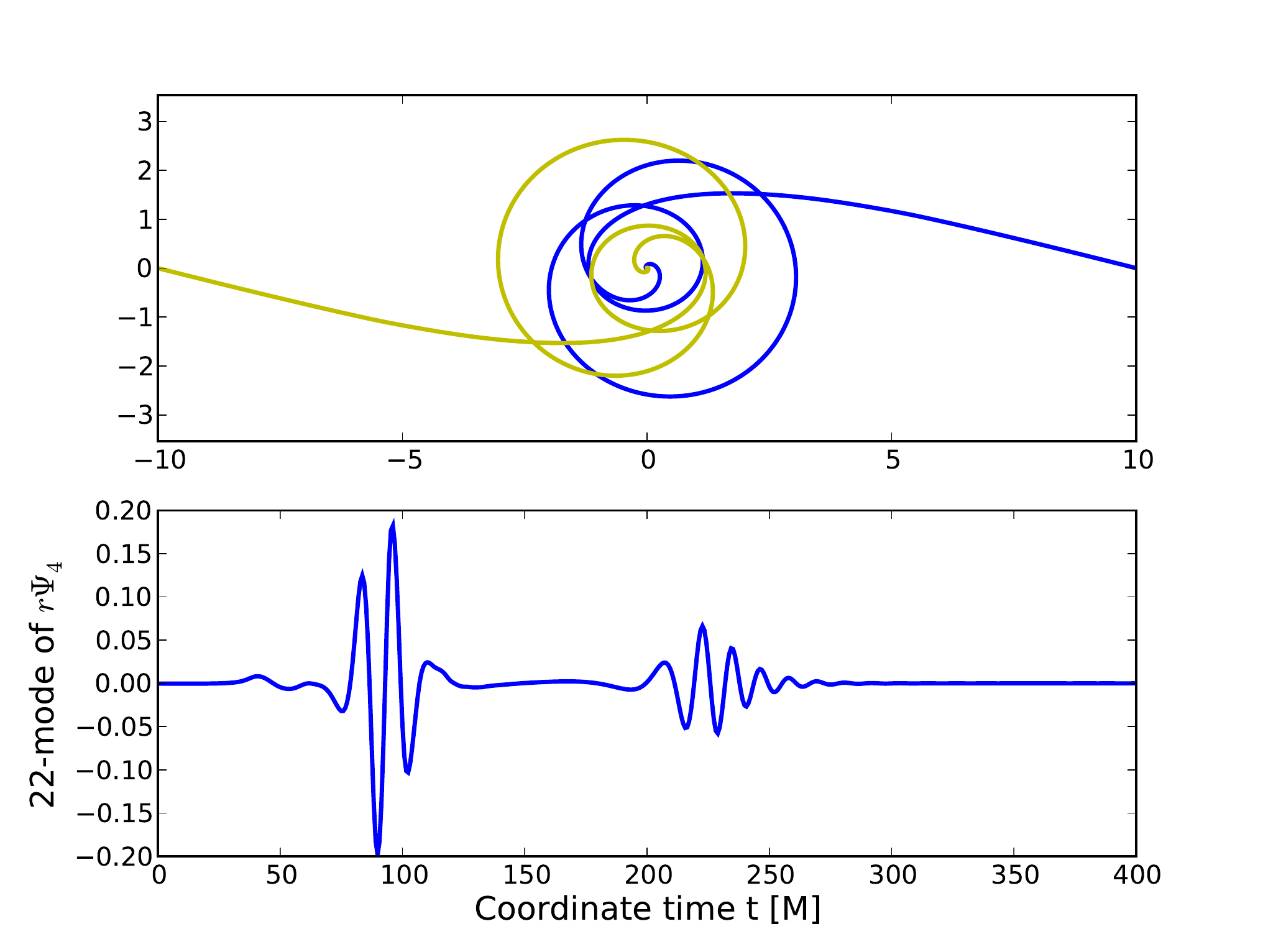}
}}
\caption{
  $P=5P_{qc}$, $\Theta=14.15^\circ$.
  Example for initial momentum which is significantly larger than that of
  quasi-circular orbits, and which can easily produce unbound orbits.
  Zoom-whirl orbits are found for much smaller shooting angles than in
  Fig.~\ref{fig:1qcA} -- \ref{fig:1qcD}. There is one whirl, and a
  short zoom followed by a merger. Due to the additional kinetic
  energy, the whirl signal increases in amplitude and exceeds the
  merger signal. }
\label{fig:5qcA}
\end{figure}

\begin{figure}[h!]
\centerline{\resizebox{8cm}{!}{
\includegraphics{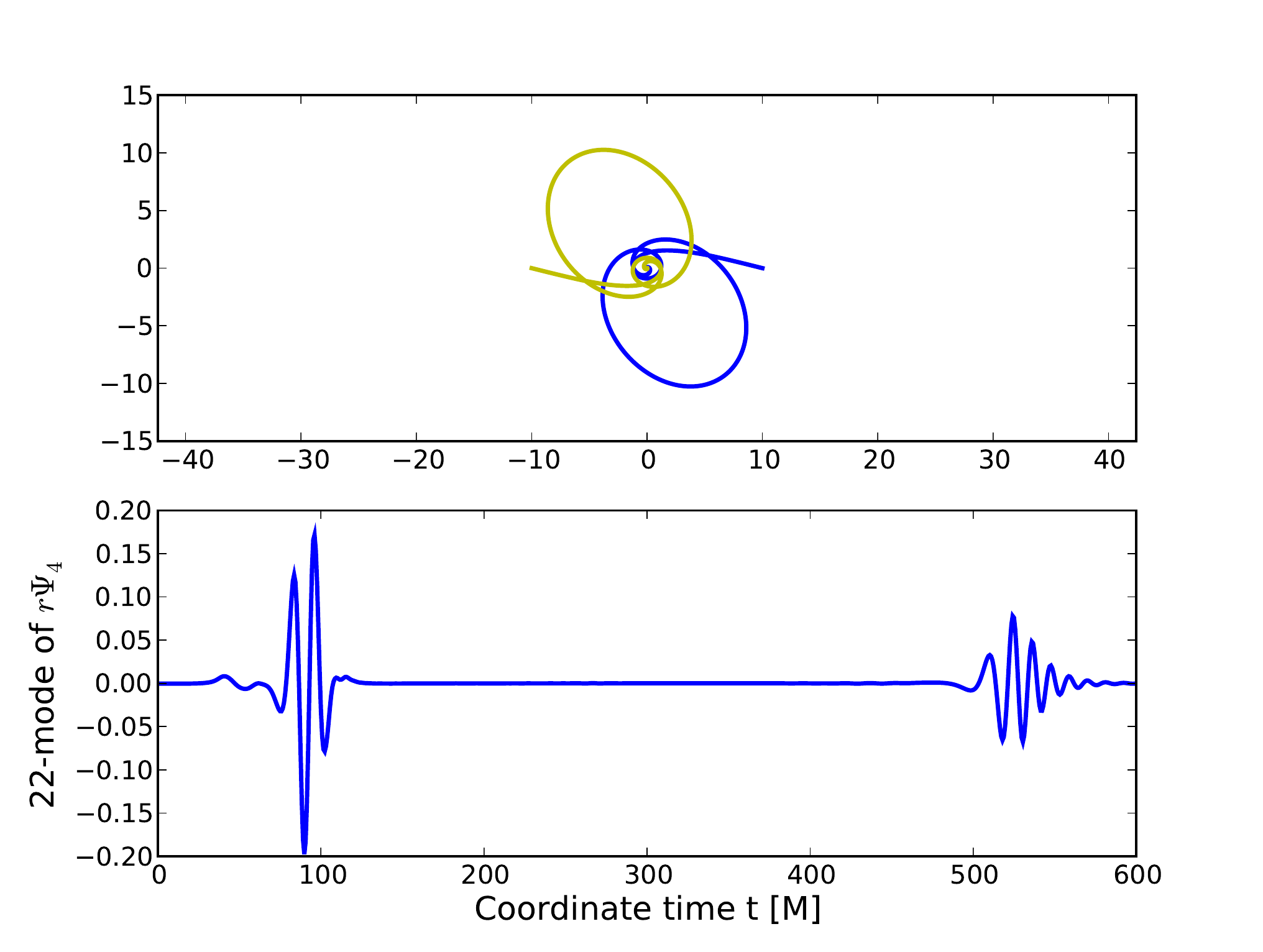}
}}
\caption{
  $P=5P_{qc}$, $\Theta=14.20^\circ$.
  The larger the momentum, the more sensitive the orbit becomes to the
  choice of the shooting angle. A small change in angle compared to 
  Fig.~\ref{fig:5qcA} leads to a much larger zoom out to an
  apocenter distance of $12M$ before it merges at the next encounter. 
  The initial whirl, however, is almost unchanged,
  highlighting the analogy to unstable circular orbits.
}
\label{fig:5qcB}
\end{figure}

\begin{figure}[h!]
\centerline{\resizebox{8cm}{!}{
\includegraphics{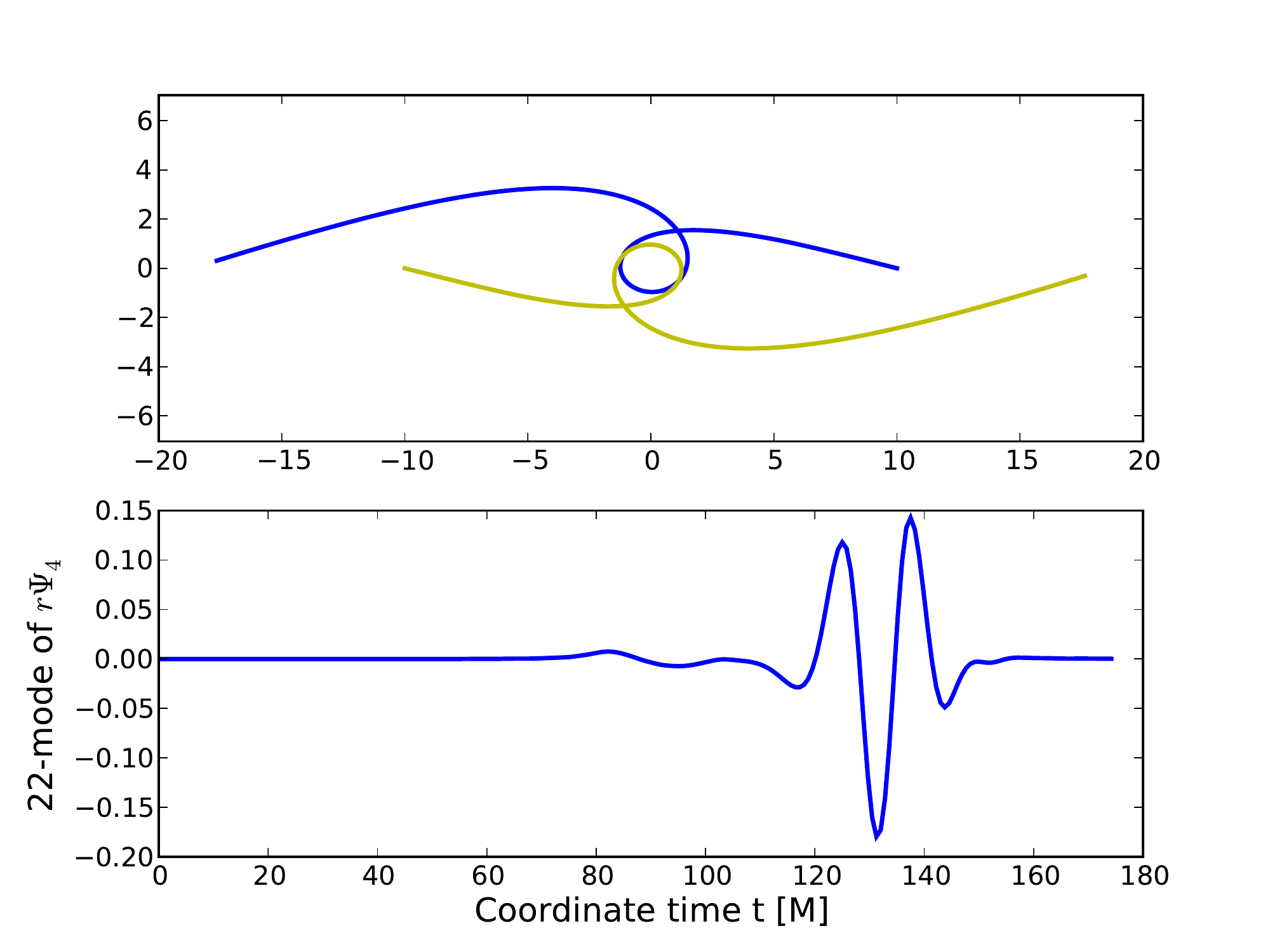}
}}
\caption{
  $P=5P_{qc}$, $\Theta=14.30^\circ$.
  Enlarging the shooting angle further compared to Fig.~\ref{fig:5qcB}
  results in a full whirl followed by a zoom to infinity (unbound orbit, no
  merger).  }
\label{fig:5qcC}
\end{figure}


\subsection{Orbital properties}
\label{Orbits}

To prime the discussion of the orbits, we first consider several
examples of puncture tracks for equal mass binaries with $P=P_{qc}$
and $P=5P_{qc}$, see Figs.\ \ref{fig:1qcA} -- \ref{fig:5qcC}. It is helpful 
to read the captions of these figures in sequence. Shown
are the puncture tracks in the $x$-$y$-coordinate plane in the upper
panels, and the $22$-mode of the waveforms in the lower panels. The
waveforms are further discussed in Sec.~\ref{Radiation}. The figures show two
sequences of runs for two momenta that explore how
the orbits change when the shooting angle is varied from small to
large.

\begin{figure}[t]
\centerline{\resizebox{10cm}{!}{
\includegraphics[width=\textwidth]{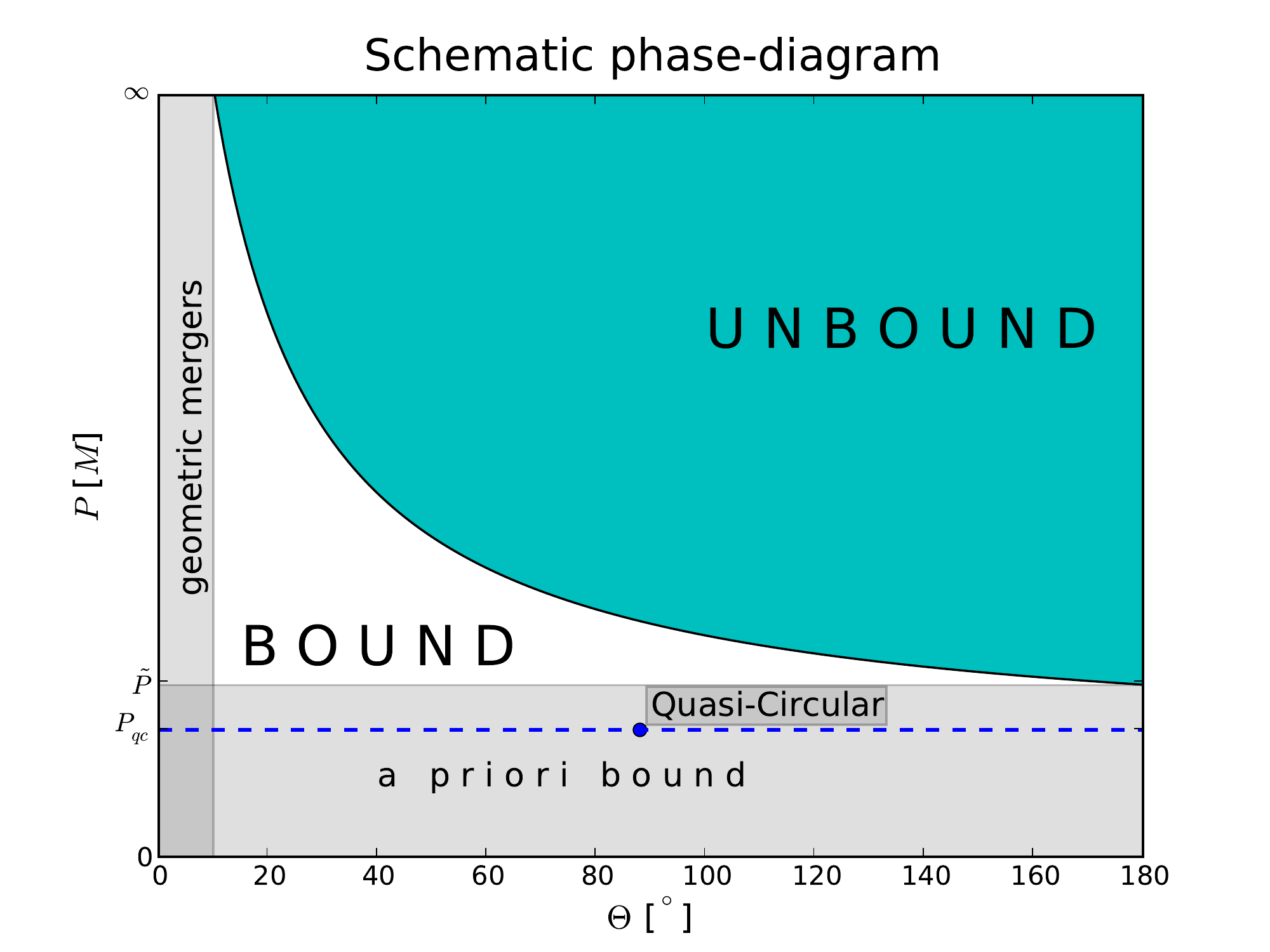}
}}
\caption{
	  This plot sketches the end state of eccentric black hole binaries 
          in the plane spanned by our parameter choice for the initial data. 
          The evolutions located in the grey shaded regions can be judged bound 
          solely based on the initial data.
}
\label{fig:PhaseDiagramSchematic}
\end{figure}

\subsubsection{Classification of orbits}
\label{Classification}

For any choice of mass ratio $\nu$ and initial separation $D$, we can
in principle fill in a ``phase-diagram'' as shown in
Fig.~\ref{fig:PhaseDiagramSchematic}, which labels orbits in a
$P$--$\Theta$ plot. The main classification is whether initial
parameters $P$ and $\Theta$ lead to orbits that are bound (implying
capture and merger) or unbound (escape to infinity). 
In Newtonian gravity, we only have to check whether the kinetic energy
exceeds the potential energy, or equivalently whether the binding
energy is positive or negative.
In general relativity, this distinction is sometimes only possible a posteriori
since the gravitational waves and the associated loss of energy and
angular momentum are only known after the Einstein equations have been
solved.  Solutions to the evolution problem define the dividing line
$P=P_{bu}(\nu,D,\Theta)$
in Fig.~\ref{fig:PhaseDiagramSchematic}. Orbits with $P>P_{bu}$ are
unbound, orbits with $P<P_{bu}$ are bound. 

A simplified, a priori upper limit on the momentum $P$ that ensures
boundedness is $\tilde{P}_{b}:=P_{bu}(\nu,D,\Theta=180^\circ)$, which
is independent of $\Theta$, cmp.\ Fig.~\ref{fig:PhaseDiagramSchematic}.
If the momentum $P$ does not suffice to escape in the direction
$\Theta=180^\circ$ (for which radiation losses are minimized), then
the orbits are bound for all $\Theta$. Here we use the assumption that
the black holes are not spinning. 
Approximating the minimal radiation loss in the ``head-off'' direction
by zero, we compute a simple estimate of $\tilde P_{b}$ based on the 
binding energy in (\ref{eqn:MADM}).
Fixing $\Theta=180^\circ$ and $D=20M$ we iteratively compute initial
data with varying $P$ to obtain the binding energy
$E_{bind}$. $\tilde{P}$ is then defined as
$\tilde{P}:=P(E_{bind}=0,\Theta=180^\circ)$ resulting in
$\tilde{P}=0.085(4\pm3)M$. There is some error
since we end the iteration at some point and since 
radiation effects are ignored.
For example, the momentum of the quasi-circular
orbit leads to bound orbits for all angles since
$P_{qc} < \tilde{P} \approx 1.377P_{qc}$.

For brevity, we will refer to the set of configurations satisfying 
$P<\tilde{P}$ as the \emph{elliptic} regime. On the other hand, orbits 
with $P>\tilde{P}$ form the \emph{hyperbolic} class. 
Note that this terminology skips over the fact that orbits in the
hyperbolic regime may still lead to a merger provided $\Theta$ is small
enough.

Inverting $P_{bu}(\nu,D,\Theta)$, we define $\Theta_{bu}(\nu,D,P)$,
the shooting angle between bound and unbound orbits, as a function of
$P$. An a priori upper limit for bound orbits is given by
$\Theta_{geom}=\Theta_{bu}(\nu,D,P=\infty)$, see
Fig.~\ref{fig:PhaseDiagramSchematic}.
In practice it is tricky to study large $P$
due to limitations in the construction of initial data.
Conceptually, however, we can think of this limit as a geometric
constraint based on the finite size of the black holes, i.e.\ the idea
is that the two black holes must merge when their event horizons
touch.  Using Euclidean geometry, $\Theta_{geom}$ is given by
$\sin(\Theta_{geom}) = d_{merger}/D$, where $d_{merger}$ is the
separation of the punctures at the time of the merger. However, the
size of the black holes depends on the gauge. The Schwarzschild radius
for a mass $m$ is $2m$ in Schwarzschild coordinates, $m/2$ for
isotropic coordinates, and depending on the moving puncture gauge
somewhere in between for the numerical evolutions.
We therefore use the numerical result for $d_{merger}$. For equal
masses we find that a common event horizon appears at a coordinate
distance of about $d_{merger} \approx 1.76$ to $1.95M$ (with a slight
drift towards smaller values with increasing momentum). For an 
initial separation of $D=20M$ This estimate leads to a geometric 
limit of $\Theta_{geom} = 10.5^\circ$ using the Euclidean formula.
This limit does not appear to be very restrictive for low momenta, but
it is not in contradiction to the runs of this study, either. All our
simulations with $\Theta < \Theta_{geom}$ end in a merger.

The determination of the ultimate fate of a system outside the above 
ranges requires a full numerical evolution. 
Here a bound system can be defined by the (future) formation of a single
event horizon, which is expensive to compute numerically. In our
evolutions we use a criterion on the lapse at the center of our grid
to determine a merger time. We justify this approach by a direct
comparison with an event horizon finder \cite{Thi08,ThiBru12}. 
The merger time $t_m$ is approximated by $t^{\alpha}_{m}$, the time by
which the lapse at the center of our grid has dropped below
$\alpha=0.3$. This is near the analytical value of a single
Schwarzschild black hole in the same and similar
gauges~\cite{HanHusPol06,ReiBru04,Alc08}.
We have chosen a moderately long evolution among the elliptic category 
and get $t_m=484.175M$ and $t_m^{\alpha}=485.524$, accurate to within 
$\Delta t_m/t_m = 0.0028$. We should mention, however,
that the lapse criterion gives worse answers when the punctures move too fast,
because the value $\alpha=0.3$ is motivated by a Schwarzschild spacetime and
hence is not well adapted to a boosted black hole. We used the lapse
criterion to estimate the merger times and list them in Tab.~\ref{tab:accuracy}.
Those values are also used in 
Figs.~\ref{fig:PunctureTracks2qc} and \ref{fig:Hist}.

Even if one performs a numerical evolution it can be difficult to determine 
whether an orbit is unbound. The absence 
of a common horizon is only a necessary but not a sufficient 
condition for unboundedness. If a merger does not
occur after a given finite time, the question is for how long the 
simulation has to be continued to settle whether the binary is bound
or unbound, and in principle this time can be infinite.
A practical, approximate criterion can be given in terms of the
initial binding energy $E_{bind}$ and the energy radiated in GWs (see
Sec.~\ref{Radiation}) during the first encounter.
Without gravitational radiation $E_{bind}$ is a constant of motion and 
the orbits are unbound for $E_{bind}>0$ and bound if $E_{bind}<0$. 
We find, unsurprisingly, that all orbits with $E_{bind}<0$ also merge 
in our evolutions. 
We judge an orbit to be dynamically captured when the energy radiated
during the first encounter exceeds the initial (positive) binding
energy. This shortens the runtime to determine whether a run is unbound 
significantly because we do not have to track the black holes to
larger and larger distances. Such a criterion is applicable close to
the threshold between bound and unbound runs, although a few 
marginally bound runs may be incorrectly labeled unbound 
(but runs are labeled bound correctly).

We conclude with remarks on the relation to periodic orbits.
Within the category of bound orbits there is a detailed classification
scheme based on periodic orbits which is complete when neglecting
radiation effects. In this classification
\cite{LevGiz08,LevGro08} one indexes all closed orbits with a
triplet of integers ($z,w,v$), where $z$ is the number of zooms within
an approximate $2\pi$ period (i.e.\ the number of ``leaves''), $v$ is
the stride over the leaves ($1 \leq v \leq z-1$), and $w$ is the
number of whirls. The total precession angle is $2\pi(w+\frac{v}{z})$.
The question is whether this classification still works in an
approximate sense for BHBs with radiation effects.
Especially near the merger of comparable
mass BHBs, the orbits shrink significantly and may not be well
represented by a single periodic orbit, but rather by a sequence of
them.
Our findings imply that the longest whirls associated with the largest
precession angles (largest $w$) occur for momenta with $P$ slightly
larger than $\tilde{P}$ and are very close to a precession of $2\pi$.
We also find that the dependence on $P$ is weak and beyond
$P\gtrsim2P_{qc}$ compatible with the statement that it only depends
on the mass ratio. Radiation damping seems to limit the
length of the whirl phase for larger $P$, although there may be 
artifacts due to the initial data. In terms of periodic tables
this means that we typically find preferred subsets of periodic orbits
that approximate our evolutions best. The number of whirl-orbits $w$
is clearly limited by the efficiency of gravitational radiation. For
equal masses $w=1$ seems to be the largest $w$ one can obtain. For
larger mass ratios $w=2$ should also become possible somewhere beyond
a mass ratio of 1:3. In the regime we are probing orbits with
$z=2,z=3$ and $v=1$ are favored. However, our data set contains too
few data points on different mass ratios to make a strong statement.

\subsubsection{Examples for orbital dynamics of BHBs}
\label{OrbitalDynamics}
 
We describe the main aspects of the orbital dynamics that we find in our
data set using the categorization introduced in the previous section.
First we consider equal mass BHBs in the elliptic regime. 
All equal mass runs start at $D=20M$ ($P=10P_{qc}$ has $D=50M$) in
such a way that $D$ shrinks. Obviously, the ensuing evolution depends
on the values of $P$ and $\Theta$.

\begin{figure}[t]
\centerline{\resizebox{9cm}{!}{\includegraphics{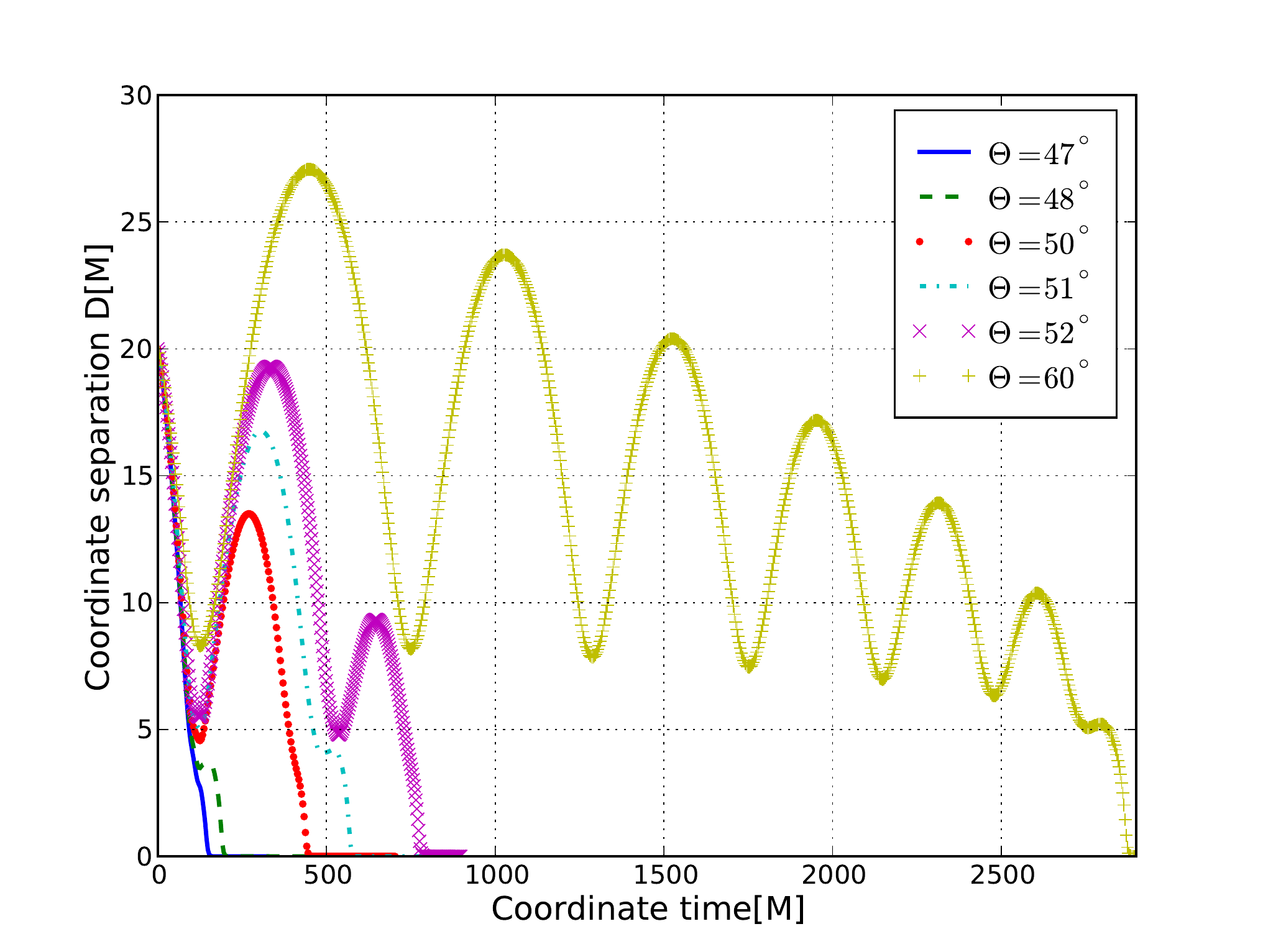}}}
\caption{Coordinate distance in the $P=1P_{qc}$ sequence as a function
  of time. For larger $\Theta$ more eccentric orbits in the inspiral 
  become possible before merger. We simulated up to $6(+1)$ orbits 
  for $\Theta=60^\circ$. 
  Note the plateaus due to whirls before merger in the 
  $48^\circ$, $51^\circ$, $60^\circ$ cases in contrast to the short 
  plunges in the other cases.
}
\label{fig:CD}
\end{figure}
%
We discuss the orbital dynamics
from low to high $\Theta$ for $P=P_{qc}$. Puncture tracks for some
values of $\Theta$ are shown in Fig.~\ref{fig:1qcA} -- \ref{fig:1qcD},
while Fig.~\ref{fig:CD} shows the coordinate distance $D$. The insets of 
Fig.~\ref{fig:Summary1qcWithOrbits} show puncture tracks for some additional 
values of $\Theta$.

At low $\Theta$ (or equivalently for high eccentricities)
$D$ monotically shrinks leading to a rather prompt merger
without completing a single orbit. The runs with larger $\Theta$ have
correspondingly higher initial orbital angular momentum and manage to
resist the strong gravitational pull for longer so that the merger
time steadily grows. For $\Theta\approx46^\circ$ the punctures
complete one orbit before merger.  At yet larger
$\Theta\gtrsim48^\circ$ the orbits begin to exhibit a circular phase
(the whirl) which is maintained for longer as $\Theta$ is increased.
However, at $\Theta\approx48.5^\circ$ the orbit leaves the circle
again towards larger radii (the zoom) delaying the merger
significantly.  In this range of $\Theta$ there is high sensitivity to
the initial data (concerning merger time as a function of $\Theta$). A
mild increase in $\Theta$ leads to a much larger $t_m$ because the BHs
slow down as they move out before falling back. In the limit
$\Theta\rightarrow 90^\circ$ the pericenter passages become shorter
while the apocenters and pericenters become increasingly degenerate. The
pericenter moves out with $\Theta$, hence the BHs do
not cross their mutual gravitational potential as deeply and
consequently not as much radiation
occurs, enabling more and more orbits before merger. 

Concerning the amount of precession, we see that although our
evolutions start somewhere beyond the 
apocenter (e.g.\ Fig.~\ref{fig:Summary1qcWithOrbits}), 
the orbits exhibit a huge precession of roughly $\pi$ and 
close to $~2\pi$ for $\Theta=48.5$ (followed by a tiny zoom).  Even
for the smallest eccentricity we studied 
$\Theta=60^\circ$ ($e\sim 0.5$) one can see that the 
ellipses still have precessions as large as
$2\pi/3$, meaning that over the course of the whole evolution the
accumulated precession amounts to more than two entire orbits. Those
values by far exceed the amounts of precession known from mildly
relativistic systems like the famous Hulse-Taylor pulsar
\cite{HulTay75} with a precession of $q=0.0037^\circ$ per orbit or the
binary pulsar \cite{Bur03} with $q=0.0044^\circ$ per orbit.
Furthermore for these systems and their correspondingly milder
gravitational radiation there is precession not only from
one apocenter to the next but also precession of the multi-leaf clover
as a whole very similar to the findings in studies of periodic orbits
\cite{LevGiz08,LevGro08} (or nearby aperiodic orbits).
This additional, peculiar precession effect is indeed small for
our evolutions as well, though not negligible.

Switching to the hyperbolic class the additional possibility
arises that the BHs just fly past each other, deflecting their trajectories and
escaping to infinity. This gives rise to the merger / fly-by threshold (see
insets in Fig.~\ref{fig:Summary_2qc}) which we will discuss later.

We again describe the orbital phenomenology from low to high values of
$\Theta$. The qualitative features of low $\Theta$ evolutions are the
same as in the elliptic category. The actual values of $\Theta$ that
lead to analogous features/characteristics (one complete orbit, a
whirl, maximum in $E_{rad}$, etc.~) decreases for increasing $P$. 
This is in agreement with the
expectation that for larger $P$ one has to shoot the BHs closer to
each other compared to the corresponding lower $P$ evolutions in order to
obtain a qualitatively similar behavior.
Again as in the elliptic category for larger values of $\Theta$ the
whirl phase is followed by a zoom. Depending on $P$ there now is a
finite range in $\Theta$ where the BHs do not escape to infinity, but
reach an outer turning point (like for elliptic orbits) and
fall back ending in a delayed merger. Beyond a certain (momentum 
dependent) value of $\Theta_{ub}$ the BHs are simply deflected or 
fly-by each other.

Inside the hyperbolic category, in our simulations we only find orbits
exhibiting whirls during the first encounter and never
thereafter. This behavior agrees with the interpretation that too much
angular momentum is radiated during the first whirl to have another
whirl episode.
Another way of explaining this is to realize that the first
pericenter distance during which dynamical capture occurs is already
within any (quasi-) stable orbit. On the next encounter the binary
will have lost additional angular momentum and will have a yet smaller
pericenter separation. The system therefore is likely to merge on the next
encounter. It is unclear whether for high momenta there can also be cases
where after a first whirl the orbital parameters fall into the narrow window
for a second whirl. 

In geodesic motion there exist solutions that escape to infinity after
a full $2\pi$ whirl. 
Like in previous studies, we were not able to find such
orbits in the comparable mass case. This is most probably due to
excessive loss of energy and angular momentum during the whirl. 
It is an open question
whether this statement holds for general momentum $P$, but if it
occurs then for momenta $1P_{qc} < P < 2P_{qc}$ or $P > 6P_{qc}$.

We proceed by analyzing precession effects and discuss resemblances to
periodic orbits. For a given $P$ the precession angle shrinks with
increasing $\Theta$ when approaching the threshold as expected. The maximal
amount of precession we find is slightly larger than in the elliptic
category.
We clearly recognize patterns known from periodic orbits.  For the
$P=2P_{qc}$ sequence we find $z=2,z=3,z=4$ orbits. The main difference to
periodic orbits is that the orbits end in a merger after the first leave has
been traversed because of the severe radiation losses.  For instance
$P=2P_{qc},\Theta=25.1^\circ$ resembles the $z=3,v=1,w=0$ orbit with $q=2\pi/3$.
When decreasing $\Theta$ by small amounts, the resulting orbits typically show
the \text{same} amount of precession (only $D_{per}$ shrinks with
$\Theta$). At some point there is a transition to another multi-leaf
clover and the precession amounts to a value of $q=\pi$ and is now
similar to the periodic orbit labeled $z=2,w=0,v=1$.

\subsubsection{Unequal mass BHB and geodesic limit}
\label{GeodesicAnalog}
Next we extend the discussion to unequal mass BHBs. By doing so we 
move towards a region in parameter space which can be increasingly 
well described by geodesics. In fact, it has been in the latter 
regime where zoom-whirl behavior was studied first~\cite{GlaKen02}. 
This begs the following question: Given a binary at a finite mass 
ratio, how far away is it from the geodesic limit?

The fact that zoom whirls can be found not only for geodesics, but
also for equal masses suggests that zoom whirls also occur for
intermediate mass ratios and adds to their expected astrophysical
relevance. Indeed we can confirm (see
also~\cite{HeaLevSho09,SteEasPre11}) the presence of zoom-whirl
behavior for mass ratios 1:2 ($\nu\approx0.2222$) and 1:3
($\nu=0.1875$) (see Fig.~\ref{fig:UEM3}). 

As the mass ratio departs from unity, gravitational radiation decreases, e.g.\ 
\cite{BerCarGon07}, which is consistent with the trend to the 
geodesic limit.
In the eccentric case we find that qualitatively a similar statement 
still holds. We point out, though, that there is a non-trivial dependence on 
$\Theta$ (or inverse eccentricity). In particular, the maximum in 
$E_{rad}/M_{ADM}$ (see Sec.~\ref{sec:Erad} and Fig.~\ref{fig:EradUEM3}) 
is close to the equal mass values for the mass ratios we have probed. 

For lower symmetric mass ratio $\nu$ we do not find significantly
longer whirl phases in our data sets. It is to be expected of course
that for some mass ratio beyond 1:3 the whirl phases eventually
\emph{will} be longer and asymptote to the geodesic limit. Highly
eccentric binaries with mass ratios up to 1:3 are in this sense still
far away from the geodesic limit.

We find evidence for the analogy of zoom-whirl dynamics and unstable
circular orbits by investigating the orbital radius during the whirl
phase for various configurations. Consistently, the whirl radius
decreases with increasing $P$, which we will refer to as the
\emph{tightening} of the whirl.  This is consistent with earlier
studies \cite{HeaLagMat09,WasHeaHer08,SpeBerCar07}, in which it was
found that the spin of the merger remnant increases with the initial
angular momentum parameter, which implies a smaller radius for the 
unstable circular orbits.

Next, we investigate geodesics to derive lower limits on the shooting
angle that separates merging from non-merging evolutions,
$\Theta_{bu}$. (The corresponding values from our evolutions can be
seen as a vertical dividing line in Figs.~\ref{fig:Summary_2qc},
\ref{fig:Summary_1-6qc}, and \ref{fig:EradUEM3}.)  Since we find that
$\Theta_{bu}$ decreases monotonically with increasing $P$, we expect
this lower limit to be most restrictive for large $P$.
The idea is analogous to the
capture/escape cavities for a photon in Schwarzschild spacetime in
\cite{Book:ShaTeu}. 

A null geodesic in the Schwarzschild spacetime on
a circular orbit is located at a radius equal to the so-called photon orbit
$r_{photon}=3m$, which leads to the limit: 
$$\Theta_{bu}^{geod} = arctan \Big(\frac{r_{photon}}{2D}\Big) \approx 16.7^\circ.$$ 
The proper computation of a null geodesic in Schwarzschild spacetime
\cite{Book:ShaTeu} leads to
$$ \Theta_{bu}^{geod} = 180^\circ - arcsin\left(\frac{3\sqrt{3}m}{D/2} 
\sqrt{1-\frac{2m}{D/2}}\right) = 27.7^\circ$$
The same calculation for marginally bound circular orbits yields
$21.8^\circ$. Fig.~\ref{fig:Summary_1-6qc} indicates
that neither of these limits apply to our evolutions, because there
are unbound orbits with $\Theta < \Theta_{bu}^{geod}$. Clearly, the
assumption of a Schwarzschild spacetime is not a good one. 

From~\cite{HeaLagMat09,WasHeaHer08} we know that the merger remnant in
our settings will settle down to a Kerr solution with spin parameters
between $0.6<a<0.823$ with only weak dependence on the initial
conditions.
Despite the fact that the Kerr metric does not describe the spacetime 
\emph{at} merger, it may be a better approximation than Schwarzschild. 
The same estimate as above for Kerr spacetime yields 
$\Theta_{bu}^{geod} = 14.9^\circ$ for $a=0.6$, 
$\Theta^{geod}_{bu} = 11.4$ for $a=0.0823$ and
$\Theta^{geod}_{bu} = 5.7^\circ$ for $a=1$. The $\Theta$-values for
$0.6\leq a \leq 0.823$ correspond rather well to the shooting angles
separating bound from unbound runs in the higher momentum cases
despite the fact that the limit from null-geodesics to (finite-size)
equal mass binaries is by no means straightforward. We will use this 
analogy in interpreting our results on the radiated energy in 
Sec.~\ref{sec:Erad} based on the tightening of the whirl orbits 
associated with a larger spin of the merger remnant.

\subsection{Radiation properties}
\label{Radiation}
\subsubsection{Waveforms}
\label{Waves}
The methods used to compute quantities characterizing the GW content 
of the spacetime are described in e.g.~\cite{BruGonHan06}.
Here we demonstrate how the orbital dynamics as described in Sec.~\ref{Results}A
are reflected in the GW signals.

The waveforms of quasi-circular binaries are rather well understood. To a 
certain extent merger waveforms as they arise from evolving quasi-circular 
binaries can be very similar to the ones seen in low eccentricity 
evolutions provided the binary circularizes before merger.
For large eccentricities it is, however, natural to expect
deviations from a quasi-circular BHB. We observe differences in the
waveforms throughout the evolution including inspiral, onset of
merger, coalescence and ring-down.
Any imprints left from the eccentric inspiral have 
to be radiated away during this process, because the final spacetime can 
be described by the Kerr metric. In fact, the merger remnant reveals a 
different signal during ring-down~\cite{GolBru09}. In particular, quite 
generically high eccentricity is correlated with an amplified ring-down 
signal.

The inspiral features show some level of agreement with~\cite{PetMat63} 
and PN models for such waveforms are known analytically to 2PN order 
\cite{TesSch10} (see the first comparison between numerical waveforms 
and Post-Newtonian ones in the eccentric regime~\cite{Hinder:2008kv}). 
However, features associated with zoom-whirl behavior 
(see Fig.~\ref{fig:1qcD} prior to merger)
are exclusive to the strong field and thus have to be dealt with using 
the tools of numerical relativity. 
These inspiral signals will be observable by future GW interferometers 
such as LISA \cite{Hou94,BabHanHus08} or eLISA/NGO \cite{eLISA:2012}, 
DECIGO \cite{Yagi:2012gb} or the ET-telescope~\cite{hannam-2009,Broeck:2010vx}.

We discuss typical waveforms of a representative subset of our evolutions. 
It is illustrative to go through Figs.~\ref{fig:1qcA}--\ref{fig:5qcC}, 
\ref{fig:Summary1qcWithOrbits} and their captions.
Our main focus is on the richness in information stored in eccentric BHB 
waveforms in contrast to quasi-circular ones because of the promising 
implications for data analysis, see \cite{Mik12}.

\begin{figure}[t]
\centerline{\resizebox{9cm}{!}{
\includegraphics{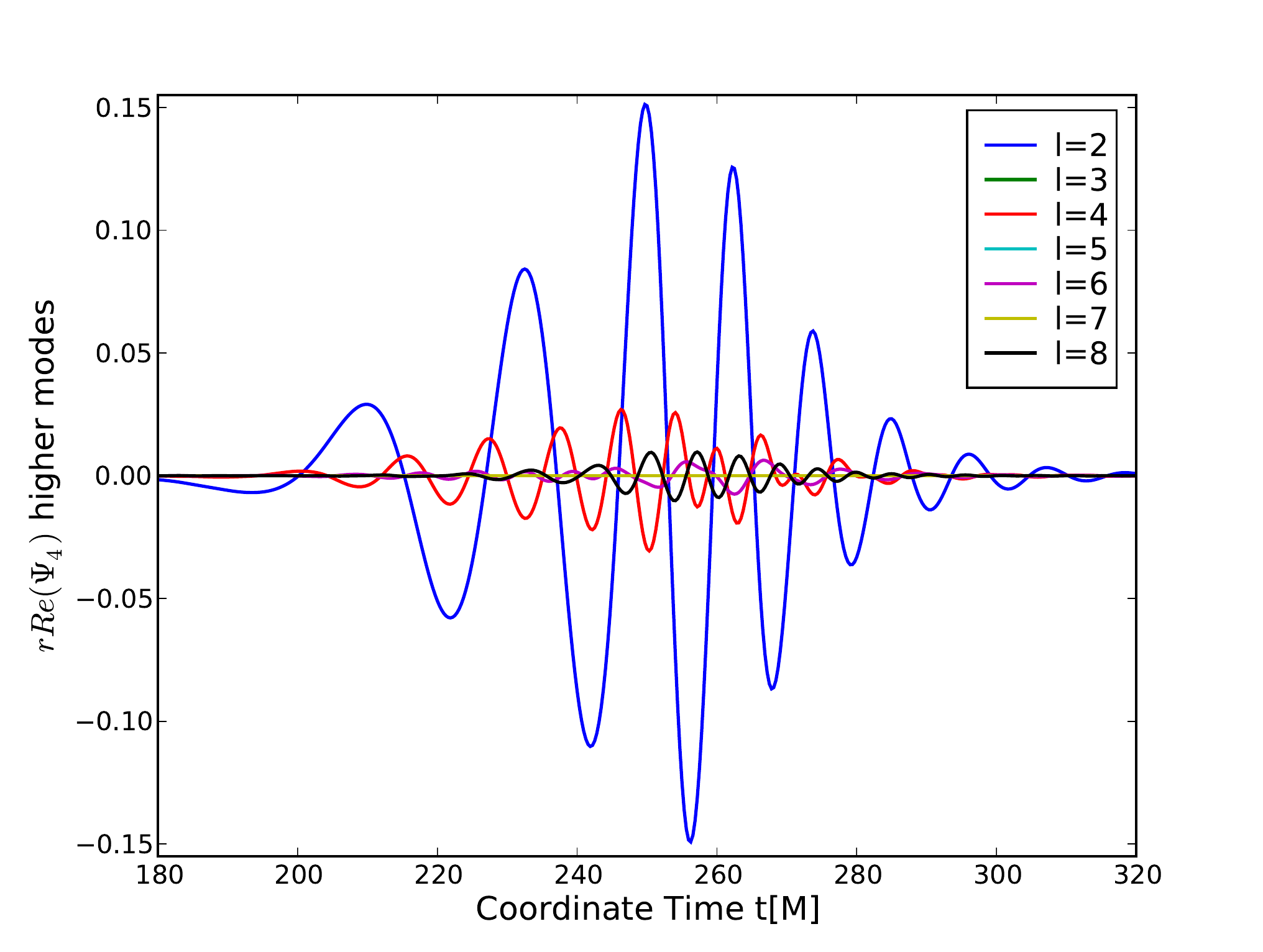}}}
\caption{
Higher modes of $r\mathcal{R}e(\Psi_4)$ summed over $m$ for the equal
mass case. Clearly one can see that $l=2$ modes are - just as in the
quasi-circular case - the dominant contribution. The plot also reveals that
higher modes exhibit a much more significant contribution from higher $l$-modes.
All odd $l$ vanish within numerical error as expected from the quadrant symmetry
of equal mass, non-spinning BHB.}
\label{fig:HigherModes1:1}
\end{figure}
Already the 22-mode shows obvious differences which become larger in
other modes. For example the $l=2$ $m=0$-mode 
of a quasi-circular orbit looks just like a smaller amplitude version of 
the $l=2$ $m=2$-mode. 
In the eccentric case they contain completely different features. We plot 
the higher $l$ modes summed over $m$ in Fig.~\ref{fig:HigherModes1:1} 
for the equal mass case, for which (without BH spin) only even $l$-modes 
contribute by symmetry. We have computed the $l\leq8$ modes and find that the $l=2$ 
is still the largest contribution, but $l=4$ has a significant 
contribution throughout the merger and $l=6$ and $l=8$ close to the 
maximum. 

As an example of how different waveforms of binaries with eccentricity
and mass ratio away from unity can be,  we show in Fig.~\ref{fig:UEM3} 
the waveform and orbital trajectories for the mass ratio 1:3.
Clearly the features induced by eccentric unequal mass BHB give rise
to waveforms which effectively break degeneracies in parameter
space \cite{Key:2010tc,Mik12}.

In this work we do not construct waveform templates.
Longer runs will be needed in order to achieve a match to a PN
waveform because of the small separations at pericenter. Performing wave
extraction at larger radii is also clearly desirable in this
context. With current codes this could be done at an acceptable
computational cost.

\subsubsection{Radiated energy}
\label{sec:Erad}
We compute the energy $E_{rad}$ radiated away in GWs and analyze 
these results together with the orbital dynamics.
For the elliptic orbits we add an estimate of the radiated energy 
of the past evolution $E^{past}_{rad} \approx -E_{bind}(t=0)$ to 
$E_{rad}$. Using this estimate we implicitly assume that the binary 
was isolated in its entire past. 
The actual value 
$E_{bind}(t=0)$ for the $P=P_{qc}$ sequence turns out to be
$E^{past}_{rad}\approx-E_{bind}(t=0)\approx 0.0057 \pm 0.0001$.
We normalize $E_{rad}$ by the ADM-mass of the initial time slice,
$M_{ADM}(t=0)$. The resulting quantity is what we call the
``efficiency'' of gravitational radiation.

\begin{figure*}[t]
\centerline{
\includegraphics[width=130mm]{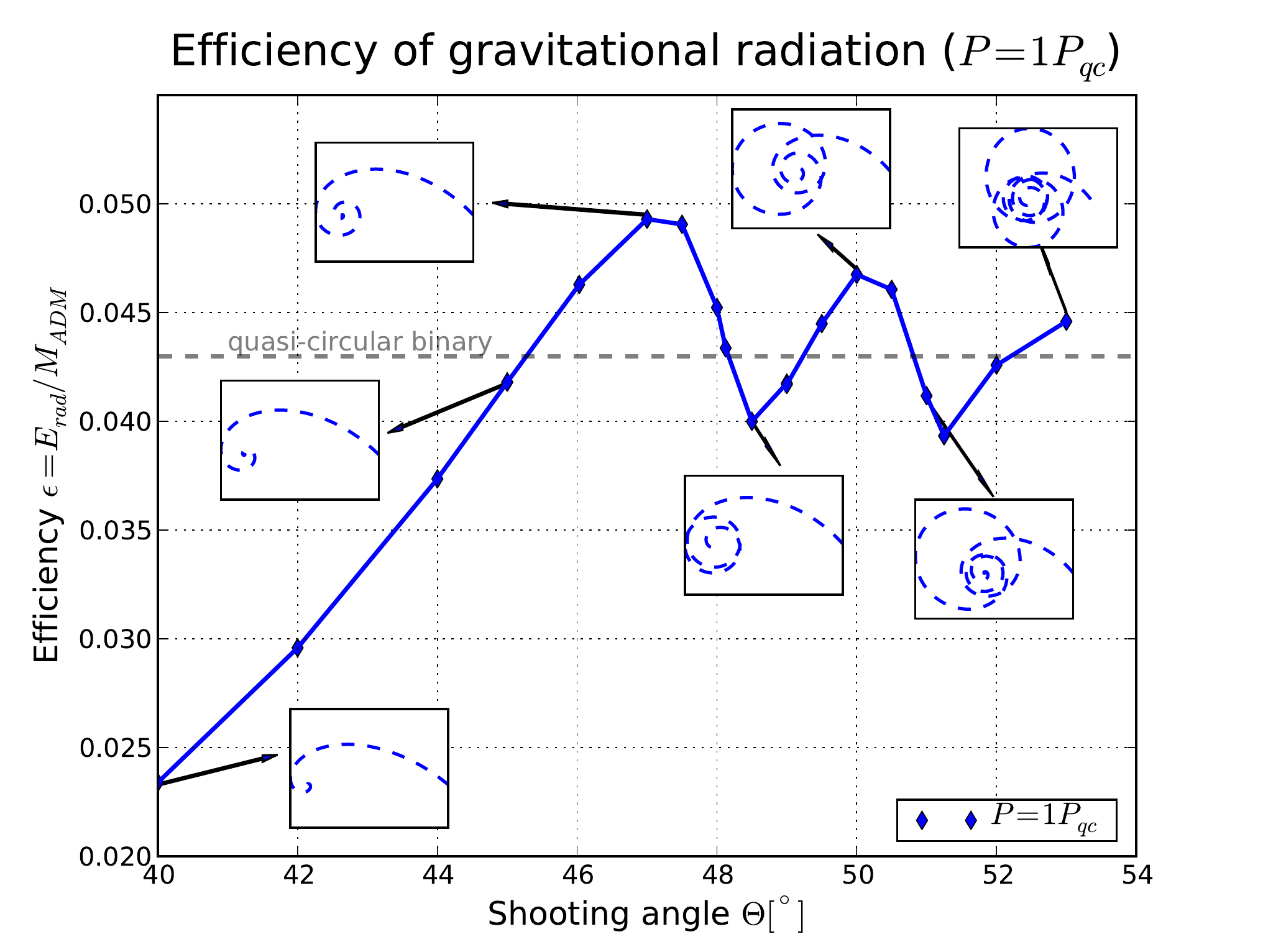}}
\caption{Radiated energy as a function of the shooting angle $\Theta$
  for the $P=1P_{qc}$ runs. 
  The small insets illustrate the corresponding orbital dynamics. One can see
  that the global maximum does \emph{not} correspond to a zoom-whirl orbit.
  Strongest zoom-whirl behavior is rather associated with the local
  \emph{minimum} in $E_{rad}$ near $\Theta=48.5^\circ$. 
  Compare with Fig.~\ref{fig:CD}.
  }
\label{fig:Summary1qcWithOrbits}
\end{figure*}
\begin{figure}[t]
\centerline{\resizebox{9cm}{!}{
\includegraphics{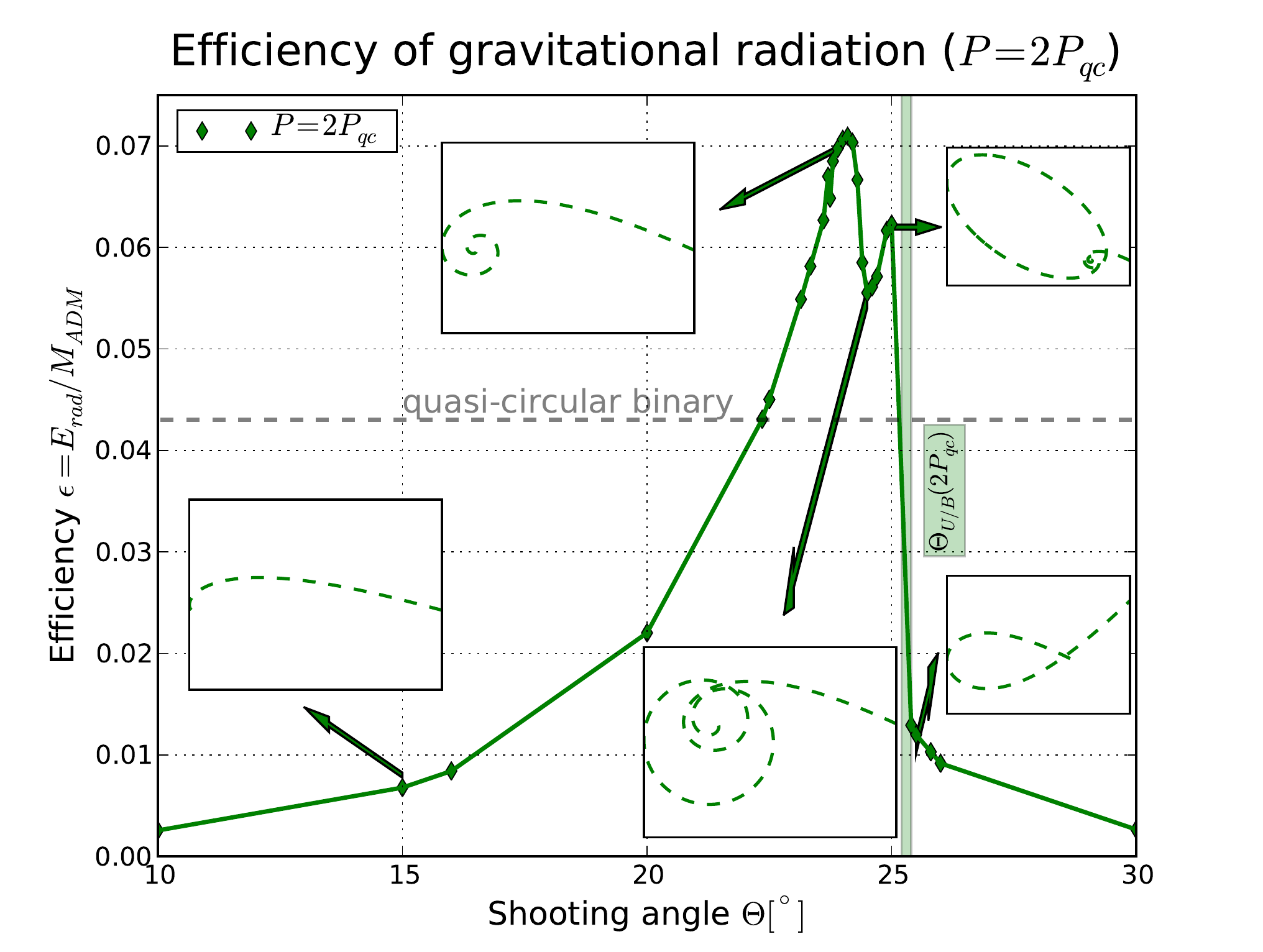}}}
\caption{Radiated energy for $P=2P_{qc}$. While in the
  $P=1P_{qc}$ case there are multiple extrema we find only two maxima
  here. Note the fine sampling around the extrema. 
  The rather large gap at $\Theta=\Theta_{bu}$ 
  reflects the problem visible in the upper right inset and in 
  Fig.~\ref{fig:PhaseSpace2qc}, namely that the BHs zoom out to very
  large distances, which implies large $t_m$ and consequently high
  computational costs (also due to the requirement of higher
  resolution).}
\label{fig:Summary_2qc}
\end{figure}
\begin{figure}[h]
\centerline{\resizebox{9cm}{!}{
\includegraphics{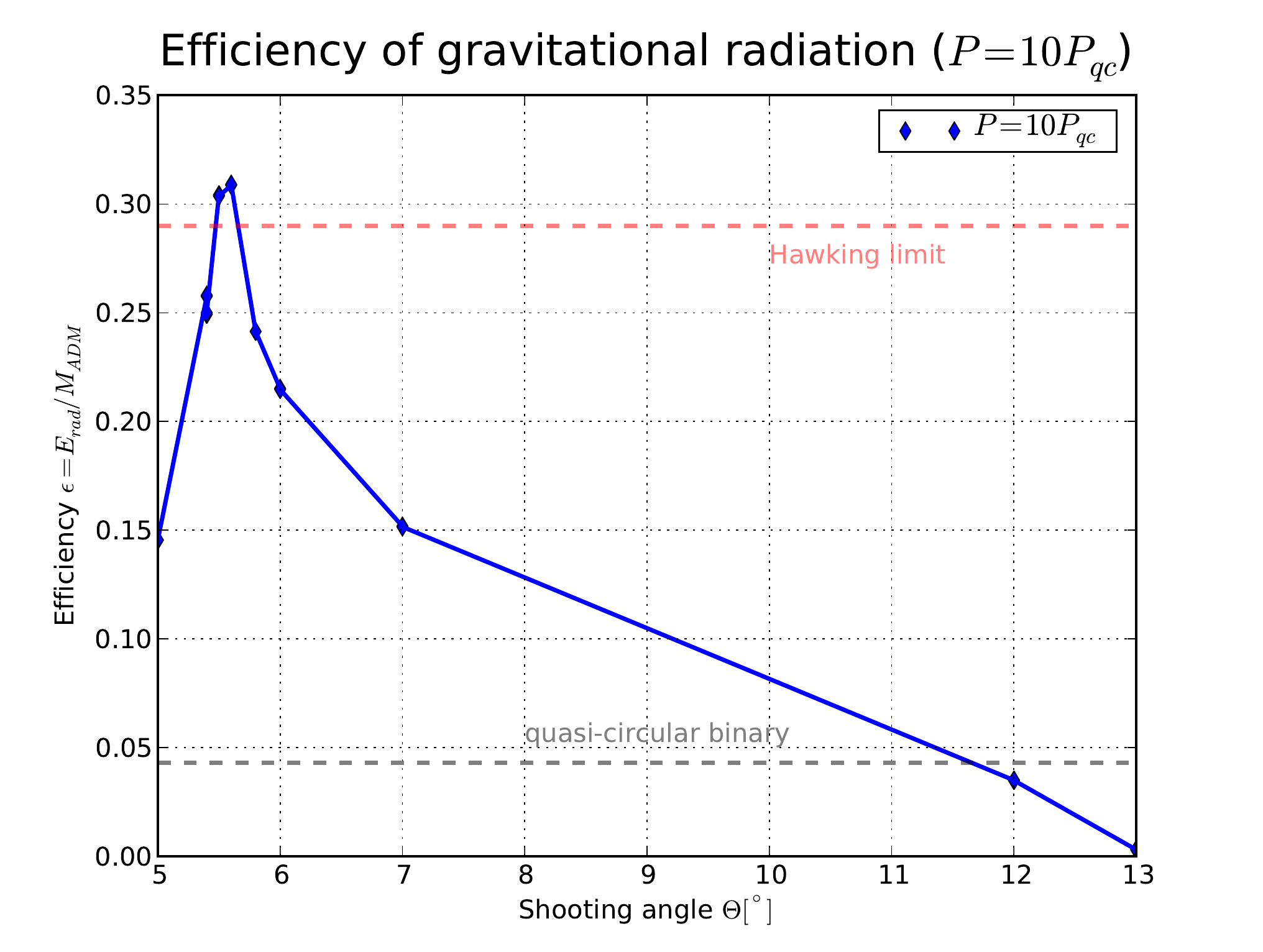}}}
\caption{Radiated energy for $P=10P_{qc}$ (i.e.~$P \approx 0.6M$). The
values for $\Theta$ are not comparable with those from the 
other $P$-sequences as
the initial separation was chosen to be $D(t=0)=50M$ due to the large junk
radiation. Only at $P=10P_{qc}$ are we able to exceed the Hawking limit 
\cite{Haw71} on the
energy release of two Schwarzschild BHs far apart without orbital angular
momentum. With $P=10P_{qc}$ the spacetime is very different from Schwarzschild. 
Note that the results are below the current maximum reported value 
of $35\pm5\%$. The contribution from junk radiation is about $0.05$ and is 
included in the data shown. 
}
\label{fig:Summary_10qc}
\end{figure}

\begin{figure*}[t]
\centerline{
\includegraphics[width=130mm]{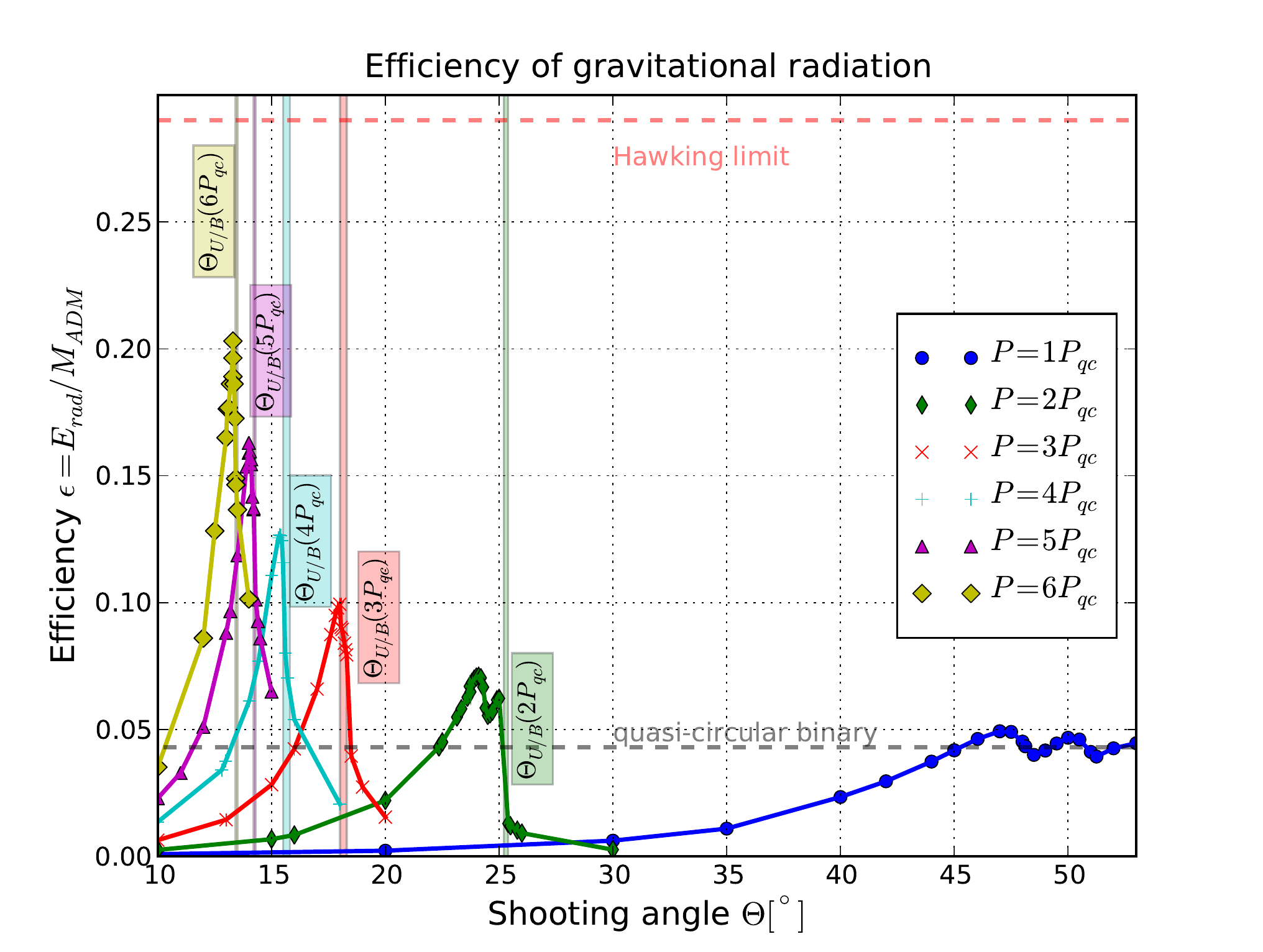}}
\caption{Radiated energy for the 1qc--6qc runs (i.e.~$0.06M \geq \, P \, \geq
  0.36M$) including the values corresponding to a quasi-circular
  binary and the Hawking limit. The vertical lines mark $\Theta_{bu}$
  for each momentum scale considered. This value is close but not
  identical to the threshold of immediate merger. As expected the
  shooting angle where zoom-whirls occur is closer to the $\Theta=0$
  (head-on) case for higher initial momentum.  The higher the initial
  momentum the more energy is radiated. In the $1P_{qc}$ sequence
  we have not included the data for $\Theta=60^\circ$, 
  but the efficiency of radiation agrees within
  plotting accuracy with the value for a quasi-circular BHB as expected from
  extrapolating the data set shown.}
\label{fig:Summary_1-6qc}
\end{figure*}

The results of all our evolutions are presented in Figs.\
\ref{fig:Summary1qcWithOrbits},
\ref{fig:Summary_2qc}, 
\ref{fig:Summary_10qc}, 
\ref{fig:Summary_1-6qc}, and
\ref{fig:UEM3}. 
The different lines (colors, symbols) in these plots correspond to
different initial momenta and each line shows the efficiency of
gravitational radiation as a function of $\Theta$.

The first global feature to notice is that gravitational radiation
becomes much more efficient for higher momenta.
We give the maximal efficiency for $7$ initial momenta $P$. 
So far the largest value $35\pm5\%$ was reported
in~\cite{ShiOkaYam08,Spe09,SpeCarPre08,SpeBerCar07} (in
which the punctures have coordinate velocities of $v=0.94$). In our
data set we come rather close to this limit (see
Fig.~\ref{fig:Summary_10qc}). The challenge in these studies arises
from the growing significance of unphysical radiation content that is 
associated with the construction of initial data. Here we did not 
intend to push this limit further, but this shows that we have 
probed part of the parameter space close to the limits 
of former investigations. As we shall demonstrate (see 
Fig.~\ref{fig:Summary_1-6qc}) the sampling is quite exhaustive 
and allows us to probe zoom-whirl behavior in a large class of 
orbits. 
In particular, one of our important findings is  
model $P=1P_{qc},\Theta=60^\circ$ (elliptic class) 
with several close encounters before merger, see Fig.~\ref{fig:1qcD}. 
The initial eccentricity is as low as $e\sim 0.5$. This is a value within
typical estimates of supermassive BHBs in galaxy merger scenarios
following star- or disk-driven hardening \cite{RoeSes11} and
also a value found for inspiraling binaries near 
galactic cores \cite{Antonini:2012ad} which are driven to 
very similar eccentricities via the Kozai mechanism.

For low momenta and the mass ratios under consideration the shooting
angles for the largest number of orbits in general neither coincide with the
maxima in $E_{rad}$ nor do they coincide with the unstable, circular
(whirl-like) orbits merging right thereafter.
Generally, the maximum in $E_{rad}$ inside the hyperbolic regime
lies close to the merger/fly-by threshold.
However, in the limit $P\rightarrow\infty$ there appears to be a
growing amount of degeneracy: the unstable-circular orbits
actually seem to coincide with the most efficient radiators.
In the next section we will give an interpretation for this behavior.

For low $\Theta$ we find, in agreement with previous studies 
\cite{SpeBerCar07}, that the radiated energy quickly drops to the 
small amounts known from head-on collisions \cite{SpeCarPre08}. This
drop can clearly be seen for every initial momentum considered in 
Fig.~\ref{fig:Summary_1-6qc}. 

In addition, the shape of the transition from large to small $\Theta$ 
is by no means trivial.
One of the key features in the radiated energy is that, especially
in the $P=1P_{qc}$ sequence but also for $P=2P_{qc}$, there appear
additional local extrema which match the number of encounters.  The
observed structure in $E_{rad}$ shows a remarkably clean periodicity
as a function of $\Theta$ and should be compared with corresponding
features in the final spin and mass in \cite{HeaLagMat09}. We find that
these features are determined entirely by the dynamics during the last
encounter. Zoom whirl effects in the $P=1P_{qc}$ sequence
\emph{minimize} radiated energy. We find that the energy is
\emph{less} than that of a quasi-circular binary in direct contrast to
\cite{PetMat63}.  We will interpret these observation in the next
section.

Looking at our findings presented in 
Fig.~\ref{fig:Summary_2qc} and \ref{fig:Summary_1-6qc}, one 
may wonder why the additional peaks, i.e.~additional 
encounters, are present in the lowest momentum sequence, 
but not in the higher momentum ones. The answer lies in 
the initial binding energy. For the large $P$ cases only 
those evolutions which radiate a lot of energy during the 
first encounter will be bound orbits (dynamical captures).
As it turns out, the radius of capture for those evolutions 
is inside the ISCO of a single Schwarzschild black hole of 
the same total mass. The capturing encounter generically 
is a whirl and thus an unstable orbit. After the BHs are 
dynamically captured they will have lost additional 
angular momentum and energy. Thus the pericenter distance 
on their next encounter will be even smaller and therefore 
always end in a merger. 

Another observation is that in the $P=2P_{qc}$ sequence 
there is a second peak next to the global maximum in contrast 
to the higher $P$ sequences. The peak arises from 
contributions during the second (and last) encounter. While 
we do observe similar orbital dynamics also for the higher 
$P$ cases, we however do not see a corresponding peak. The 
reason is obvious once one compares the GW amplitudes during 
the capturing first encounter with the amplitude during 
merger, see Fig.~\ref{fig:5qcA} and \ref{fig:5qcB}. 
For the large $P$ evolutions the mergers on the second 
encounter only have a negligible contribution to the 
radiated energy, but the whirly, capturing encounter 
dominates the energy loss.

\begin{figure}[!here]
\centerline{\resizebox{9cm}{!}{
\includegraphics{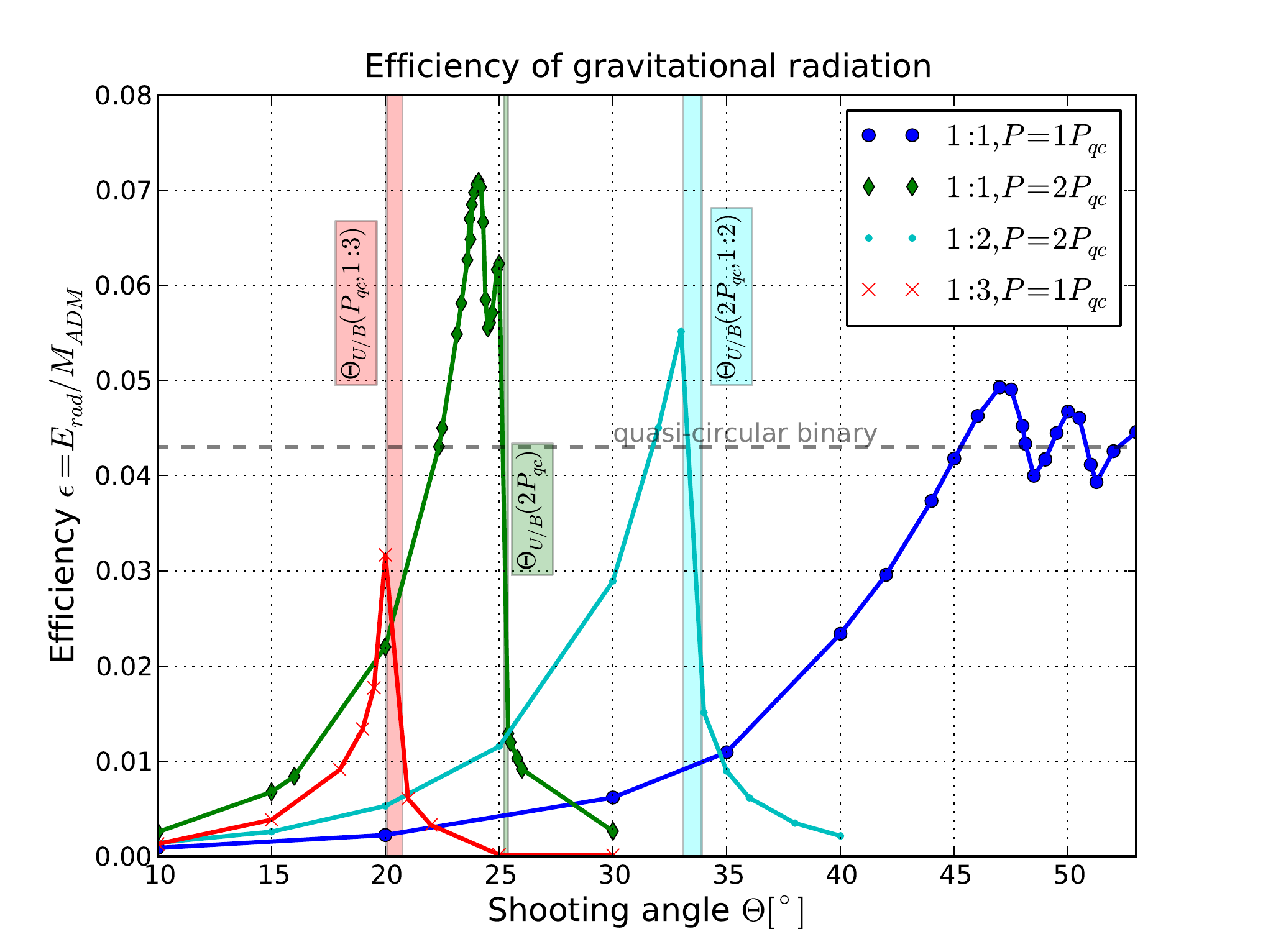}}}
\caption{ 
  Radiated energy $E_{rad}$ for mass ratios 1:1, 1:2 and
  1:3 and some particular values of $P$. 
  Qualitatively the results for unequal masses are similar to
  those for equal masses. For example, note the drop in $E_{rad}$ after a global
  maximum, the global maximum corresponds to an orbit that roughly
  completes 1 orbit, and zoom-whirl behavior for low $P$ is associated
  with inefficient radiation as in the equal mass regime.
  For $P=2P_{qc}$, zoom-whirl behavior for mass ratio 1:2 occurs for
  larger $\Theta$ than in the 1:1 case.
}
\label{fig:EradUEM3}
\end{figure}

Results for unequal mass runs are shown in Fig.~\ref{fig:EradUEM3}.
According to our findings the scaling of
radiated energy with mass ratio is eccentricity-dependent. 
Comparing the maxima in
$E_{rad}$ between equal mass and unequal mass runs we find that a mass
ratio 1:2 still gives a maximal efficiency which is not too far away
from the corresponding equal mass run with the same $P/M_{ADM}$. This
result is in contradiction to our expectation from quasi-circular
binaries where $E_{rad}$ decreases steeply with mass ratio. 
Also our results for mass ratios 1:3 show a similar 
trend suggesting that a such mass ratios still are (in the above 
sense) far away from the geodesic limit. Our results suggest further 
parameter studies to analyze the scaling in the eccentric regime 
along the mass ratio axis. Clearly, $E_{rad}$ is much more sensitive
to $P$ rather than $\nu$.

\subsection{New diagnostics}
\label{sec:PhaseSpace}
Many interesting questions about BHB cannot be tackled by just looking at
gauge-invariant quantities. In this section we suggest new diagnostics 
that are helpful to interpret these spacetimes. 
 
\begin{figure}[!here]
\centerline{\resizebox{9cm}{!}{
\includegraphics{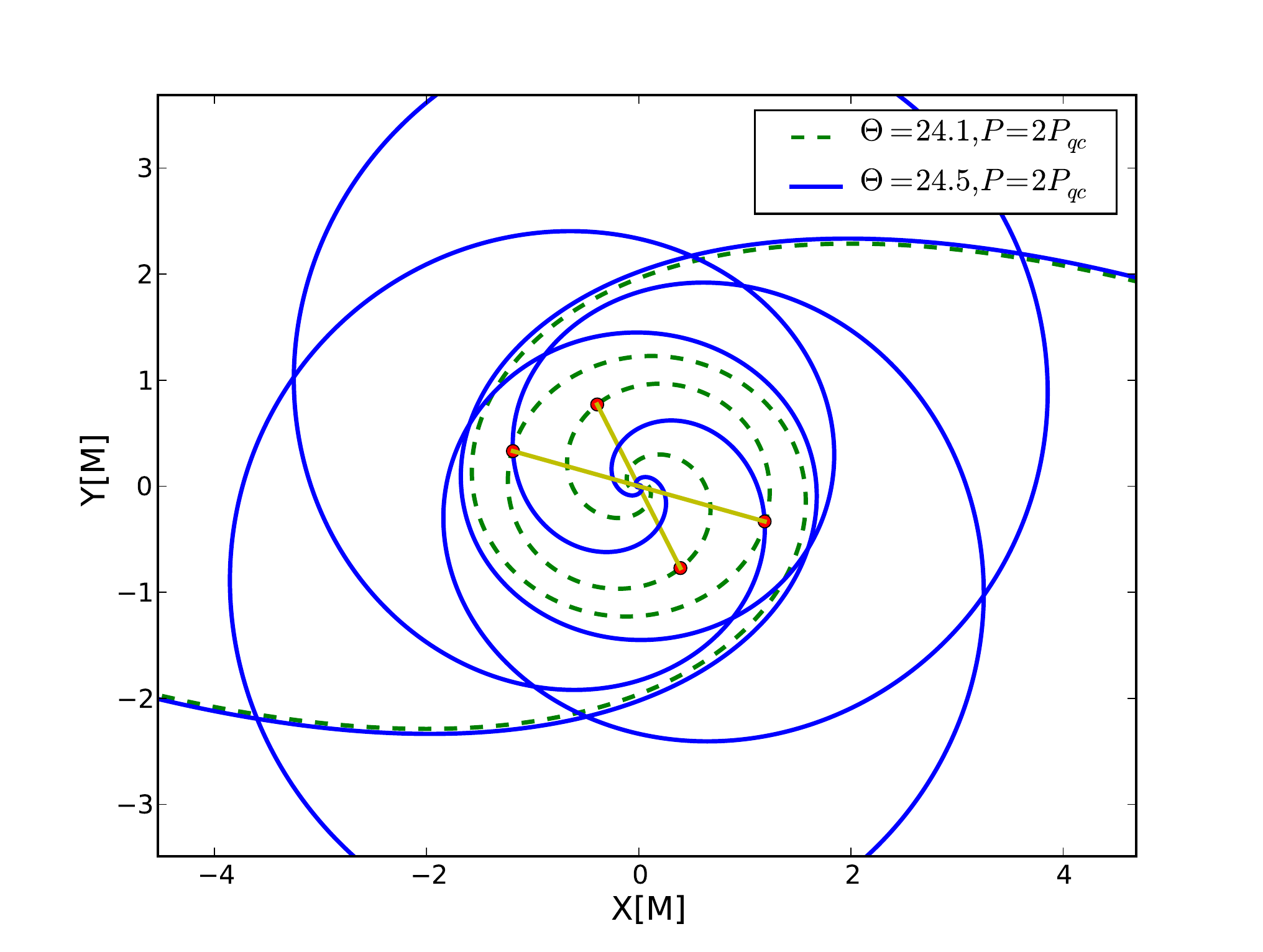}}}
\caption{
  The inner region of puncture tracks from two different runs.  
  The $P=2P_{qc}$, $\Theta=24.1^\circ$ run (blue solid line) corresponds
  to the global maximum in $E_{rad}$.  The yellow straight lines
  through the origin represent the separation vectors $\vec{D}$ 
  at the time when a common  
  horizon forms. The result \cite{GolBru09} that most efficient
  mergers occur when the tangent vectors of the orbits are closest to
  being orthogonal to $\vec{D}(t_m)$ carries over to larger $P$.}
\label{fig:PunctureTracks2qc}
\end{figure}

A first example is the observation in \cite{GolBru09} that maxima in
$E_{rad}$ coincide with a particular orbital configuration at the time
of merger. Whenever the angle between the tangent vector of the puncture
and the separation vector $\vec{D}$ at the time of
merger is largest, the radiated energy is maximized. Here, we confirm this
behavior also for $P=2P_{qc}$ orbits in
Fig.~\ref{fig:PunctureTracks2qc} demonstrating the robustness of our
gauge-dependent conclusions in \cite{GolBru09}.

We interpret this empirical finding in the following way. Maximizing 
the above mentioned angle translates into maximizing 
$\vec{L} = \vec{D} \times \vec{P}$, the Newtonian expression for the 
angular momentum of two point masses. We therefore conjecture based 
on our data set that the strongest ring-down signals are caused by 
those evolutions which maximize the angular momentum at the moment
of merger. 

\begin{figure}[t]
\centerline{\resizebox{9cm}{!}{
\includegraphics{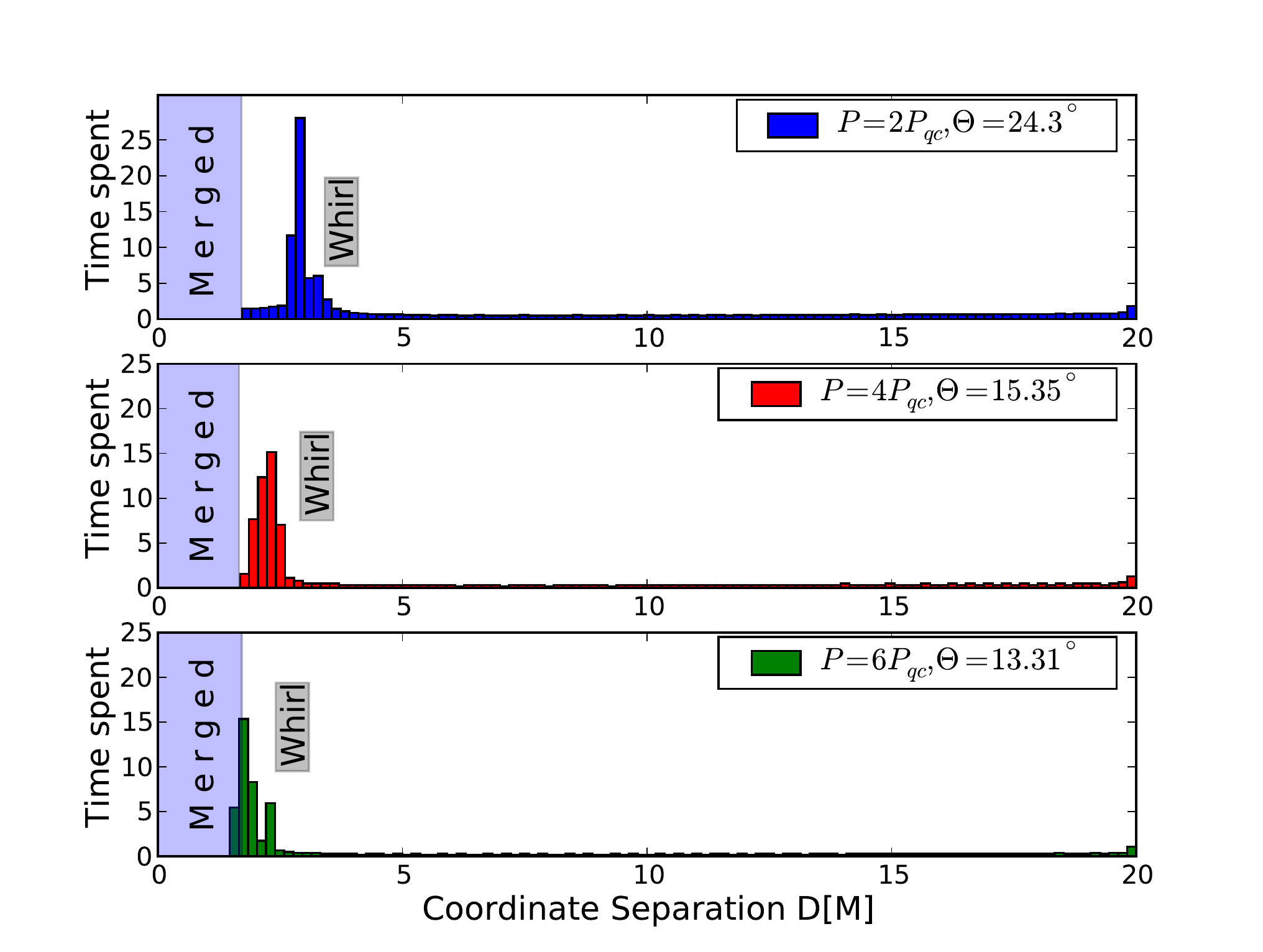}}}
\caption{Histogram of the coordinate separation $D$ for three
  different evolutions with $P=2$,$4$,$6P_{qc}$. This plot shows how
  much time a binary has spent at a given separation $D$. All runs
  shown here correspond to the longest whirl phase found at each fixed
  momentum. Clearly visible is the tightening of the whirl radius for larger
  $P$ (compare with \cite{GolBru09}). The overlap of the shaded region in
  $P=6P_{qc}$ with the histogram happens because during the whirl the
  separation is indeed shorter than at the onset of merger. 
}
\label{fig:Hist}
\end{figure}

Another useful diagnostic is the
histogram of $D(t)$ (see Fig.~\ref{fig:Hist} and \cite{GolBru09}). 
It measures the time the binary spends within an interval
$D \pm \Delta D$ of coordinate separation.
We focus on two important conclusions drawn from this plot.
First, as already reported in \cite{GolBru09} the whirls show up as a 
sharp and well-defined peak allowing us to measure the radii of unstable 
circular orbits in these highly non-linear spacetimes. 
Second, the whirl radii are becoming systematically tighter as $P$
increases (see Fig.~\ref{fig:Hist}) which is related to a higher
Kerr-parameter of the merger remnant\cite{SpeBerCar07,HeaLevSho09}.
We checked the coordinate separation as a function of time
separately to exclude a possible issue with our merger time estimate
which is not well suited for large $P$.

For low momenta \cite{HeaLagMat09,WasHeaHer08} showed that the final spin
parameter lies within $0.6 < a < 0.0832$ with the tendency that the
spin parameter grows with the initial momentum. Thus the
resulting background spacetime will have a tighter ISCO. 
As the binary spends considerably more
time at the whirl radius than at a Newtonian pericenter at the 
same distance (see Fig.~\ref{fig:Hist}) the binary radiates much more
efficiently if this whirl occurs at a smaller radius. 
Thus the geodesic analog together with our gauge-dependent diagnostics
give a natural explanation for our earlier observations:
In the high momentum case zoom-whirl orbits do coincide
with the most efficient radiators where the whirls are tight, while 
this is not the case in the low momentum regime, where the whirl 
radii are significantly larger.

\begin{figure}[!here]
\centerline{\resizebox{9cm}{!}{
\includegraphics{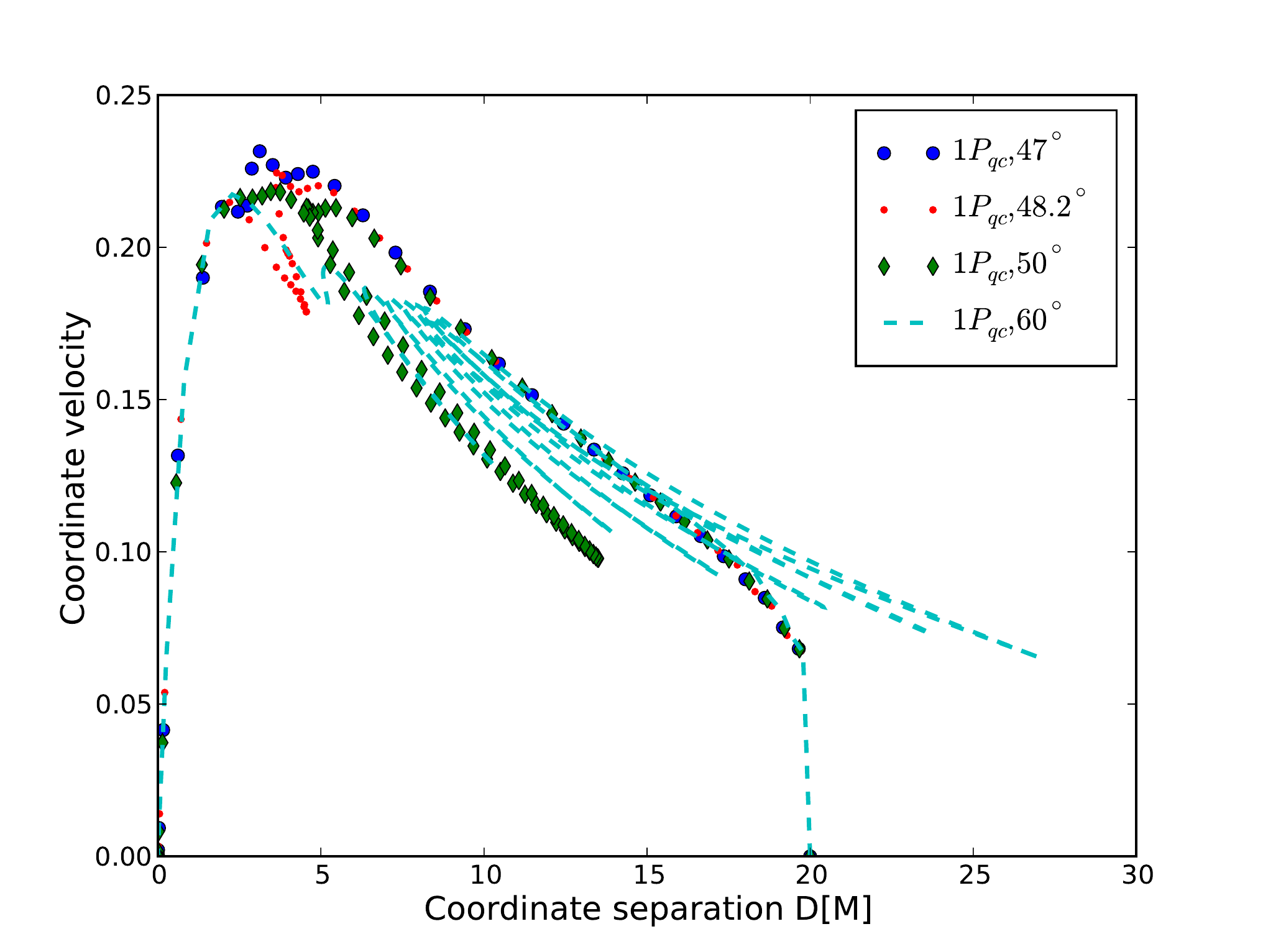}}}
\caption{Phase space for $P=1P_{qc}$ sequence. The $\Theta=48.2$ 
evolution takes a detour in phase space thereby avoiding the region where 
radiation is most efficient. The $\Theta=47$ evolution radiates more
efficiently because it reaches further towards the upper left and at the same
time spends considerable time at low $D$ (see Fig.~6 in \cite{GolBru09}).
}
\label{fig:PhaseSpace1qc}
\end{figure}
\begin{figure}[!here]
\centerline{\resizebox{9cm}{!}{
\includegraphics{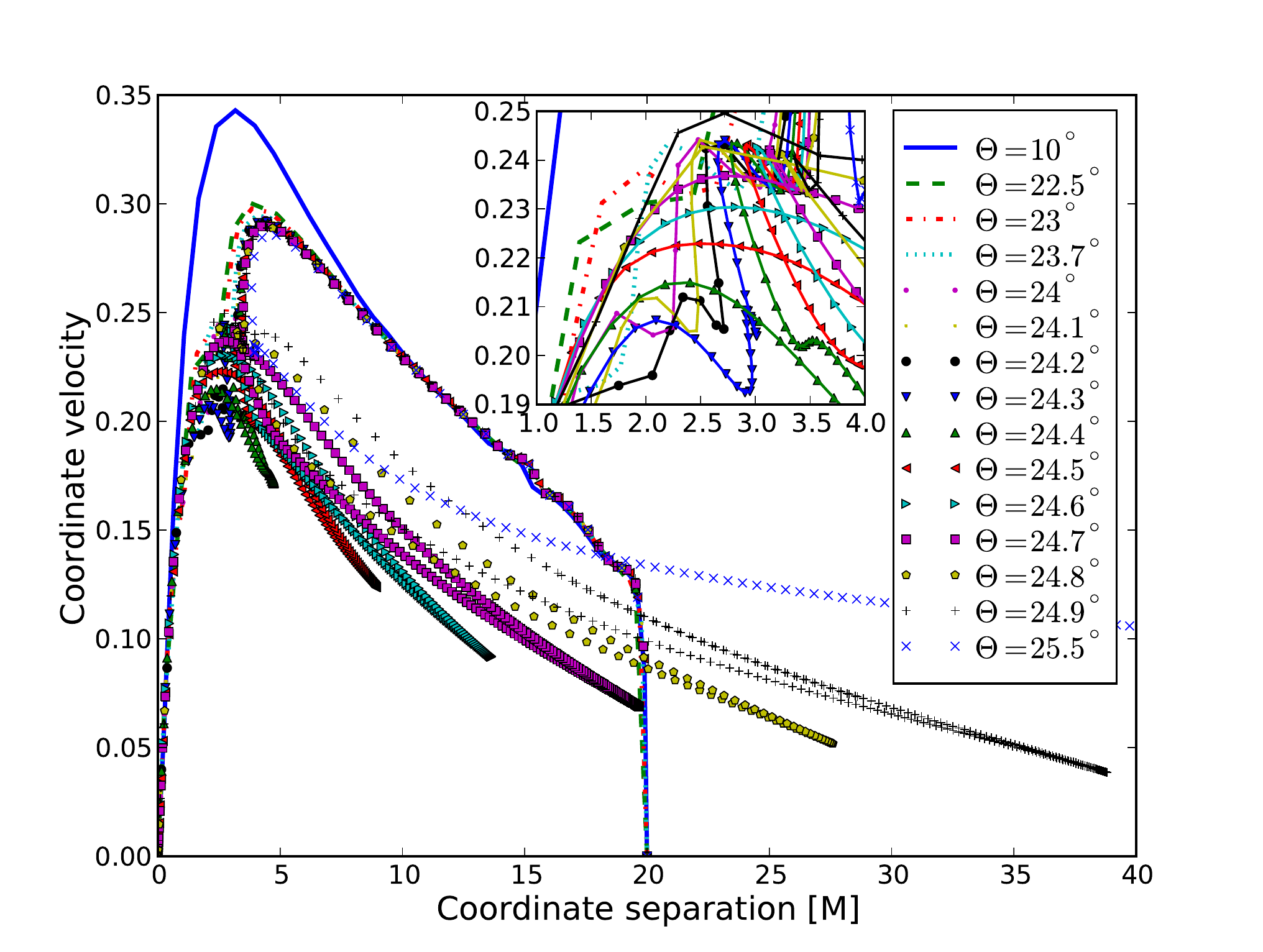}}}
\caption{Phase space for $P=2P_{qc}$ sequence. All runs except
  $\Theta=25.5^\circ$ are bound.  Near the threshold of immediate
  merger the zooms ($24.3^\circ \lesssim \Theta \lesssim 24.9^\circ$)
  may extend far out. The inset shows a close-up
  of the inner region where whirls
  occur (the point density is enlarged there). 
  }
\label{fig:PhaseSpace2qc}
\end{figure}

\begin{figure}[!here]
\centerline{\resizebox{9cm}{!}{
\includegraphics{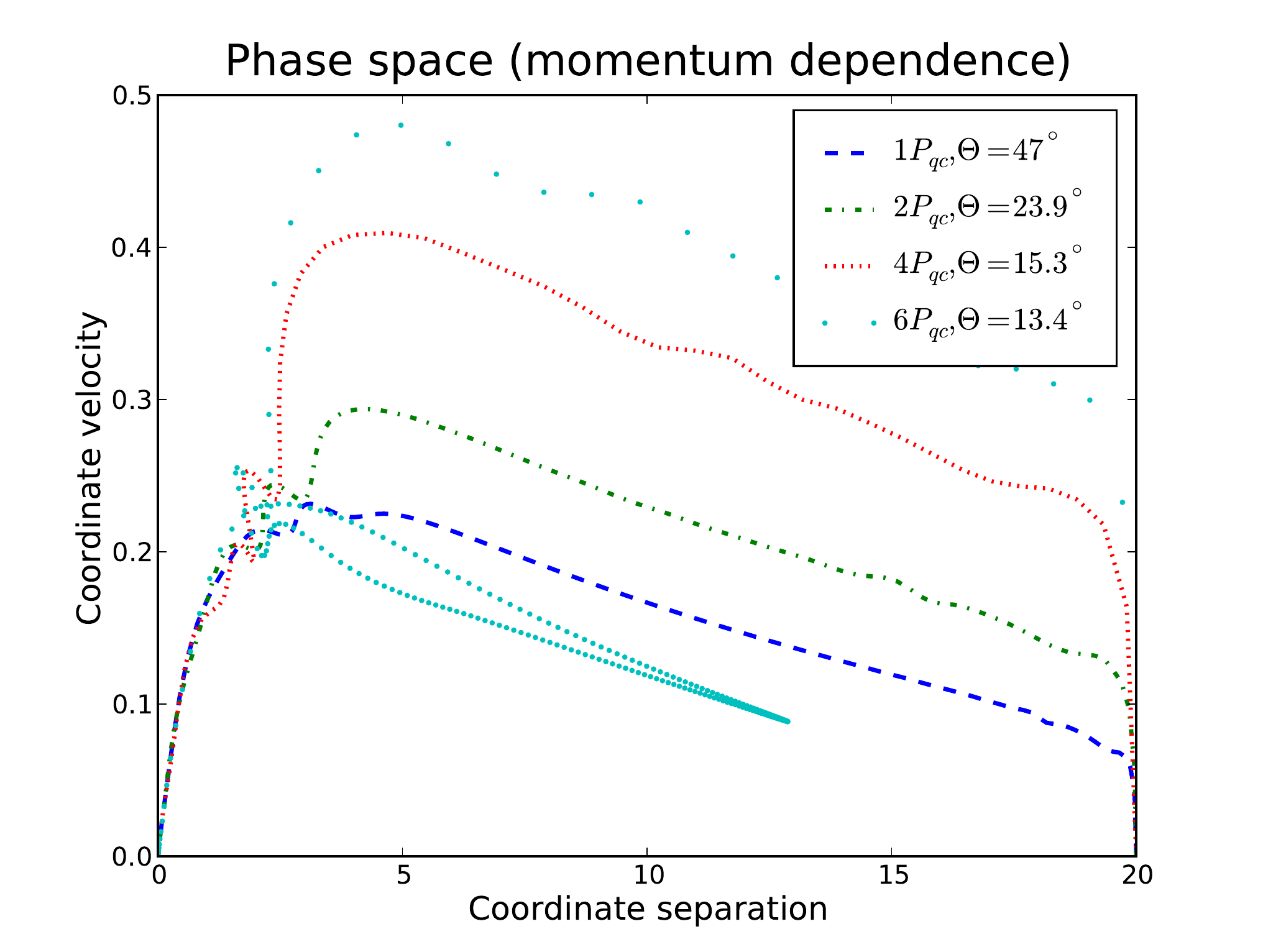}}}
\caption{
  Phase space for different initial momenta $P$ with angles chosen
  for near maximal $E_{rad}$.
  Interestingly, close to the merger \emph{all} binaries
  reach the same coordinate velocity independent of the large
  differences in initial conditions. This effect appears as a blurred
  "focal" point in phase space. Hence the whirl
  itself becomes more important for $E_{rad}$ than the actual
  merger. The deceleration in the $6P_{qc}$ whirl phase is larger than during
  the  merger.
}
\label{fig:PhaseSpacePdependence}
\end{figure}
%
\begin{figure}[t]
\centerline{\resizebox{9cm}{!}{
\includegraphics{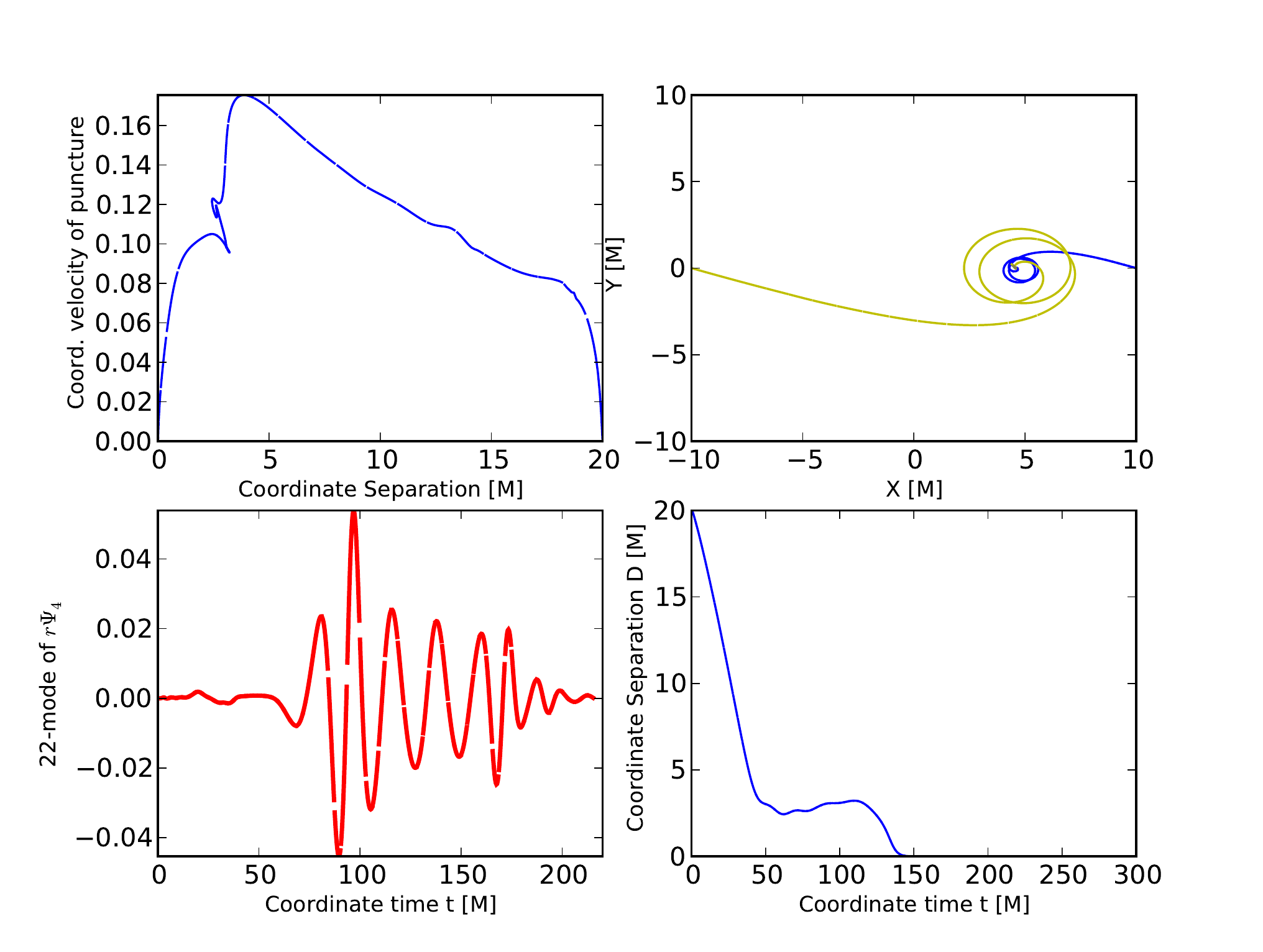}}}
\caption{Mass ratio 1:3. The whirl phase is slightly but not
  significantly longer than for the equal mass configuration. This
  difference can be understood by comparing radiation efficiencies of
  these systems. A comparison of the 22-mode to the equal mass runs suggests
  strong sensitivity on the mass ratio. Note that the merger is not 
  the strongest signal in $\Psi_4$.}
\label{fig:UEM3}
\end{figure}

As a final diagnostic we present trajectories of the binaries through phase 
space, see Figs.\ 
\ref{fig:PhaseSpace1qc}, 
\ref{fig:PhaseSpace2qc}, 
\ref{fig:PhaseSpacePdependence}, and
\ref{fig:UEM3}. 
We choose $D(t)$ and the coordinate
velocity $v(t)$ of the punctures as generalized coordinates. 
This construction is explicitly coordinate dependent, and 
switching to another gauge will lead to different trajectories.
However, previous investigations led to the conclusion that the moving
puncture gauge leads to puncture tracks that correspond rather well to
what an observer sees from infinity,
e.g.\ \cite{CamLouMar05,HanHusPol06,DieHerPol05}. In particular, for
orbiting motion one can argue based on the shift condition that this
should be the case \cite{HanHusPol06}, although for linear motion the
situation is different.
Hence, one has to keep the gauge issue in mind, but the moving
puncture gauge leads to rather robust features in the phase space
trajectories as we will discuss next.

To familiarize oneself with the trajectories in phase space consider a
circular motion with constant velocity.  This motion corresponds to a
single point in a $D$-$v$ phase space diagram.  A Kepler ellipse
corresponds to a line, which is curved according to Kepler's third
law. 

In Fig.~\ref{fig:PhaseSpace1qc} we show runs for $P=1P_{qc}$ for four
different $\Theta$, while Fig.~\ref{fig:PhaseSpace2qc} gives a global
impression of many different angles for $P=2P_{qc}$ (each for equal
masses).
These simulations start at the lower right corner at $D=20M$, $v=0$ and
quickly rise due to the initial gauge adjustment, so that the
coordinate velocity of each BH approaches the value given expected from 
the intial data.
Then the trajectory moves towards the upper left as the 
orbits shrink and the punctures move faster.
When the black holes merge the trajectory ends in the origin at
$D=0$, $v=0$.  Whirls or parts of tight circular orbits are
indicated by approximately constant $D$ but decreasing $v$, 
with $D\approx2M$--$5M$. 
Zooms follow roughly the shape of elliptic orbits, with $D$ varying
between $5M$--$10M$ at pericenter out to an apocenter at $15M$--$27M$ in
Fig.~\ref{fig:PhaseSpace1qc} or up to $40M$ in
Fig.~\ref{fig:PhaseSpace2qc}.  Fig.~\ref{fig:PhaseSpace2qc} also shows
one orbit for $P=2P_{qc}$ which escapes to infinity.

A head-on collision in phase space looks very much like an eccentric
binary starting on the ${v}=0$ line at the value of $D$ that
corresponds to the same total energy. We note that a head-on collision
always constitutes an upper envelope in the phase space, i.e.\ fixing
$P$ and comparing $\Theta=0$ runs with $\Theta\neq0$ evolutions we
always find ${v}(t,\Theta=0)>{v}(t,\Theta\neq0)$ (at least for
$D(t)\leq D(t=0)$).

We note that the motion of the punctures at the onset of merger is 
still rather mildly relativistic. This finding turns
out to be surprisingly insensitive to the initial momentum. 
Fig.~\ref{fig:PhaseSpacePdependence} shows runs for different $P$ with
angles chosen for maximum radiation efficency.
Increasing the initial momentum of the punctures leads to a motion
which is rather relativistic when entering the whirl phase, up to
$v \sim 0.5$ and Lorentz factor $W \sim 1.155$ for $P=6P_{qc}$ 
(see Fig.~\ref{fig:PhaseSpacePdependence}), but in the whirl they
decelerate by large amounts (for $P=6P_{qc}$ the decrease in velocity
is as large as during the merger). At merger however \emph{all} equal
mass evolutions approach 
$v = 0.22 \pm 0.02$ or $W = 1.025\pm0.005$.  

For a detailed look at the 
transition zone between bound and unbound evolutions, see 
Fig.~\ref{fig:PhaseSpace2qc} for the $P=2P_{qc}$
sequence varying $\Theta$.  First, we recognize the
two extreme cases of merger and unbound motion. Inbetween we find
a very complex transition which is governed by several additional 
(non-closed) loops corresponding to orbits that zoom out after a first
whirl phase thereby slowing down (i.e.\ move to the lower right) before
returning towards the upper left. Note, that during the next approach
the binary must follow a path further to the lower left because the
system is dissipative.
The set of apocenters from all runs in this sequence forms a lower 
envelope (as the upper envelope mentioned before) that is 
never crossed by any of our evolutions at the same initial $D$. 
Note how far the zooms may extend when the binary approaches the
merger / fly-by threshold resulting in an ever larger runtime. We face
difficulties evolving orbits near $\Theta \rightarrow
\Theta^{merger}_{fly-by}$ because those orbits need very long
evolutions (decrease of accuracy) and the black holes may reach
distances close to or even beyond the wave extraction sphere, thereby
producing artificial features in $E_{rad}$.

An investigation of the phase space trajectories of unequal mass BHBs
reveals that the general shape of the trajectories is quite robust
with respect to the mass ratio. An example is given in Fig.~\ref{fig:UEM3}.
There are two basic differences: (1) Unequal mass orbits move
systematically slower compared to a corresponding equal mass binary
with comparable initial momentum. (2) Unequal mass mergers start their
final plunge from an inreasingly larger coordinate separation than
equal mass mergers. With regard to gravitational radiation, unequal
mass binaries do not tap as deeply into the gravitational potential as
equal mass binaries, thus cannot extract as much energy from the
spacetime. Furthermore, the deceleration is milder for higher mass
ratios, again suggesting weaker signals from higher mass ratios.

\section{Conclusions}
\label{Conclusion}

Numerical relativity has confirmed the existence of zoom-whirl 
orbits beyond the geodesic and PN regime
\cite{PreKhu07,SpeBerCar07,HinVaiHer07,WasHeaHer08,HeaLevSho09,GolBru09,HeaHerHin08}, thereby 
emphasizing their general relativistic origin. 
Previous studies explored the rich extension to the 
phenomenology of the GR two-body problem offered by zoom-whirl 
dynamics in various ways.
In this work we performed numerical relativity simulations to
investigate the parameter space of comparable mass, non-spinning,
eccentric BHB for low and intermediate momenta
more comprehensively than before. We 
explored zoom-whirl behavior in both the hyperbolic and elliptic 
regime carrying out more than $100$ numerical evolutions 
in order to obtain a decent sampling of the 
underlying parameter space. We discussed various features of the orbits and 
characterized the corresponding GW emission, and 
we developed new diagnostics to analyze binary spacetimes by using
phase space trajectories and a histogram of the coordinate separation.

%
%
For elliptic orbits, 
we discover zoom-whirls with imprints in the GWs
that are comparable in amplitude to the merger 
waveform for eccentricities as low as $e \sim 0.5$. 
This is an important finding for the astrophysical relevance 
of zoom-whirl orbits. In particular, such values are within 
expected eccentricities of supermassive BHB that have resulted 
from galaxy mergers and subsequent star- or gas-driven 
hardening \cite{RoeSes11}.
For low momenta, there occur several minima and maxima in the
radiated energy when varying the shooting angle from head-on to quasi-circular orbits.
We demonstrate that zoom-whirl dynamics may actually 
\emph{minimize} the radiated energy in sharp contrast to 
\cite{Pet64}.

In the elliptic regime whirls are found only during
their last encounter. 
They emerge in \emph{disjunct} intervals of the initial 
angular momentum (i.e.~shooting angle). Apparently, as 
long as the binaries are not circularized just prior to 
plunge, zoom-whirls can always be found during the last 
encounter by a very modest amount of fine-tuning.
%
%
In the hyperbolic regime we find that all evolutions 
that lead to dynamical capture reveal whirl features during 
the capturing encounter and then simply plunge during the 
following encounter, potentially from a large separation.

High-momentum zoom-whirls maximize the radiated energy. The Kerr 
spacetime that the merger remnant will settle down to exhibits a
larger spin parameter. This translates into a tighter unstable 
circular orbit (i.e.~the whirl radius) resulting in more pronounced
dynamics of the mass quadrupole. Especially the first, capturing 
close-encounter burst in high momentum evolutions can easily 
overwhelm the merger signal.

In the unequal mass case eccentric BHB are found to be more efficient
radiators than expected from quasi-cirular studies of unequal mass
BHB. As a consequence the whirls are not significantly longer for the
mass ratios under consideration. We note that numerical relativity is
seeing improvements in dealing with large mass ratios
\cite{LouZlo10,GonSpeBru08}. More detailed studies for unequal masses
and also the inclusion of spin are promising directions for eccentric
BHB simulations in the future.

The present work as well as other studies strongly suggest to include
eccentricity in the waveform templates used in the data analysis of GW
detectors, since eccentricity effectively breaks
degeneracies in parameter space.

%
\acknowledgments
We are grateful to David Hilditch, Sebastiano Bernuzzi, Ulrich Sperhake, 
Frans Pretorius, Luis Lehner, Deirdre Shoemaker, Doreen M\"uller, Norbert Lages, and Marcus Thierfelder 
for discussions, and we also thank Marcus Thierfelder for giving access 
to his event horizon finder.
This work was supported in part by grants DFG GK 1523, DFG
SFB/Transregio~7, and DLR LISA Germany at the Friedrich-Schiller
University Jena as well as NSF Grant PHY-0963136, and
NASA Grant NNX11AE11G at the University of Illinois at
Urbana-Champaign.
Computations were performed at LRZ Munich. 




\bibliography{refs,refsextra}

\begin{thebibliography}{100}%
\makeatletter
\providecommand \@ifxundefined [1]{%
 \ifx #1\undefined \expandafter \@firstoftwo
 \else \expandafter \@secondoftwo
\fi
}%
\providecommand \@ifnum [1]{%
 \ifnum #1\expandafter \@firstoftwo
 \else \expandafter \@secondoftwo
\fi
}%
\providecommand \enquote [1]{``#1''}%
\providecommand \bibnamefont  [1]{#1}%
\providecommand \bibfnamefont [1]{#1}%
\providecommand \citenamefont [1]{#1}%
\providecommand\href[0]{\@sanitize\@href}%
\providecommand\@href[1]{\endgroup\@@startlink{#1}\endgroup\@@href}%
\providecommand\@@href[1]{#1\@@endlink}%
\providecommand \@sanitize [0]{\begingroup\catcode`\&12\catcode`\#12\relax}%
\@ifxundefined \pdfoutput {\@firstoftwo}{%
 \@ifnum{\z@=\pdfoutput}{\@firstoftwo}{\@secondoftwo}%
}{%
 \providecommand\@@startlink[1]{\leavevmode\special{html:<a href="#1">}}%
 \providecommand\@@endlink[0]{\special{html:</a>}}%
}{%
 \providecommand\@@startlink[1]{%
  \leavevmode
  \pdfstartlink
   attr{/Border[0 0 1 ]/H/I/C[0 1 1]}%
   user{/Subtype/Link/A<</Type/Action/S/URI/URI(#1)>>}%
  \relax
 }%
 \providecommand\@@endlink[0]{\pdfendlink}%
}%
\providecommand \url  [0]{\begingroup\@sanitize \@url }%
\providecommand \@url [1]{\endgroup\@href {#1}{\urlprefix}}%
\providecommand \urlprefix [0]{URL }%
\providecommand \Eprint[0]{\href }%
\@ifxundefined \urlstyle {%
  \providecommand \doi [1]{doi:\discretionary{}{}{}#1}%
}{%
  \providecommand \doi [0]{doi:\discretionary{}{}{}\begingroup
  \urlstyle{rm}\Url }%
}%
\providecommand \doibase [0]{http://dx.doi.org/}%
\providecommand \Doi[1]{\href{\doibase#1}}%
\providecommand \bibAnnote [3]{%
  \BibitemShut{#1}%
  \begin{quotation}\noindent
    \textsc{Key:}\ #2\\\textsc{Annotation:}\ #3%
  \end{quotation}%
}%
\providecommand \bibAnnoteFile [2]{%
  \IfFileExists{#2}{\bibAnnote {#1} {#2} {\input{#2}}}{}%
}%
\providecommand \typeout [0]{\immediate \write \m@ne }%
\providecommand \selectlanguage [0]{\@gobble}%
\providecommand \bibinfo [0]{\@secondoftwo}%
\providecommand \bibfield [0]{\@secondoftwo}%
\providecommand \translation [1]{[#1]}%
\providecommand \BibitemOpen[0]{}%
\providecommand \bibitemStop [0]{}%
\providecommand \bibitemNoStop [0]{.\EOS\space}%
\providecommand \EOS [0]{\spacefactor3000\relax}%
\providecommand \BibitemShut [1]{\csname bibitem#1\endcsname}%
\bibitem{GlaKen02}%
  \BibitemOpen
  \bibfield{author}{%
  \bibinfo {author} {\bibfnamefont{K.}~\bibnamefont{Glampedakis}}\ and\
  \bibinfo {author} {\bibfnamefont{D.}~\bibnamefont{Kennefick}},\ }%
  \bibfield{journal}{%
  \Doi{10.1103/PhysRevD.66.044002}{\bibinfo {journal} {Phys. Rev.}}\ }%
  \textbf{\bibinfo {volume} {D66}},\ \bibinfo {pages} {044002} (\bibinfo {year}
  {2002}),\ \Eprint{http://arxiv.org/abs/gr-qc/0203086}{arXiv:gr-qc/0203086}%
  \bibAnnoteFile{NoStop}{GlaKen02}%
\bibitem{Note1}%
  \BibitemOpen
  \bibinfo {note} {The term zoom-whirl is attributed to C. Cutler and E.
  Poisson (who in turn mention it was suggested by K. Thorne).}%
  \bibAnnoteFile{Stop}{Note1}%
\bibitem{HulTay75}%
  \BibitemOpen
  \bibfield{author}{%
  \bibinfo {author} {\bibfnamefont{R.~A.}\ \bibnamefont{Hulse}}\ and\ \bibinfo
  {author} {\bibfnamefont{J.~H.}\ \bibnamefont{Taylor}},\ }%
  \bibfield{journal}{%
  \bibinfo {journal} {ApJ}\ }%
  \textbf{\bibinfo {volume} {195}},\ \bibinfo {pages} {L51} (\bibinfo {year}
  {1975})%
  \bibAnnoteFile{NoStop}{HulTay75}%
\bibitem{Bur03}%
  \BibitemOpen
  \bibfield{author}{%
  \bibinfo {author} {\bibfnamefont{M.}~\bibnamefont{Burgay}}, \bibinfo {author}
  {\bibfnamefont{N.}~\bibnamefont{D'Amico}}, \bibinfo {author}
  {\bibfnamefont{A.}~\bibnamefont{Possenti}}, \bibinfo {author}
  {\bibfnamefont{R.}~\bibnamefont{Manchester}}, \bibinfo {author}
  {\bibfnamefont{A.}~\bibnamefont{Lyne}}, \bibinfo {author}
  {\bibfnamefont{B.}~\bibnamefont{Joshi}}, \bibinfo {author}
  {\bibfnamefont{M.}~\bibnamefont{McLaughlin}}, \bibinfo {author}
  {\bibfnamefont{M.}~\bibnamefont{Kramer}}, \bibinfo {author}
  {\bibfnamefont{J.}~\bibnamefont{Sarkissian}}, \bibinfo {author}
  {\bibfnamefont{F.}~\bibnamefont{Camilo}}, \bibinfo {author}
  {\bibfnamefont{V.}~\bibnamefont{Kalogera}}, \bibinfo {author}
  {\bibfnamefont{C.}~\bibnamefont{Kim}},\ and\ \bibinfo {author}
  {\bibfnamefont{D.}~\bibnamefont{Lorimer}},\ }%
  \bibfield{journal}{%
  \Doi{10.1038/nature02124}{\bibinfo {journal} {Nature}}\ }%
  \textbf{\bibinfo {volume} {426}},\ \bibinfo {pages} {531} (\bibinfo {year}
  {2003}),\ \Eprint{http://arxiv.org/abs/astro-ph/0312071}{astro-ph/0312071}%
  \bibAnnoteFile{NoStop}{Bur03}%
\bibitem{KraStaMan06}%
  \BibitemOpen
  \bibfield{author}{%
  \bibinfo {author} {\bibfnamefont{M.}~\bibnamefont{Kramer}}, \bibinfo {author}
  {\bibfnamefont{I.}~\bibnamefont{Stairs}}, \bibinfo {author}
  {\bibfnamefont{R.}~\bibnamefont{Manchester}}, \bibinfo {author}
  {\bibfnamefont{M.}~\bibnamefont{McLaughlin}}, \bibinfo {author}
  {\bibfnamefont{A.}~\bibnamefont{Lyne}}, \bibinfo {author}
  {\bibfnamefont{R.}~\bibnamefont{Ferdman}}, \bibinfo {author}
  {\bibfnamefont{M.}~\bibnamefont{Burgay}}, \bibinfo {author}
  {\bibfnamefont{D.}~\bibnamefont{Lorimer}}, \bibinfo {author}
  {\bibfnamefont{A.}~\bibnamefont{Possenti}}, \bibinfo {author}
  {\bibfnamefont{N.}~\bibnamefont{D'Amico}}, \bibinfo {author}
  {\bibfnamefont{J.}~\bibnamefont{Sarkissian}}, \bibinfo {author}
  {\bibfnamefont{G.}~\bibnamefont{Hobbs}}, \bibinfo {author}
  {\bibfnamefont{J.}~\bibnamefont{Reynolds}}, \bibinfo {author}
  {\bibfnamefont{P.}~\bibnamefont{Freire}},\ and\ \bibinfo {author}
  {\bibfnamefont{F.}~\bibnamefont{Camilo}},\ }%
  \bibfield{journal}{%
  \bibinfo {journal} {Science}\ }%
  \textbf{\bibinfo {volume} {314}},\ \bibinfo {pages} {97} (\bibinfo {year}
  {2006}),\
  \Eprint{http://arxiv.org/abs/astro-ph/0609417}{arXiv:astro-ph/0609417
  [astro-ph]}%
  \bibAnnoteFile{NoStop}{KraStaMan06}%
\bibitem{Sillanpaa:1988zz}%
  \BibitemOpen
  \bibfield{author}{%
  \bibinfo {author} {\bibfnamefont{A.}~\bibnamefont{Sillanpaa}}, \bibinfo
  {author} {\bibfnamefont{S.}~\bibnamefont{Haarala}}, \bibinfo {author}
  {\bibfnamefont{M.}~\bibnamefont{Valtonen}}, \bibinfo {author}
  {\bibfnamefont{B.}~\bibnamefont{Sundelius}},\ and\ \bibinfo {author}
  {\bibfnamefont{G.}~\bibnamefont{Byrd}},\ }%
  \bibfield{journal}{%
  \bibinfo {journal} {Astrophys.J.}\ }%
  \textbf{\bibinfo {volume} {325}},\ \bibinfo {pages} {628} (\bibinfo {year}
  {1988})%
  \bibAnnoteFile{NoStop}{Sillanpaa:1988zz}%
\bibitem{valtonen-2009}%
  \BibitemOpen
  \bibfield{author}{%
  \bibinfo {author} {\bibfnamefont{M.}~\bibnamefont{Valtonen}}, \bibinfo
  {author} {\bibfnamefont{S.}~\bibnamefont{Mikkola}}, \bibinfo {author}
  {\bibfnamefont{D.}~\bibnamefont{Merritt}}, \bibinfo {author}
  {\bibfnamefont{A.}~\bibnamefont{Gopakumar}}, \bibinfo {author}
  {\bibfnamefont{H.}~\bibnamefont{Lehto}}, \emph{et~al.},\ }%
  \bibfield{journal}{%
  \Doi{10.1088/0004-637X/709/2/725}{\bibinfo {journal} {Astrophys.J.}}\ }%
  \textbf{\bibinfo {volume} {709}},\ \bibinfo {pages} {725} (\bibinfo {year}
  {2010}),\ \Eprint{http://arxiv.org/abs/0912.1209}{arXiv:0912.1209
  [astro-ph.HE]}%
  \bibAnnoteFile{NoStop}{valtonen-2009}%
\bibitem{LevCon10}%
  \BibitemOpen
  \bibfield{author}{%
  \bibinfo {author} {\bibfnamefont{J.}~\bibnamefont{Levin}}\ and\ \bibinfo
  {author} {\bibfnamefont{H.}~\bibnamefont{Contreras}},\ }%
  \bibfield{journal}{%
  \Doi{10.1088/0264-9381/28/17/175001}{\bibinfo {journal} {Class.Quant.Grav.}}\
  }%
  \textbf{\bibinfo {volume} {28}},\ \bibinfo {pages} {175001} (\bibinfo {year}
  {2011}),\ \Eprint{http://arxiv.org/abs/1009.2533}{arXiv:1009.2533 [gr-qc]}%
  \bibAnnoteFile{NoStop}{LevCon10}%
\bibitem{Cha83}%
  \BibitemOpen
  \bibfield{author}{%
  \bibinfo {author} {\bibfnamefont{S.}~\bibnamefont{Chandrasekhar}},\ }%
  \emph{\bibinfo {title} {The {M}athematical {T}heory of {B}lack {H}oles}}\
  (\bibinfo {publisher} {Oxford University Press, USA},\ \bibinfo {year}
  {1983})%
  \bibAnnoteFile{NoStop}{Cha83}%
\bibitem{Mar03}%
  \BibitemOpen
  \bibfield{author}{%
  \bibinfo {author} {\bibfnamefont{K.}~\bibnamefont{Martel}},\ }%
  \bibfield{journal}{%
  \bibinfo {journal} {Phys. Rev.}\ }%
  \textbf{\bibinfo {volume} {D69}},\ \bibinfo {pages} {044025} (\bibinfo {year}
  {2004}),\ \Eprint{http://arxiv.org/abs/gr-qc/0311017}{arXiv:gr-qc/0311017
  [gr-qc]}%
  \bibAnnoteFile{NoStop}{Mar03}%
\bibitem{LevGiz08}%
  \BibitemOpen
  \bibfield{author}{%
  \bibinfo {author} {\bibfnamefont{J.}~\bibnamefont{Levin}}\ and\ \bibinfo
  {author} {\bibfnamefont{G.}~\bibnamefont{Perez-Giz}},\ }%
  \bibfield{journal}{%
  \bibinfo {journal} {Phys. Rev.}\ }%
  \textbf{\bibinfo {volume} {D77}},\ \bibinfo {pages} {103005} (\bibinfo {year}
  {2008}),\ \Eprint{http://arxiv.org/abs/0802.0459}{arXiv:0802.0459 [gr-qc]}%
  \bibAnnoteFile{NoStop}{LevGiz08}%
\bibitem{DraHug05}%
  \BibitemOpen
  \bibfield{author}{%
  \bibinfo {author} {\bibfnamefont{S.}~\bibnamefont{Drasco}}\ and\ \bibinfo
  {author} {\bibfnamefont{S.~A.}\ \bibnamefont{Hughes}},\ }%
  \bibfield{journal}{%
  \bibinfo {journal} {Phys. Rev.}\ }%
  \textbf{\bibinfo {volume} {D73}},\ \bibinfo {pages} {024027} (\bibinfo {year}
  {2006}),\ \Eprint{http://arxiv.org/abs/gr-qc/0509101}{arXiv:gr-qc/0509101
  [gr-qc]}%
  \bibAnnoteFile{NoStop}{DraHug05}%
\bibitem{Haas:2007kz}%
  \BibitemOpen
  \bibfield{author}{%
  \bibinfo {author} {\bibfnamefont{R.}~\bibnamefont{Haas}},\ }%
  \bibfield{journal}{%
  \Doi{10.1103/PhysRevD.75.124011}{\bibinfo {journal} {Phys. Rev.}}\ }%
  \textbf{\bibinfo {volume} {D75}},\ \bibinfo {pages} {124011} (\bibinfo {year}
  {2007}),\ \Eprint{http://arxiv.org/abs/0704.0797}{arXiv:0704.0797 [gr-qc]}%
  \bibAnnoteFile{NoStop}{Haas:2007kz}%
\bibitem{LevGro08}%
  \BibitemOpen
  \bibfield{author}{%
  \bibinfo {author} {\bibfnamefont{J.}~\bibnamefont{Levin}}\ and\ \bibinfo
  {author} {\bibfnamefont{B.}~\bibnamefont{Grossman}},\ }%
  \bibfield{journal}{%
  \bibinfo {journal} {Phys. Rev.}\ }%
  \textbf{\bibinfo {volume} {D79}},\ \bibinfo {pages} {043016} (\bibinfo {year}
  {2009}),\ \Eprint{http://arxiv.org/abs/0809.3838}{arXiv:0809.3838 [gr-qc]}%
  \bibAnnoteFile{NoStop}{LevGro08}%
\bibitem{GroLev08}%
  \BibitemOpen
  \bibfield{author}{%
  \bibinfo {author} {\bibfnamefont{R.}~\bibnamefont{Grossman}}\ and\ \bibinfo
  {author} {\bibfnamefont{J.}~\bibnamefont{Levin}},\ }%
  \bibfield{journal}{%
  \bibinfo {journal} {Phys. Rev.}\ }%
  \textbf{\bibinfo {volume} {D79}},\ \bibinfo {pages} {043017} (\bibinfo {year}
  {2009}),\ \Eprint{http://arxiv.org/abs/0811.3798}{arXiv:0811.3798 [gr-qc]}%
  \bibAnnoteFile{NoStop}{GroLev08}%
\bibitem{hughes-2005-94}%
  \BibitemOpen
  \bibfield{author}{%
  \bibinfo {author} {\bibfnamefont{S.~A.}\ \bibnamefont{Hughes}}, \bibinfo
  {author} {\bibfnamefont{S.}~\bibnamefont{Drasco}}, \bibinfo {author}
  {\bibfnamefont{E.~E.}\ \bibnamefont{Flanagan}},\ and\ \bibinfo {author}
  {\bibfnamefont{J.}~\bibnamefont{Franklin}},\ }%
  \bibfield{journal}{%
  \Doi{10.1103/PhysRevLett.94.221101}{\bibinfo {journal} {Phys. Rev. Lett.}}\
  }%
  \textbf{\bibinfo {volume} {94}},\ \bibinfo {pages} {221101} (\bibinfo {month}
  {Jun}\ \bibinfo {year} {2005})%
  \bibAnnoteFile{NoStop}{hughes-2005-94}%
\bibitem{PatWil02}%
  \BibitemOpen
  \bibfield{author}{%
  \bibinfo {author} {\bibfnamefont{M.~E.}\ \bibnamefont{Pati}}\ and\ \bibinfo
  {author} {\bibfnamefont{C.~M.}\ \bibnamefont{Will}},\ }%
  \bibfield{journal}{%
  \bibinfo {journal} {Phys. Rev.}\ }%
  \textbf{\bibinfo {volume} {D65}},\ \bibinfo {pages} {104008} (\bibinfo {year}
  {2002})%
  \bibAnnoteFile{NoStop}{PatWil02}%
\bibitem{PreKhu07}%
  \BibitemOpen
  \bibfield{author}{%
  \bibinfo {author} {\bibfnamefont{F.}~\bibnamefont{Pretorius}}\ and\ \bibinfo
  {author} {\bibfnamefont{D.}~\bibnamefont{Khurana}},\ }%
  \bibfield{journal}{%
  \Doi{10.1088/0264-9381/24/12/S07}{\bibinfo {journal} {Class. Quant. Grav.}}\
  }%
  \textbf{\bibinfo {volume} {24}},\ \bibinfo {pages} {S83} (\bibinfo {year}
  {2007}),\ \Eprint{http://arxiv.org/abs/gr-qc/0702084}{arXiv:gr-qc/0702084}%
  \bibAnnoteFile{NoStop}{PreKhu07}%
\bibitem{SpeBerCar07}%
  \BibitemOpen
  \bibfield{author}{%
  \bibinfo {author} {\bibfnamefont{U.}~\bibnamefont{Sperhake}}, \bibinfo
  {author} {\bibfnamefont{E.}~\bibnamefont{Berti}}, \bibinfo {author}
  {\bibfnamefont{V.}~\bibnamefont{Cardoso}}, \bibinfo {author}
  {\bibfnamefont{J.~A.}\ \bibnamefont{Gonz{\'a}lez}}, \bibinfo {author}
  {\bibfnamefont{B.}~\bibnamefont{Br{\"u}gmann}},\ and\ \bibinfo {author}
  {\bibfnamefont{M.}~\bibnamefont{Ansorg}},\ }%
  \bibfield{journal}{%
  \Doi{10.1103/PhysRevD.78.064069}{\bibinfo {journal} {Phys. Rev.}}\ }%
  \textbf{\bibinfo {volume} {D78}},\ \bibinfo {pages} {064069} (\bibinfo {year}
  {2008}),\ \Eprint{http://arxiv.org/abs/0710.3823}{arXiv:0710.3823 [gr-qc]}%
  \bibAnnoteFile{NoStop}{SpeBerCar07}%
\bibitem{HinVaiHer07}%
  \BibitemOpen
  \bibfield{author}{%
  \bibinfo {author} {\bibfnamefont{I.}~\bibnamefont{Hinder}}, \bibinfo {author}
  {\bibfnamefont{B.}~\bibnamefont{Vaishnav}}, \bibinfo {author}
  {\bibfnamefont{F.}~\bibnamefont{Herrmann}}, \bibinfo {author}
  {\bibfnamefont{D.}~\bibnamefont{Shoemaker}},\ and\ \bibinfo {author}
  {\bibfnamefont{P.}~\bibnamefont{Laguna}},\ }%
  \bibfield{journal}{%
  \Doi{10.1103/PhysRevD.77.081502}{\bibinfo {journal} {Phys. Rev.}}\ }%
  \textbf{\bibinfo {volume} {D77}},\ \bibinfo {pages} {081502} (\bibinfo {year}
  {2008}),\ \Eprint{http://arxiv.org/abs/0710.5167}{arXiv:0710.5167 [gr-qc]}%
  \bibAnnoteFile{NoStop}{HinVaiHer07}%
\bibitem{WasHeaHer08}%
  \BibitemOpen
  \bibfield{author}{%
  \bibinfo {author} {\bibfnamefont{M.~C.}\ \bibnamefont{Washik}}, \bibinfo
  {author} {\bibfnamefont{J.}~\bibnamefont{Healy}}, \bibinfo {author}
  {\bibfnamefont{F.}~\bibnamefont{Herrmann}}, \bibinfo {author}
  {\bibfnamefont{I.}~\bibnamefont{Hinder}}, \bibinfo {author}
  {\bibfnamefont{D.~M.}\ \bibnamefont{Shoemaker}}, \bibinfo {author}
  {\bibfnamefont{P.}~\bibnamefont{Laguna}},\ and\ \bibinfo {author}
  {\bibfnamefont{R.~A.}\ \bibnamefont{Matzner}},\ }%
  \bibfield{journal}{%
  \bibinfo {journal} {Phys. Rev. Lett.}\ }%
  \textbf{\bibinfo {volume} {101}},\ \bibinfo {pages} {061102} (\bibinfo {year}
  {2008}),\ \Eprint{http://arxiv.org/abs/0802.2520}{arXiv:0802.2520 [gr-qc]}%
  \bibAnnoteFile{NoStop}{WasHeaHer08}%
\bibitem{GolBru09}%
  \BibitemOpen
  \bibfield{author}{%
  \bibinfo {author} {\bibfnamefont{R.}~\bibnamefont{Gold}}\ and\ \bibinfo
  {author} {\bibfnamefont{B.}~\bibnamefont{Br{\"u}gmann}},\ }%
  \bibfield{journal}{%
  \bibinfo {journal} {Class.Quant.Grav.}\ }%
  \textbf{\bibinfo {volume} {27}},\ \bibinfo {pages} {084035} (\bibinfo {year}
  {2010}),\ \Eprint{http://arxiv.org/abs/0911.3862}{arXiv:0911.3862 [gr-qc]}%
  \bibAnnoteFile{NoStop}{GolBru09}%
\bibitem{HeaLagMat09}%
  \BibitemOpen
  \bibfield{author}{%
  \bibinfo {author} {\bibfnamefont{J.}~\bibnamefont{Healy}}, \bibinfo {author}
  {\bibfnamefont{P.}~\bibnamefont{Laguna}}, \bibinfo {author}
  {\bibfnamefont{R.~A.}\ \bibnamefont{Matzner}},\ and\ \bibinfo {author}
  {\bibfnamefont{D.~M.}\ \bibnamefont{Shoemaker}},\ }%
  \bibfield{journal}{%
  \bibinfo {journal} {Phys. Rev.}\ }%
  \textbf{\bibinfo {volume} {D81}},\ \bibinfo {pages} {081501} (\bibinfo {year}
  {2010}),\ \Eprint{http://arxiv.org/abs/0905.3914}{arXiv:0905.3914 [gr-qc]}%
  \bibAnnoteFile{NoStop}{HeaLagMat09}%
\bibitem{HeaLevSho09}%
  \BibitemOpen
  \bibfield{author}{%
  \bibinfo {author} {\bibfnamefont{J.}~\bibnamefont{Healy}}, \bibinfo {author}
  {\bibfnamefont{J.}~\bibnamefont{Levin}},\ and\ \bibinfo {author}
  {\bibfnamefont{D.}~\bibnamefont{Shoemaker}},\ }%
  \bibfield{journal}{%
  \Doi{10.1103/PhysRevLett.103.131101}{\bibinfo {journal} {Phys. Rev. Lett.}}\
  }%
  \textbf{\bibinfo {volume} {103}},\ \bibinfo {pages} {131101} (\bibinfo {year}
  {2009}),\ \Eprint{http://arxiv.org/abs/0907.0671}{arXiv:0907.0671 [gr-qc]}%
  \bibAnnoteFile{NoStop}{HeaLevSho09}%
\bibitem{PhysRevD50-3816}%
  \BibitemOpen
  \bibfield{author}{%
  \bibinfo {author} {\bibfnamefont{C.}~\bibnamefont{Cutler}}, \bibinfo {author}
  {\bibfnamefont{D.}~\bibnamefont{Kennefick}},\ and\ \bibinfo {author}
  {\bibfnamefont{E.}~\bibnamefont{Poisson}},\ }%
  \bibfield{journal}{%
  \Doi{10.1103/PhysRevD.50.3816}{\bibinfo {journal} {Phys. Rev. D}}\ }%
  \textbf{\bibinfo {volume} {50}},\ \bibinfo {pages} {3816} (\bibinfo {month}
  {Sep}\ \bibinfo {year} {1994})%
  \bibAnnoteFile{NoStop}{PhysRevD50-3816}%
\bibitem{HeaHerHin08}%
  \BibitemOpen
  \bibfield{author}{%
  \bibinfo {author} {\bibfnamefont{J.}~\bibnamefont{Healy}}, \bibinfo {author}
  {\bibfnamefont{F.}~\bibnamefont{Herrmann}}, \bibinfo {author}
  {\bibfnamefont{I.}~\bibnamefont{Hinder}}, \bibinfo {author}
  {\bibfnamefont{D.~M.}\ \bibnamefont{Shoemaker}}, \bibinfo {author}
  {\bibfnamefont{P.}~\bibnamefont{Laguna}},\ and\ \bibinfo {author}
  {\bibfnamefont{R.~A.}\ \bibnamefont{Matzner}},\ }%
  \bibfield{journal}{%
  \bibinfo {journal} {Phys. Rev. Lett.}\ }%
  \textbf{\bibinfo {volume} {102}},\ \bibinfo {pages} {041101} (\bibinfo {year}
  {2009}),\ \Eprint{http://arxiv.org/abs/0807.3292}{arXiv:0807.3292 [gr-qc]}%
  \bibAnnoteFile{NoStop}{HeaHerHin08}%
\bibitem{martel-1999-60}%
  \BibitemOpen
  \bibfield{author}{%
  \bibinfo {author} {\bibfnamefont{K.}~\bibnamefont{Martel}}\ and\ \bibinfo
  {author} {\bibfnamefont{E.}~\bibnamefont{Poisson}},\ }%
  \bibfield{journal}{%
  \bibinfo {journal} {Phys. Rev.}\ }%
  \textbf{\bibinfo {volume} {D60}},\ \bibinfo {pages} {124008} (\bibinfo {year}
  {1999})%
  \bibAnnoteFile{NoStop}{martel-1999-60}%
\bibitem{Mar99}%
  \BibitemOpen
  \bibfield{author}{%
  \bibinfo {author} {\bibfnamefont{K.}~\bibnamefont{Martel}},\ }%
  \bibfield{journal}{%
  \bibinfo {journal} {AIP Conf. Proc.}\ }%
  \textbf{\bibinfo {volume} {493}},\ \bibinfo {pages} {48} (\bibinfo {year}
  {1999}),\ \Eprint{http://arxiv.org/abs/gr-qc/9908043}{arXiv:gr-qc/9908043
  [gr-qc]}%
  \bibAnnoteFile{NoStop}{Mar99}%
\bibitem{BroZim09}%
  \BibitemOpen
  \bibfield{author}{%
  \bibinfo {author} {\bibfnamefont{D.~A.}\ \bibnamefont{Brown}}\ and\ \bibinfo
  {author} {\bibfnamefont{P.~J.}\ \bibnamefont{Zimmerman}},\ }%
  \bibfield{journal}{%
  \bibinfo {journal} {Phys. Rev.}\ }%
  \textbf{\bibinfo {volume} {D81}},\ \bibinfo {pages} {024007} (\bibinfo {year}
  {2010}),\ \Eprint{http://arxiv.org/abs/0909.0066}{arXiv:0909.0066 [gr-qc]}%
  \bibAnnoteFile{NoStop}{BroZim09}%
\bibitem{Cokelaer:2009hj}%
  \BibitemOpen
  \bibfield{author}{%
  \bibinfo {author} {\bibfnamefont{T.}~\bibnamefont{Cokelaer}}\ and\ \bibinfo
  {author} {\bibfnamefont{D.}~\bibnamefont{Pathak}},\ }%
  \bibfield{journal}{%
  \bibinfo {journal} {Class. Quant. Grav.}\ }%
  \textbf{\bibinfo {volume} {26}},\ \bibinfo {pages} {045013} (\bibinfo {year}
  {2009}),\ \Eprint{http://arxiv.org/abs/0903.4791}{arXiv:0903.4791 [gr-qc]}%
  \bibAnnoteFile{NoStop}{Cokelaer:2009hj}%
\bibitem{VaiHinSho09}%
  \BibitemOpen
  \bibfield{author}{%
  \bibinfo {author} {\bibfnamefont{B.}~\bibnamefont{Vaishnav}}, \bibinfo
  {author} {\bibfnamefont{I.}~\bibnamefont{Hinder}}, \bibinfo {author}
  {\bibfnamefont{D.}~\bibnamefont{Shoemaker}},\ and\ \bibinfo {author}
  {\bibfnamefont{F.}~\bibnamefont{Herrman}},\ }%
  \bibfield{journal}{%
  \bibinfo {journal} {Classical and Quantum Gravity}\ }%
  \textbf{\bibinfo {volume} {26}},\ \bibinfo {pages} {204008} (\bibinfo {year}
  {2009})%
  \bibAnnoteFile{NoStop}{VaiHinSho09}%
\bibitem{Mik12}%
  \BibitemOpen
  \bibfield{author}{%
  \bibinfo {author} {\bibfnamefont{B.}~\bibnamefont{Mikoczi}}, \bibinfo
  {author} {\bibfnamefont{B.}~\bibnamefont{Kocsis}}, \bibinfo {author}
  {\bibfnamefont{P.}~\bibnamefont{Forgacs}},\ and\ \bibinfo {author}
  {\bibfnamefont{M.}~\bibnamefont{Vasuth}}}%
   (\bibinfo {year} {2012}),\
  \Eprint{http://arxiv.org/abs/1206.5786}{arXiv:1206.5786 [gr-qc]}%
  \bibAnnoteFile{NoStop}{Mik12}%
\bibitem{East:2012ww}%
  \BibitemOpen
  \bibfield{author}{%
  \bibinfo {author} {\bibfnamefont{W.~E.}\ \bibnamefont{East}}\ and\ \bibinfo
  {author} {\bibfnamefont{F.}~\bibnamefont{Pretorius}}}%
   (\bibinfo {year} {2012}),\
  \Eprint{http://arxiv.org/abs/1208.5279}{arXiv:1208.5279 [astro-ph.HE]}%
  \bibAnnoteFile{NoStop}{East:2012ww}%
\bibitem{EasPreSte11}%
  \BibitemOpen
  \bibfield{author}{%
  \bibinfo {author} {\bibfnamefont{W.~E.}\ \bibnamefont{East}}, \bibinfo
  {author} {\bibfnamefont{F.}~\bibnamefont{Pretorius}},\ and\ \bibinfo {author}
  {\bibfnamefont{B.~C.}\ \bibnamefont{Stephens}},\ }%
  \bibfield{journal}{%
  \Doi{10.1103/PhysRevD.85.124009}{\bibinfo {journal} {Phys. Rev.}}\ }%
  \textbf{\bibinfo {volume} {D85}},\ \bibinfo {pages} {124009} (\bibinfo {year}
  {2012}),\ \Eprint{http://arxiv.org/abs/1111.3055}{arXiv:1111.3055
  [astro-ph.HE]}%
  \bibAnnoteFile{NoStop}{EasPreSte11}%
\bibitem{GolBerThi11}%
  \BibitemOpen
  \bibfield{author}{%
  \bibinfo {author} {\bibfnamefont{R.}~\bibnamefont{Gold}}, \bibinfo {author}
  {\bibfnamefont{S.}~\bibnamefont{Bernuzzi}}, \bibinfo {author}
  {\bibfnamefont{M.}~\bibnamefont{Thierfelder}}, \bibinfo {author}
  {\bibfnamefont{B.}~\bibnamefont{Br{\"u}gmann}},\ and\ \bibinfo {author}
  {\bibfnamefont{F.}~\bibnamefont{Pretorius}}}%
   (\bibinfo {year} {2011}),\
  \Eprint{http://arxiv.org/abs/1109.5128}{arXiv:1109.5128 [gr-qc]}%
  \bibAnnoteFile{NoStop}{GolBerThi11}%
\bibitem{SteEasPre11}%
  \BibitemOpen
  \bibfield{author}{%
  \bibinfo {author} {\bibfnamefont{B.~C.}\ \bibnamefont{Stephens}}, \bibinfo
  {author} {\bibfnamefont{W.~E.}\ \bibnamefont{East}},\ and\ \bibinfo {author}
  {\bibfnamefont{F.}~\bibnamefont{Pretorius}},\ }%
  \bibfield{journal}{%
  \bibinfo {journal} {Astrophys. J.}\ }%
  \textbf{\bibinfo {volume} {737}},\ \bibinfo {pages} {L5} (\bibinfo {year}
  {2011}),\ \Eprint{http://arxiv.org/abs/1105.3175}{arXiv:1105.3175
  [astro-ph.HE]}%
  \bibAnnoteFile{NoStop}{SteEasPre11}%
\bibitem{ShiOkaYam08}%
  \BibitemOpen
  \bibfield{author}{%
  \bibinfo {author} {\bibfnamefont{M.}~\bibnamefont{Shibata}}, \bibinfo
  {author} {\bibfnamefont{H.}~\bibnamefont{Okawa}},\ and\ \bibinfo {author}
  {\bibfnamefont{T.}~\bibnamefont{Yamamoto}},\ }%
  \bibfield{journal}{%
  \Doi{10.1103/PhysRevD.78.101501}{\bibinfo {journal} {Phys. Rev.}}\ }%
  \textbf{\bibinfo {volume} {D78}},\ \bibinfo {pages} {101501} (\bibinfo {year}
  {2008}),\ \Eprint{http://arxiv.org/abs/0810.4735}{arXiv:0810.4735 [gr-qc]}%
  \bibAnnoteFile{NoStop}{ShiOkaYam08}%
\bibitem{Sperhake:2009jz}%
  \BibitemOpen
  \bibfield{author}{%
  \bibinfo {author} {\bibfnamefont{U.}~\bibnamefont{Sperhake}}, \bibinfo
  {author} {\bibfnamefont{V.}~\bibnamefont{Cardoso}}, \bibinfo {author}
  {\bibfnamefont{F.}~\bibnamefont{Pretorius}}, \bibinfo {author}
  {\bibfnamefont{E.}~\bibnamefont{Berti}}, \bibinfo {author}
  {\bibfnamefont{T.}~\bibnamefont{Hinderer}}, \emph{et~al.},\ }%
  \bibfield{journal}{%
  \Doi{10.1103/PhysRevLett.103.131102}{\bibinfo {journal} {Phys. Rev. Lett.}}\
  }%
  \textbf{\bibinfo {volume} {103}},\ \bibinfo {pages} {131102} (\bibinfo {year}
  {2009}),\ \Eprint{http://arxiv.org/abs/0907.1252}{arXiv:0907.1252 [gr-qc]}%
  \bibAnnoteFile{NoStop}{Sperhake:2009jz}%
\bibitem{SpeCarPre08}%
  \BibitemOpen
  \bibfield{author}{%
  \bibinfo {author} {\bibfnamefont{U.}~\bibnamefont{Sperhake}}, \bibinfo
  {author} {\bibfnamefont{V.}~\bibnamefont{Cardoso}}, \bibinfo {author}
  {\bibfnamefont{F.}~\bibnamefont{Pretorius}}, \bibinfo {author}
  {\bibfnamefont{E.}~\bibnamefont{Berti}},\ and\ \bibinfo {author}
  {\bibfnamefont{J.~A.}\ \bibnamefont{Gonz{\'a}lez}},\ }%
  \bibfield{journal}{%
  \bibinfo {journal} {Phys.Rev.Lett.}\ }%
  \textbf{\bibinfo {volume} {101}},\ \bibinfo {pages} {161101} (\bibinfo {year}
  {2008}),\ \Eprint{http://arxiv.org/abs/0806.1738}{arXiv:0806.1738 [gr-qc]}%
  \bibAnnoteFile{NoStop}{SpeCarPre08}%
\bibitem{Berti:2010ce}%
  \BibitemOpen
  \bibfield{author}{%
  \bibinfo {author} {\bibfnamefont{E.}~\bibnamefont{Berti}}, \bibinfo {author}
  {\bibfnamefont{V.}~\bibnamefont{Cardoso}}, \bibinfo {author}
  {\bibfnamefont{T.}~\bibnamefont{Hinderer}}, \bibinfo {author}
  {\bibfnamefont{M.}~\bibnamefont{Lemos}}, \bibinfo {author}
  {\bibfnamefont{F.}~\bibnamefont{Pretorius}}, \emph{et~al.},\ }%
  \bibfield{journal}{%
  \Doi{10.1103/PhysRevD.81.104048}{\bibinfo {journal} {Phys. Rev.}}\ }%
  \textbf{\bibinfo {volume} {D81}},\ \bibinfo {pages} {104048} (\bibinfo {year}
  {2010}),\ \Eprint{http://arxiv.org/abs/1003.0812}{arXiv:1003.0812 [gr-qc]}%
  \bibAnnoteFile{NoStop}{Berti:2010ce}%
\bibitem{Sperhake:2010dn}%
  \BibitemOpen
  \bibfield{author}{%
  \bibinfo {author} {\bibfnamefont{U.}~\bibnamefont{Sperhake}}, \bibinfo
  {author} {\bibfnamefont{V.}~\bibnamefont{Cardoso}}, \bibinfo {author}
  {\bibfnamefont{F.}~\bibnamefont{Pretorius}}, \bibinfo {author}
  {\bibfnamefont{E.}~\bibnamefont{Berti}}, \bibinfo {author}
  {\bibfnamefont{T.}~\bibnamefont{Hinderer}}, \emph{et~al.}}%
   (\bibinfo {year} {2010}),\
  \Eprint{http://arxiv.org/abs/1003.0882}{arXiv:1003.0882 [gr-qc]}%
  \bibAnnoteFile{NoStop}{Sperhake:2010dn}%
\bibitem{Pet64}%
  \BibitemOpen
  \bibfield{author}{%
  \bibinfo {author} {\bibfnamefont{P.~C.}\ \bibnamefont{Peters}},\ }%
  \bibfield{journal}{%
  \bibinfo {journal} {Phys. Rev.}\ }%
  \textbf{\bibinfo {volume} {136}},\ \bibinfo {pages} {B1224} (\bibinfo {year}
  {1964})%
  \bibAnnoteFile{NoStop}{Pet64}%
\bibitem{SesRoeRey11}%
  \BibitemOpen
  \bibfield{author}{%
  \bibinfo {author} {\bibfnamefont{A.}~\bibnamefont{Sesana}}, \bibinfo {author}
  {\bibfnamefont{C.}~\bibnamefont{Roedig}}, \bibinfo {author}
  {\bibfnamefont{M.}~\bibnamefont{Reynolds}},\ and\ \bibinfo {author}
  {\bibfnamefont{M.}~\bibnamefont{Dotti}}}%
   (\bibinfo {year} {2011}),\
  \Eprint{http://arxiv.org/abs/1107.2927}{arXiv:1107.2927 [astro-ph.CO]}%
  \bibAnnoteFile{NoStop}{SesRoeRey11}%
\bibitem{Khan:2011gi}%
  \BibitemOpen
  \bibfield{author}{%
  \bibinfo {author} {\bibfnamefont{F.}~\bibnamefont{Khan}}, \bibinfo {author}
  {\bibfnamefont{A.}~\bibnamefont{Just}},\ and\ \bibinfo {author}
  {\bibfnamefont{D.}~\bibnamefont{Merritt}},\ }%
  \bibfield{journal}{%
  \Doi{10.1088/0004-637X/732/2/89}{\bibinfo {journal} {Astrophys.J.}}\ }%
  \textbf{\bibinfo {volume} {732}},\ \bibinfo {pages} {89} (\bibinfo {year}
  {2011}),\ \Eprint{http://arxiv.org/abs/1103.0272}{arXiv:1103.0272
  [astro-ph.CO]}%
  \bibAnnoteFile{NoStop}{Khan:2011gi}%
\bibitem{Preto:2011gu}%
  \BibitemOpen
  \bibfield{author}{%
  \bibinfo {author} {\bibfnamefont{M.}~\bibnamefont{Preto}}, \bibinfo {author}
  {\bibfnamefont{I.}~\bibnamefont{Berentzen}}, \bibinfo {author}
  {\bibfnamefont{P.}~\bibnamefont{Berczik}},\ and\ \bibinfo {author}
  {\bibfnamefont{R.}~\bibnamefont{Spurzem}}}%
   (\bibinfo {year} {2011}),\
  \Eprint{http://arxiv.org/abs/1102.4855}{arXiv:1102.4855 [astro-ph.GA]}%
  \bibAnnoteFile{NoStop}{Preto:2011gu}%
\bibitem{PapNelMas01}%
  \BibitemOpen
  \bibfield{author}{%
  \bibinfo {author} {\bibfnamefont{J.~C.~B.}\ \bibnamefont{Papaloizou}},
  \bibinfo {author} {\bibfnamefont{R.~P.}\ \bibnamefont{{Nelson}}},\ and\
  \bibinfo {author} {\bibfnamefont{F.}~\bibnamefont{{Masset}}},\ }%
  \bibfield{journal}{%
  \Doi{10.1051/0004-6361:20000011}{\bibinfo {journal} {Astronomy and
  Astrophysics}}\ }%
  \textbf{\bibinfo {volume} {366}},\ \bibinfo {pages} {263} (\bibinfo {month}
  {jan}\ \bibinfo {year} {2001})%
  \bibAnnoteFile{NoStop}{PapNelMas01}%
\bibitem{Armitage:2005xq}%
  \BibitemOpen
  \bibfield{author}{%
  \bibinfo {author} {\bibfnamefont{P.~J.}\ \bibnamefont{Armitage}}\ and\
  \bibinfo {author} {\bibfnamefont{P.}~\bibnamefont{Natarajan}},\ }%
  \bibfield{journal}{%
  \Doi{10.1086/497108}{\bibinfo {journal} {Astrophys.J.}}\ }%
  \textbf{\bibinfo {volume} {634}},\ \bibinfo {pages} {921} (\bibinfo {year}
  {2005}),\
  \Eprint{http://arxiv.org/abs/astro-ph/0508493}{arXiv:astro-ph/0508493
  [astro-ph]}%
  \bibAnnoteFile{NoStop}{Armitage:2005xq}%
\bibitem{Cuadra:2008xn}%
  \BibitemOpen
  \bibfield{author}{%
  \bibinfo {author} {\bibfnamefont{J.}~\bibnamefont{Cuadra}}, \bibinfo {author}
  {\bibfnamefont{P.}~\bibnamefont{Armitage}}, \bibinfo {author}
  {\bibfnamefont{R.}~\bibnamefont{Alexander}},\ and\ \bibinfo {author}
  {\bibfnamefont{M.}~\bibnamefont{Begelman}}}%
   (\bibinfo {year} {2008}),\
  \Eprint{http://arxiv.org/abs/0809.0311}{arXiv:0809.0311 [astro-ph]}%
  \bibAnnoteFile{NoStop}{Cuadra:2008xn}%
\bibitem{RoeSes11}%
  \BibitemOpen
  \bibfield{author}{%
  \bibinfo {author} {\bibfnamefont{C.}~\bibnamefont{Roedig}}\ and\ \bibinfo
  {author} {\bibfnamefont{A.}~\bibnamefont{Sesana}},\ }%
  \bibfield{journal}{%
  \Doi{10.1088/1742-6596/363/1/012035}{\bibinfo {journal} {J.Phys.Conf.Ser.}}\
  }%
  \textbf{\bibinfo {volume} {363}},\ \bibinfo {pages} {012035} (\bibinfo {year}
  {2012}),\ \Eprint{http://arxiv.org/abs/1111.3742}{arXiv:1111.3742
  [astro-ph.CO]}%
  \bibAnnoteFile{NoStop}{RoeSes11}%
\bibitem{zwart-1999}%
  \BibitemOpen
  \bibfield{author}{%
  \bibinfo {author} {\bibfnamefont{S.~F.}\ \bibnamefont{Portegies~Zwart}}\ and\
  \bibinfo {author} {\bibfnamefont{S.}~\bibnamefont{McMillan}}}%
   (\bibinfo {year} {1999}),\
  \Eprint{http://arxiv.org/abs/astro-ph/9910061}{arXiv:astro-ph/9910061
  [astro-ph]}%
  \bibAnnoteFile{NoStop}{zwart-1999}%
\bibitem{benacquista-2001-18}%
  \BibitemOpen
  \bibfield{author}{%
  \bibinfo {author} {\bibfnamefont{M.}~\bibnamefont{Benacquista}}, \bibinfo
  {author} {\bibfnamefont{S.~F.}\ \bibnamefont{Portegies~Zwart}},\ and\
  \bibinfo {author} {\bibfnamefont{F.}~\bibnamefont{Rasio}},\ }%
  \bibfield{journal}{%
  \Doi{10.1088/0264-9381/18/19/308}{\bibinfo {journal} {Class.Quant.Grav.}}\ }%
  \textbf{\bibinfo {volume} {18}},\ \bibinfo {pages} {4025} (\bibinfo {year}
  {2001}),\ \Eprint{http://arxiv.org/abs/gr-qc/0010020}{arXiv:gr-qc/0010020
  [gr-qc]}%
  \bibAnnoteFile{NoStop}{benacquista-2001-18}%
\bibitem{wen-2003-598}%
  \BibitemOpen
  \bibfield{author}{%
  \bibinfo {author} {\bibfnamefont{L.}~\bibnamefont{Wen}},\ }%
  \bibfield{journal}{%
  \Doi{10.1086/378794}{\bibinfo {journal} {Astrophys.J.}}\ }%
  \textbf{\bibinfo {volume} {598}},\ \bibinfo {pages} {419} (\bibinfo {year}
  {2003}),\
  \Eprint{http://arxiv.org/abs/astro-ph/0211492}{arXiv:astro-ph/0211492
  [astro-ph]}%
  \bibAnnoteFile{NoStop}{wen-2003-598}%
\bibitem{pooley-2003}%
  \BibitemOpen
  \bibfield{author}{%
  \bibinfo {author} {\bibfnamefont{D.}~\bibnamefont{Pooley}}, \bibinfo {author}
  {\bibfnamefont{W.}~\bibnamefont{Lewin}}, \bibinfo {author}
  {\bibfnamefont{S.}~\bibnamefont{Anderson}}, \bibinfo {author}
  {\bibfnamefont{H.}~\bibnamefont{Baumgardt}}, \bibinfo {author}
  {\bibfnamefont{A.}~\bibnamefont{Filippenko}}, \emph{et~al.},\ }%
  \bibfield{journal}{%
  \Doi{10.1086/377074}{\bibinfo {journal} {Astrophys.J.}}\ }%
  \textbf{\bibinfo {volume} {591}},\ \bibinfo {pages} {L131} (\bibinfo {year}
  {2003}),\
  \Eprint{http://arxiv.org/abs/astro-ph/0305003}{arXiv:astro-ph/0305003
  [astro-ph]}%
  \bibAnnoteFile{NoStop}{pooley-2003}%
\bibitem{miller-2002}%
  \BibitemOpen
  \bibfield{author}{%
  \bibinfo {author} {\bibfnamefont{M.~C.}\ \bibnamefont{Miller}}\ and\ \bibinfo
  {author} {\bibfnamefont{D.~P.}\ \bibnamefont{Hamilton}},\ }%
  \bibfield{journal}{%
  \bibinfo {journal} {Astrophys.J.Lett.}}%
   (\bibinfo {year} {2002}),\
  \Eprint{http://arxiv.org/abs/astro-ph/0202298}{arXiv:astro-ph/0202298
  [astro-ph]}%
  \bibAnnoteFile{NoStop}{miller-2002}%
\bibitem{LeeRamVen09}%
  \BibitemOpen
  \bibfield{author}{%
  \bibinfo {author} {\bibfnamefont{W.~H.}\ \bibnamefont{Lee}}, \bibinfo
  {author} {\bibfnamefont{E.}~\bibnamefont{Ramirez-Ruiz}},\ and\ \bibinfo
  {author} {\bibfnamefont{G.}~\bibnamefont{van~de Ven}},\ }%
  \bibfield{journal}{%
  \Doi{10.1088/0004-637X/720/1/953}{\bibinfo {journal} {Astrophys. J.}}\ }%
  \textbf{\bibinfo {volume} {720}},\ \bibinfo {pages} {953} (\bibinfo {year}
  {2010}),\ \Eprint{http://arxiv.org/abs/0909.2884}{arXiv:0909.2884
  [astro-ph.HE]}%
  \bibAnnoteFile{NoStop}{LeeRamVen09}%
\bibitem{oleary-2008b}%
  \BibitemOpen
  \bibfield{author}{%
  \bibinfo {author} {\bibfnamefont{R.~M.}\ \bibnamefont{O'Leary}}, \bibinfo
  {author} {\bibfnamefont{B.}~\bibnamefont{Kocsis}},\ and\ \bibinfo {author}
  {\bibfnamefont{A.}~\bibnamefont{Loeb}}}%
   (\bibinfo {year} {2008}),\
  \Eprint{http://arxiv.org/abs/0807.2638}{arXiv:0807.2638 [astro-ph]}%
  \bibAnnoteFile{NoStop}{oleary-2008b}%
\bibitem{brown-2008ns}%
  \BibitemOpen
  \bibfield{author}{%
  \bibinfo {author} {\bibfnamefont{W.~R.}\ \bibnamefont{Brown}}}%
   (\bibinfo {year} {2008}),\
  \Eprint{http://arxiv.org/abs/0811.0571}{arXiv:0811.0571 [astro-ph]}%
  \bibAnnoteFile{NoStop}{brown-2008ns}%
\bibitem{zwart-2005}%
  \BibitemOpen
  \bibfield{author}{%
  \bibinfo {author} {\bibfnamefont{S.~F.}\ \bibnamefont{Portegies~Zwart}},
  \bibinfo {author} {\bibfnamefont{H.}~\bibnamefont{Baumgardt}}, \bibinfo
  {author} {\bibfnamefont{S.~L.}\ \bibnamefont{McMillan}}, \bibinfo {author}
  {\bibfnamefont{J.}~\bibnamefont{Makino}}, \bibinfo {author}
  {\bibfnamefont{P.}~\bibnamefont{Hut}}, \emph{et~al.},\ }%
  \bibfield{journal}{%
  \bibinfo {journal} {Astrophys.J.}}%
   (\bibinfo {year} {2005}),\
  \Eprint{http://arxiv.org/abs/astro-ph/0511397}{arXiv:astro-ph/0511397
  [astro-ph]}%
  \bibAnnoteFile{NoStop}{zwart-2005}%
\bibitem{blaes-2002-578}%
  \BibitemOpen
  \bibfield{author}{%
  \bibinfo {author} {\bibfnamefont{O.}~\bibnamefont{Blaes}}, \bibinfo {author}
  {\bibfnamefont{M.~H.}\ \bibnamefont{Lee}},\ and\ \bibinfo {author}
  {\bibfnamefont{A.}~\bibnamefont{Socrates}},\ }%
  \bibfield{journal}{%
  \Doi{10.1086/342655}{\bibinfo {journal} {Astrophys.J.}}\ }%
  \textbf{\bibinfo {volume} {578}},\ \bibinfo {pages} {775} (\bibinfo {year}
  {2002}),\
  \Eprint{http://arxiv.org/abs/astro-ph/0203370}{arXiv:astro-ph/0203370
  [astro-ph]}%
  \bibAnnoteFile{NoStop}{blaes-2002-578}%
\bibitem{HopAle05}%
  \BibitemOpen
  \bibfield{author}{%
  \bibinfo {author} {\bibfnamefont{C.}~\bibnamefont{Hopman}}\ and\ \bibinfo
  {author} {\bibfnamefont{T.}~\bibnamefont{Alexander}},\ }%
  \bibfield{journal}{%
  \bibinfo {journal} {Astrophys.J.}\ }%
  \textbf{\bibinfo {volume} {629}},\ \bibinfo {pages} {362} (\bibinfo {year}
  {2005}),\
  \Eprint{http://arxiv.org/abs/astro-ph/0503672}{arXiv:astro-ph/0503672
  [astro-ph]}%
  \bibAnnoteFile{NoStop}{HopAle05}%
\bibitem{ManBroGai07}%
  \BibitemOpen
  \bibfield{author}{%
  \bibinfo {author} {\bibfnamefont{I.}~\bibnamefont{Mandel}}, \bibinfo {author}
  {\bibfnamefont{D.~A.}\ \bibnamefont{Brown}}, \bibinfo {author}
  {\bibfnamefont{J.~R.}\ \bibnamefont{Gair}},\ and\ \bibinfo {author}
  {\bibfnamefont{M.~C.}\ \bibnamefont{Miller}},\ }%
  \bibfield{journal}{%
  \bibinfo {journal} {Astrophys.J.}\ }%
  \textbf{\bibinfo {volume} {681}},\ \bibinfo {pages} {1431} (\bibinfo {year}
  {2008}),\ \Eprint{http://arxiv.org/abs/0705.0285}{arXiv:0705.0285
  [astro-ph]}%
  \bibAnnoteFile{NoStop}{ManBroGai07}%
\bibitem{iwasawa-2005}%
  \BibitemOpen
  \bibfield{author}{%
  \bibinfo {author} {\bibfnamefont{M.}~\bibnamefont{Iwasawa}}, \bibinfo
  {author} {\bibfnamefont{Y.}~\bibnamefont{Funato}},\ and\ \bibinfo {author}
  {\bibfnamefont{J.}~\bibnamefont{Makino}},\ }%
  \bibfield{journal}{%
  \Doi{10.1086/507473}{\bibinfo {journal} {Astrophys.J.}}\ }%
  \textbf{\bibinfo {volume} {651}},\ \bibinfo {pages} {1059} (\bibinfo {year}
  {2006}),\
  \Eprint{http://arxiv.org/abs/astro-ph/0511391}{arXiv:astro-ph/0511391
  [astro-ph]}%
  \bibAnnoteFile{NoStop}{iwasawa-2005}%
\bibitem{volonteri-2003-582}%
  \BibitemOpen
  \bibfield{author}{%
  \bibinfo {author} {\bibfnamefont{M.}~\bibnamefont{Volonteri}}, \bibinfo
  {author} {\bibfnamefont{F.}~\bibnamefont{Haardt}},\ and\ \bibinfo {author}
  {\bibfnamefont{P.}~\bibnamefont{Madau}},\ }%
  \bibfield{journal}{%
  \Doi{10.1086/344675}{\bibinfo {journal} {Astrophys.J.}}\ }%
  \textbf{\bibinfo {volume} {582}},\ \bibinfo {pages} {559} (\bibinfo {year}
  {2003}),\
  \Eprint{http://arxiv.org/abs/astro-ph/0207276}{arXiv:astro-ph/0207276
  [astro-ph]}%
  \bibAnnoteFile{NoStop}{volonteri-2003-582}%
\bibitem{Antonini:2012ad}%
  \BibitemOpen
  \bibfield{author}{%
  \bibinfo {author} {\bibfnamefont{F.}~\bibnamefont{Antonini}}\ and\ \bibinfo
  {author} {\bibfnamefont{H.}~\bibnamefont{Perets}},\ }%
  \bibfield{journal}{%
  \Doi{10.1088/0004-637X/757/1/27}{\bibinfo {journal} {Astrophys.J.}}\ }%
  \textbf{\bibinfo {volume} {757}},\ \bibinfo {pages} {27} (\bibinfo {year}
  {2012}),\ \Eprint{http://arxiv.org/abs/1203.2938}{arXiv:1203.2938
  [astro-ph.GA]}%
  \bibAnnoteFile{NoStop}{Antonini:2012ad}%
\bibitem{Kowalska:2010qg}%
  \BibitemOpen
  \bibfield{author}{%
  \bibinfo {author} {\bibfnamefont{I.}~\bibnamefont{Kowalska}}, \bibinfo
  {author} {\bibfnamefont{T.}~\bibnamefont{Bulik}}, \bibinfo {author}
  {\bibfnamefont{K.}~\bibnamefont{Belczynski}}, \bibinfo {author}
  {\bibfnamefont{M.}~\bibnamefont{Dominik}},\ and\ \bibinfo {author}
  {\bibfnamefont{D.}~\bibnamefont{Gondek-Rosinska}},\ }%
  \bibfield{journal}{%
  \bibinfo {journal} {Astron.Astrophys.}\ }%
  \textbf{\bibinfo {volume} {527}},\ \bibinfo {pages} {A70} (\bibinfo {year}
  {2011}),\ \Eprint{http://arxiv.org/abs/1010.0511}{arXiv:1010.0511
  [astro-ph.CO]}%
  \bibAnnoteFile{NoStop}{Kowalska:2010qg}%
\bibitem{Kocsis:2011ch}%
  \BibitemOpen
  \bibfield{author}{%
  \bibinfo {author} {\bibfnamefont{B.}~\bibnamefont{Kocsis}}, \bibinfo {author}
  {\bibfnamefont{A.}~\bibnamefont{Ray}},\ and\ \bibinfo {author}
  {\bibfnamefont{S.~P.}\ \bibnamefont{Zwart}}}%
   (\bibinfo {year} {2011}),\
  \Eprint{http://arxiv.org/abs/1110.6172}{arXiv:1110.6172 [astro-ph.GA]}%
  \bibAnnoteFile{NoStop}{Kocsis:2011ch}%
\bibitem{BruGonHan06}%
  \BibitemOpen
  \bibfield{author}{%
  \bibinfo {author} {\bibfnamefont{B.}~\bibnamefont{Br{\"u}gmann}}, \bibinfo
  {author} {\bibfnamefont{J.~A.}\ \bibnamefont{Gonz{\'a}lez}}, \bibinfo
  {author} {\bibfnamefont{M.}~\bibnamefont{Hannam}}, \bibinfo {author}
  {\bibfnamefont{S.}~\bibnamefont{Husa}}, \bibinfo {author}
  {\bibfnamefont{U.}~\bibnamefont{Sperhake}},\ and\ \bibinfo {author}
  {\bibfnamefont{W.}~\bibnamefont{Tichy}},\ }%
  \bibfield{journal}{%
  \Doi{10.1103/PhysRevD.77.024027}{\bibinfo {journal} {Phys. Rev.}}\ }%
  \textbf{\bibinfo {volume} {D77}},\ \bibinfo {pages} {024027} (\bibinfo {year}
  {2008}),\ \Eprint{http://arxiv.org/abs/gr-qc/0610128}{arXiv:gr-qc/0610128}%
  \bibAnnoteFile{NoStop}{BruGonHan06}%
\bibitem{BruTicJan03}%
  \BibitemOpen
  \bibfield{author}{%
  \bibinfo {author} {\bibfnamefont{B.}~\bibnamefont{Br{\"u}gmann}}, \bibinfo
  {author} {\bibfnamefont{W.}~\bibnamefont{Tichy}},\ and\ \bibinfo {author}
  {\bibfnamefont{N.}~\bibnamefont{Jansen}},\ }%
  \bibfield{journal}{%
  \bibinfo {journal} {Phys. Rev. Lett.}\ }%
  \textbf{\bibinfo {volume} {92}},\ \bibinfo {pages} {211101} (\bibinfo {year}
  {2004}),\ \Eprint{http://arxiv.org/abs/gr-qc/0312112}{gr-qc/0312112}%
  \bibAnnoteFile{NoStop}{BruTicJan03}%
\bibitem{HusGonHan07}%
  \BibitemOpen
  \bibfield{author}{%
  \bibinfo {author} {\bibfnamefont{S.}~\bibnamefont{Husa}}, \bibinfo {author}
  {\bibfnamefont{J.~A.}\ \bibnamefont{Gonz{\'a}lez}}, \bibinfo {author}
  {\bibfnamefont{M.}~\bibnamefont{Hannam}}, \bibinfo {author}
  {\bibfnamefont{B.}~\bibnamefont{Br{\"u}gmann}},\ and\ \bibinfo {author}
  {\bibfnamefont{U.}~\bibnamefont{Sperhake}},\ }%
  \bibfield{journal}{%
  \bibinfo {journal} {Class. Quant. Grav.}\ }%
  \textbf{\bibinfo {volume} {25}},\ \bibinfo {pages} {105006} (\bibinfo {year}
  {2008}),\ \Eprint{http://arxiv.org/abs/0706.0740}{arXiv:0706.0740 [gr-qc]}%
  \bibAnnoteFile{NoStop}{HusGonHan07}%
\bibitem{BraBru97}%
  \BibitemOpen
  \bibfield{author}{%
  \bibinfo {author} {\bibfnamefont{S.}~\bibnamefont{Brandt}}\ and\ \bibinfo
  {author} {\bibfnamefont{B.}~\bibnamefont{Br{\"u}gmann}},\ }%
  \bibfield{journal}{%
  \bibinfo {journal} {Phys. Rev. Lett.}\ }%
  \textbf{\bibinfo {volume} {78}},\ \bibinfo {pages} {3606} (\bibinfo {year}
  {1997}),\ \Eprint{http://arxiv.org/abs/gr-qc/9703066}{gr-qc/9703066}%
  \bibAnnoteFile{NoStop}{BraBru97}%
\bibitem{AnsBruTic04}%
  \BibitemOpen
  \bibfield{author}{%
  \bibinfo {author} {\bibfnamefont{M.}~\bibnamefont{Ansorg}}, \bibinfo {author}
  {\bibfnamefont{B.}~\bibnamefont{Br{\"u}gmann}},\ and\ \bibinfo {author}
  {\bibfnamefont{W.}~\bibnamefont{Tichy}},\ }%
  \bibfield{journal}{%
  \bibinfo {journal} {Phys. Rev.}\ }%
  \textbf{\bibinfo {volume} {D70}},\ \bibinfo {pages} {064011} (\bibinfo {year}
  {2004}),\ \Eprint{http://arxiv.org/abs/gr-qc/0404056}{gr-qc/0404056}%
  \bibAnnoteFile{NoStop}{AnsBruTic04}%
\bibitem{CamLouMar05}%
  \BibitemOpen
  \bibfield{author}{%
  \bibinfo {author} {\bibfnamefont{M.}~\bibnamefont{Campanelli}}, \bibinfo
  {author} {\bibfnamefont{C.~O.}\ \bibnamefont{Lousto}}, \bibinfo {author}
  {\bibfnamefont{P.}~\bibnamefont{Marronetti}},\ and\ \bibinfo {author}
  {\bibfnamefont{Y.}~\bibnamefont{Zlochower}},\ }%
  \bibfield{journal}{%
  \bibinfo {journal} {Phys. Rev. Lett.}\ }%
  \textbf{\bibinfo {volume} {96}},\ \bibinfo {pages} {111101} (\bibinfo {year}
  {2006}),\ \Eprint{http://arxiv.org/abs/gr-qc/0511048}{gr-qc/0511048}%
  \bibAnnoteFile{NoStop}{CamLouMar05}%
\bibitem{BakCenCho05}%
  \BibitemOpen
  \bibfield{author}{%
  \bibinfo {author} {\bibfnamefont{J.~G.}\ \bibnamefont{Baker}}, \bibinfo
  {author} {\bibfnamefont{J.}~\bibnamefont{Centrella}}, \bibinfo {author}
  {\bibfnamefont{D.-I.}\ \bibnamefont{Choi}}, \bibinfo {author}
  {\bibfnamefont{M.}~\bibnamefont{Koppitz}},\ and\ \bibinfo {author}
  {\bibfnamefont{J.}~\bibnamefont{van Meter}},\ }%
  \bibfield{journal}{%
  \bibinfo {journal} {Phys. Rev. Lett.}\ }%
  \textbf{\bibinfo {volume} {96}},\ \bibinfo {pages} {111102} (\bibinfo {year}
  {2006}),\ \Eprint{http://arxiv.org/abs/gr-qc/0511103}{gr-qc/0511103}%
  \bibAnnoteFile{NoStop}{BakCenCho05}%
\bibitem{ShiNak95}%
  \BibitemOpen
  \bibfield{author}{%
  \bibinfo {author} {\bibfnamefont{M.}~\bibnamefont{Shibata}}\ and\ \bibinfo
  {author} {\bibfnamefont{T.}~\bibnamefont{Nakamura}},\ }%
  \bibfield{journal}{%
  \bibinfo {journal} {Phys. Rev.}\ }%
  \textbf{\bibinfo {volume} {D52}},\ \bibinfo {pages} {5428} (\bibinfo {year}
  {1995})%
  \bibAnnoteFile{NoStop}{ShiNak95}%
\bibitem{BauSha98}%
  \BibitemOpen
  \bibfield{author}{%
  \bibinfo {author} {\bibfnamefont{T.~W.}\ \bibnamefont{Baumgarte}}\ and\
  \bibinfo {author} {\bibfnamefont{S.~L.}\ \bibnamefont{Shapiro}},\ }%
  \bibfield{journal}{%
  \bibinfo {journal} {Phys. Rev.}\ }%
  \textbf{\bibinfo {volume} {D59}},\ \bibinfo {pages} {024007} (\bibinfo {year}
  {1998}),\ \Eprint{http://arxiv.org/abs/gr-qc/9810065}{gr-qc/9810065}%
  \bibAnnoteFile{NoStop}{BauSha98}%
\bibitem{Sch10}%
  \BibitemOpen
  \bibfield{author}{%
  \bibinfo {author} {\bibfnamefont{E.}~\bibnamefont{Schnetter}},\ }%
  \bibfield{journal}{%
  \Doi{10.1088/0264-9381/27/16/167001}{\bibinfo {journal} {Class. Quant.
  Grav.}}\ }%
  \textbf{\bibinfo {volume} {27}},\ \bibinfo {pages} {167001} (\bibinfo {year}
  {2010}),\ \Eprint{http://arxiv.org/abs/1003.0859}{arXiv:1003.0859 [gr-qc]}%
  \bibAnnoteFile{NoStop}{Sch10}%
\bibitem{WalBruMue09}%
  \BibitemOpen
  \bibfield{author}{%
  \bibinfo {author} {\bibfnamefont{B.}~\bibnamefont{Walther}}, \bibinfo
  {author} {\bibfnamefont{B.}~\bibnamefont{Br{\"u}gmann}},\ and\ \bibinfo
  {author} {\bibfnamefont{D.}~\bibnamefont{M{\"u}ller}},\ }%
  \bibfield{journal}{%
  \bibinfo {journal} {Phys. Rev.}\ }%
  \textbf{\bibinfo {volume} {D79}},\ \bibinfo {pages} {124040} (\bibinfo {year}
  {2009}),\ \Eprint{http://arxiv.org/abs/0901.0993}{arXiv:0901.0993 [gr-qc]}%
  \bibAnnoteFile{NoStop}{WalBruMue09}%
\bibitem{Note2}%
  \BibitemOpen
  \bibinfo {note} {Generally speaking, if such an orbit is unbound then the
  black holes separate without close encounter. If bound, then after reaching
  the first apocenter the orbit becomes inward-bound, which in principle is
  already included for $\Theta \in [0,90^\circ ]$.}%
  \bibAnnoteFile{Stop}{Note2}%
\bibitem{PreFlaTeu92}%
  \BibitemOpen
  \bibfield{author}{%
  \bibinfo {author} {\bibfnamefont{W.~H.}\ \bibnamefont{Press}}, \bibinfo
  {author} {\bibfnamefont{B.~P.}\ \bibnamefont{Flannery}}, \bibinfo {author}
  {\bibfnamefont{S.~A.}\ \bibnamefont{Teukolsky}},\ and\ \bibinfo {author}
  {\bibfnamefont{W.~T.}\ \bibnamefont{Vetterling}},\ }%
  \emph{\bibinfo {title} {Numerical Recipes in {C}}},\ \bibinfo {edition}
  {2nd}\ ed.\ (\bibinfo {publisher} {Cambridge University Press},\ \bibinfo
  {address} {New York},\ \bibinfo {year} {1992})%
  \bibAnnoteFile{NoStop}{PreFlaTeu92}%
\bibitem{memmesheimer-2005-71}%
  \BibitemOpen
  \bibfield{author}{%
  \bibinfo {author} {\bibfnamefont{R.-M.}\ \bibnamefont{Memmesheimer}}\ and\
  \bibinfo {author} {\bibfnamefont{G.}~\bibnamefont{Sch{\"a}fer}},\ }%
  \bibfield{journal}{%
  \Doi{10.1103/PhysRevD.71.044021}{\bibinfo {journal} {Phys. Rev.}}\ }%
  \textbf{\bibinfo {volume} {D71}},\ \bibinfo {pages} {044021} (\bibinfo {year}
  {2005}),\ \Eprint{http://arxiv.org/abs/gr-qc/0411057}{arXiv:gr-qc/0411057
  [gr-qc]}%
  \bibAnnoteFile{NoStop}{memmesheimer-2005-71}%
\bibitem{MueBru09}%
  \BibitemOpen
  \bibfield{author}{%
  \bibinfo {author} {\bibfnamefont{D.}~\bibnamefont{M{\"u}ller}}\ and\ \bibinfo
  {author} {\bibfnamefont{B.}~\bibnamefont{Br{\"u}gmann}},\ }%
  \bibfield{journal}{%
  \Doi{10.1088/0264-9381/27/11/114008}{\bibinfo {journal} {Class. Quant.
  Grav.}}\ }%
  \textbf{\bibinfo {volume} {27}},\ \bibinfo {pages} {114008} (\bibinfo {year}
  {2010}),\ \Eprint{http://arxiv.org/abs/0912.3125}{arXiv:0912.3125 [gr-qc]}%
  \bibAnnoteFile{NoStop}{MueBru09}%
\bibitem{MueGriBru10}%
  \BibitemOpen
  \bibfield{author}{%
  \bibinfo {author} {\bibfnamefont{D.}~\bibnamefont{M\"uller}}, \bibinfo
  {author} {\bibfnamefont{J.}~\bibnamefont{Grigsby}},\ and\ \bibinfo {author}
  {\bibfnamefont{B.}~\bibnamefont{Br\"ugmann}},\ }%
  \bibfield{journal}{%
  \bibinfo {journal} {Phys. Rev.}\ }%
  \textbf{\bibinfo {volume} {D82}},\ \bibinfo {pages} {064004} (\bibinfo {year}
  {2010}),\ \Eprint{http://arxiv.org/abs/1003.4681}{arXiv:1003.4681 [gr-qc]}%
  \bibAnnoteFile{NoStop}{MueGriBru10}%
\bibitem{Alic:2010wu}%
  \BibitemOpen
  \bibfield{author}{%
  \bibinfo {author} {\bibfnamefont{D.}~\bibnamefont{Alic}}, \bibinfo {author}
  {\bibfnamefont{L.}~\bibnamefont{Rezzolla}}, \bibinfo {author}
  {\bibfnamefont{I.}~\bibnamefont{Hinder}},\ and\ \bibinfo {author}
  {\bibfnamefont{P.}~\bibnamefont{Mosta}},\ }%
  \bibfield{journal}{%
  \Doi{10.1088/0264-9381/27/24/245023}{\bibinfo {journal} {Class.Quant.Grav.}}\
  }%
  \textbf{\bibinfo {volume} {27}},\ \bibinfo {pages} {245023} (\bibinfo {year}
  {2010}),\ \Eprint{http://arxiv.org/abs/1008.2212}{arXiv:1008.2212 [gr-qc]}%
  \bibAnnoteFile{NoStop}{Alic:2010wu}%
\bibitem{Thi08}%
  \BibitemOpen
  \bibfield{author}{%
  \bibinfo {author} {\bibfnamefont{M.}~\bibnamefont{Thierfelder}},\ }%
  \enquote{\bibinfo {title} {Event horizon finder for black holes},}\
  (\bibinfo {year} {2008}),\ \bibinfo {note} {{D}iploma {T}hesis, {U}niversity
  of {J}ena}%
  \bibAnnoteFile{NoStop}{Thi08}%
\bibitem{ThiBru12}%
  \BibitemOpen
  \bibfield{author}{%
  \bibinfo {author} {\bibfnamefont{M.}~\bibnamefont{Thierfelder}}\ and\
  \bibinfo {author} {\bibfnamefont{B.}~\bibnamefont{Br\"ugmann}},\ }%
  \bibfield{journal}{%
  \bibinfo {journal} {Phys. Rev.}\ }%
  \textbf{\bibinfo {volume} {D}} (\bibinfo {year} {to be submitted})%
  \bibAnnoteFile{NoStop}{ThiBru12}%
\bibitem{HanHusPol06}%
  \BibitemOpen
  \bibfield{author}{%
  \bibinfo {author} {\bibfnamefont{M.}~\bibnamefont{Hannam}}, \bibinfo {author}
  {\bibfnamefont{S.}~\bibnamefont{Husa}}, \bibinfo {author}
  {\bibfnamefont{D.}~\bibnamefont{Pollney}}, \bibinfo {author}
  {\bibfnamefont{B.}~\bibnamefont{Br{\"u}gmann}},\ and\ \bibinfo {author}
  {\bibfnamefont{N.}~\bibnamefont{{\'O~Murchadha}}},\ }%
  \bibfield{journal}{%
  \bibinfo {journal} {Phys. Rev. Lett.}\ }%
  \textbf{\bibinfo {volume} {99}},\ \bibinfo {pages} {241102} (\bibinfo {year}
  {2007}),\ \Eprint{http://arxiv.org/abs/gr-qc/0606099}{gr-qc/0606099}%
  \bibAnnoteFile{NoStop}{HanHusPol06}%
\bibitem{ReiBru04}%
  \BibitemOpen
  \bibfield{author}{%
  \bibinfo {author} {\bibfnamefont{B.}~\bibnamefont{Reimann}}\ and\ \bibinfo
  {author} {\bibfnamefont{B.}~\bibnamefont{Br{\"u}gmann}},\ }%
  \bibfield{journal}{%
  \bibinfo {journal} {Phys. Rev.}\ }%
  \textbf{\bibinfo {volume} {D69}},\ \bibinfo {pages} {124009} (\bibinfo {year}
  {2004}),\ \Eprint{http://arxiv.org/abs/gr-qc/0401098}{gr-qc/0401098}%
  \bibAnnoteFile{NoStop}{ReiBru04}%
\bibitem{Alc08}%
  \BibitemOpen
  \bibfield{author}{%
  \bibinfo {author} {\bibfnamefont{M.}~\bibnamefont{Alcubierre}},\ }%
  \emph{\bibinfo {title} {Introduction to 3+1 Numerical Relativity}}\ (\bibinfo
  {publisher} {Oxford University Press, USA},\ \bibinfo {year} {2008})%
  \bibAnnoteFile{NoStop}{Alc08}%
\bibitem{BerCarGon07}%
  \BibitemOpen
  \bibfield{author}{%
  \bibinfo {author} {\bibfnamefont{E.}~\bibnamefont{Berti}}, \bibinfo {author}
  {\bibfnamefont{V.}~\bibnamefont{Cardoso}}, \bibinfo {author}
  {\bibfnamefont{J.~A.}\ \bibnamefont{Gonz{\'a}lez}}, \bibinfo {author}
  {\bibfnamefont{U.}~\bibnamefont{Sperhake}},\ and\ \bibinfo {author}
  {\bibfnamefont{B.}~\bibnamefont{Br{\"u}gmann}},\ }%
  \bibfield{journal}{%
  \Doi{10.1088/0264-9381/25/11/114035}{\bibinfo {journal} {Class. Quant.
  Grav.}}\ }%
  \textbf{\bibinfo {volume} {25}},\ \bibinfo {pages} {114035} (\bibinfo {year}
  {2008}),\ \Eprint{http://arxiv.org/abs/0711.1097}{arXiv:0711.1097 [gr-qc]}%
  \bibAnnoteFile{NoStop}{BerCarGon07}%
\bibitem{Book:ShaTeu}%
  \BibitemOpen
  \bibfield{author}{%
  \bibinfo {author} {\bibfnamefont{S.}~\bibnamefont{Shapiro}}\ and\ \bibinfo
  {author} {\bibfnamefont{S.}~\bibnamefont{Teukolsky}},\ }%
  \emph{\bibinfo {title} {Blackholes, White Dwarfs and Neutron Stars: The
  physics of compact objects}}\ (\bibinfo {publisher} {John Wiley \& Sons},\
  \bibinfo {year} {1983})%
  \bibAnnoteFile{NoStop}{Book:ShaTeu}%
\bibitem{PetMat63}%
  \BibitemOpen
  \bibfield{author}{%
  \bibinfo {author} {\bibfnamefont{P.~C.}\ \bibnamefont{Peters}}\ and\ \bibinfo
  {author} {\bibfnamefont{J.}~\bibnamefont{Mathews}},\ }%
  \bibfield{journal}{%
  \bibinfo {journal} {Phys. Rev.}\ }%
  \textbf{\bibinfo {volume} {131}},\ \bibinfo {pages} {435} (\bibinfo {year}
  {1963})%
  \bibAnnoteFile{NoStop}{PetMat63}%
\bibitem{TesSch10}%
  \BibitemOpen
  \bibfield{author}{%
  \bibinfo {author} {\bibfnamefont{M.}~\bibnamefont{Tessmer}}\ and\ \bibinfo
  {author} {\bibfnamefont{G.}~\bibnamefont{Sch{\"a}fer}},\ }%
  \bibfield{journal}{%
  \Doi{10.1002/andp.201100007}{\bibinfo {journal} {Annalen Phys.}}\ }%
  \textbf{\bibinfo {volume} {523}},\ \bibinfo {pages} {813} (\bibinfo {year}
  {2011}),\ \Eprint{http://arxiv.org/abs/1012.3894}{arXiv:1012.3894 [gr-qc]}%
  \bibAnnoteFile{NoStop}{TesSch10}%
\bibitem{Hinder:2008kv}%
  \BibitemOpen
  \bibfield{author}{%
  \bibinfo {author} {\bibfnamefont{I.}~\bibnamefont{Hinder}}, \bibinfo {author}
  {\bibfnamefont{F.}~\bibnamefont{Herrmann}}, \bibinfo {author}
  {\bibfnamefont{P.}~\bibnamefont{Laguna}},\ and\ \bibinfo {author}
  {\bibfnamefont{D.}~\bibnamefont{Shoemaker}},\ }%
  \bibfield{journal}{%
  \Doi{10.1103/PhysRevD.82.024033}{\bibinfo {journal} {Phys. Rev.}}\ }%
  \textbf{\bibinfo {volume} {D82}},\ \bibinfo {pages} {024033} (\bibinfo {year}
  {2010}),\ \Eprint{http://arxiv.org/abs/0806.1037}{arXiv:0806.1037 [gr-qc]}%
  \bibAnnoteFile{NoStop}{Hinder:2008kv}%
\bibitem{Hou94}%
  \BibitemOpen
  \bibfield{author}{%
  \bibinfo {author} {\bibfnamefont{J.}~\bibnamefont{Hough}}}%
   (\bibinfo {year} {1994}),\ \bibinfo {note} {prepared for {E}doardo {A}maldi
  Meeting on Gravitational Wave Experiments, {R}ome, {I}taly, 14-17 Jun 1994}%
  \bibAnnoteFile{NoStop}{Hou94}%
\bibitem{BabHanHus08}%
  \BibitemOpen
  \bibfield{author}{%
  \bibinfo {author} {\bibfnamefont{S.}~\bibnamefont{Babak}}, \bibinfo {author}
  {\bibfnamefont{M.}~\bibnamefont{Hannam}}, \bibinfo {author}
  {\bibfnamefont{S.}~\bibnamefont{Husa}},\ and\ \bibinfo {author}
  {\bibfnamefont{B.}~\bibnamefont{Schutz}},\ }%
  \enquote{\bibinfo {title} {{Resolving Super Massive Black Holes with
  LISA}},}\  (\bibinfo {year} {2008}),\
  \Eprint{http://arxiv.org/abs/0806.1591}{arXiv:0806.1591 [gr-qc]}%
  \bibAnnoteFile{NoStop}{BabHanHus08}%
\bibitem{eLISA:2012}%
  \BibitemOpen
  \bibfield{author}{%
  \bibinfo {author} {\bibfnamefont{P.}~\bibnamefont{Amaro-Seoane}}, \bibinfo
  {author} {\bibfnamefont{S.}~\bibnamefont{Aoudia}}, \bibinfo {author}
  {\bibfnamefont{S.}~\bibnamefont{Babak}}, \bibinfo {author}
  {\bibfnamefont{P.}~\bibnamefont{Binetruy}}, \bibinfo {author}
  {\bibfnamefont{E.}~\bibnamefont{Berti}}, \emph{et~al.}}%
   (\bibinfo {year} {2012}),\
  \Eprint{http://arxiv.org/abs/1201.3621}{arXiv:1201.3621 [astro-ph.CO]}%
  \bibAnnoteFile{NoStop}{eLISA:2012}%
\bibitem{Yagi:2012gb}%
  \BibitemOpen
  \bibfield{author}{%
  \bibinfo {author} {\bibfnamefont{K.}~\bibnamefont{Yagi}},\ }%
  \bibfield{journal}{%
  \Doi{10.1088/0264-9381/29/7/075005}{\bibinfo {journal} {Class.Quant.Grav.}}\
  }%
  \textbf{\bibinfo {volume} {29}},\ \bibinfo {pages} {075005} (\bibinfo {year}
  {2012}),\ \Eprint{http://arxiv.org/abs/1202.3512}{arXiv:1202.3512
  [astro-ph.CO]}%
  \bibAnnoteFile{NoStop}{Yagi:2012gb}%
\bibitem{hannam-2009}%
  \BibitemOpen
  \bibfield{author}{%
  \bibinfo {author} {\bibfnamefont{M.}~\bibnamefont{Hannam}}\ and\ \bibinfo
  {author} {\bibfnamefont{I.}~\bibnamefont{Hawke}},\ }%
  \bibfield{journal}{%
  \Doi{10.1007/s10714-010-1008-2}{\bibinfo {journal} {Gen.Rel.Grav.}}\ }%
  \textbf{\bibinfo {volume} {43}},\ \bibinfo {pages} {465} (\bibinfo {year}
  {2011}),\ \Eprint{http://arxiv.org/abs/0908.3139}{arXiv:0908.3139 [gr-qc]}%
  \bibAnnoteFile{NoStop}{hannam-2009}%
\bibitem{Broeck:2010vx}%
  \BibitemOpen
  \bibfield{author}{%
  \bibinfo {author} {\bibfnamefont{C.}~\bibnamefont{Van Den~Broeck}},\ }%
  \bibfield{journal}{%
  \Doi{10.1142/9789814374552_0302}{\bibinfo {journal} {MG12 Proc.}},\ \bibinfo
  {pages} {1682}}%
   (\bibinfo {year} {2010}),\
  \Eprint{http://arxiv.org/abs/1003.1386}{arXiv:1003.1386 [gr-qc]}%
  \bibAnnoteFile{NoStop}{Broeck:2010vx}%
\bibitem{Key:2010tc}%
  \BibitemOpen
  \bibfield{author}{%
  \bibinfo {author} {\bibfnamefont{J.~S.}\ \bibnamefont{Key}}\ and\ \bibinfo
  {author} {\bibfnamefont{N.~J.}\ \bibnamefont{Cornish}},\ }%
  \bibfield{journal}{%
  \Doi{10.1103/PhysRevD.83.083001}{\bibinfo {journal} {Phys. Rev.}}\ }%
  \textbf{\bibinfo {volume} {D83}},\ \bibinfo {pages} {083001} (\bibinfo {year}
  {2011}),\ \Eprint{http://arxiv.org/abs/1006.3759}{arXiv:1006.3759 [gr-qc]}%
  \bibAnnoteFile{NoStop}{Key:2010tc}%
\bibitem{Haw71}%
  \BibitemOpen
  \bibfield{author}{%
  \bibinfo {author} {\bibfnamefont{S.}~\bibnamefont{Hawking}},\ }%
  \bibfield{journal}{%
  \bibinfo {journal} {Phys. Rev. Lett.}\ }%
  \textbf{\bibinfo {volume} {26}},\ \bibinfo {pages} {1344} (\bibinfo {year}
  {1971})%
  \bibAnnoteFile{NoStop}{Haw71}%
\bibitem{Spe09}%
  \BibitemOpen
  \bibfield{author}{%
  \bibinfo {author} {\bibfnamefont{U.}~\bibnamefont{Sperhake}},\ }%
  \bibfield{journal}{%
  \Doi{10.1007/978-3-540-88460-6_4}{\bibinfo {journal} {Lect. Notes Phys.}}\ }%
  \textbf{\bibinfo {volume} {769}},\ \bibinfo {pages} {125} (\bibinfo {year}
  {2009})%
  \bibAnnoteFile{NoStop}{Spe09}%
\bibitem{DieHerPol05}%
  \BibitemOpen
  \bibfield{author}{%
  \bibinfo {author} {\bibfnamefont{P.}~\bibnamefont{Diener}}, \bibinfo {author}
  {\bibfnamefont{F.}~\bibnamefont{Herrmann}}, \bibinfo {author}
  {\bibfnamefont{D.}~\bibnamefont{Pollney}}, \bibinfo {author}
  {\bibfnamefont{E.}~\bibnamefont{Schnetter}}, \bibinfo {author}
  {\bibfnamefont{E.}~\bibnamefont{Seidel}}, \bibinfo {author}
  {\bibfnamefont{R.}~\bibnamefont{Takahashi}}, \bibinfo {author}
  {\bibfnamefont{J.}~\bibnamefont{Thornburg}},\ and\ \bibinfo {author}
  {\bibfnamefont{J.}~\bibnamefont{Ventrella}},\ }%
  \bibfield{journal}{%
  \bibinfo {journal} {Phys. Rev. Lett.}\ }%
  \textbf{\bibinfo {volume} {96}},\ \bibinfo {pages} {121101} (\bibinfo {year}
  {2006}),\ \Eprint{http://arxiv.org/abs/gr-qc/0512108}{gr-qc/0512108}%
  \bibAnnoteFile{NoStop}{DieHerPol05}%
\bibitem{LouZlo10}%
  \BibitemOpen
  \bibfield{author}{%
  \bibinfo {author} {\bibfnamefont{C.~O.}\ \bibnamefont{Lousto}}\ and\ \bibinfo
  {author} {\bibfnamefont{Y.}~\bibnamefont{Zlochower}},\ }%
  \bibfield{journal}{%
  \Doi{10.1103/PhysRevLett.106.041101}{\bibinfo {journal} {Phys. Rev. Lett.}}\
  }%
  \textbf{\bibinfo {volume} {106}},\ \bibinfo {pages} {041101} (\bibinfo {year}
  {2011}),\ \Eprint{http://arxiv.org/abs/1009.0292}{arXiv:1009.0292 [gr-qc]}%
  \bibAnnoteFile{NoStop}{LouZlo10}%
\bibitem{GonSpeBru08}%
  \BibitemOpen
  \bibfield{author}{%
  \bibinfo {author} {\bibfnamefont{J.~A.}\ \bibnamefont{Gonz{\'a}lez}},
  \bibinfo {author} {\bibfnamefont{U.}~\bibnamefont{Sperhake}},\ and\ \bibinfo
  {author} {\bibfnamefont{B.}~\bibnamefont{Br{\"u}gmann}},\ }%
  \bibfield{journal}{%
  \Doi{10.1103/PhysRevD.79.124006}{\bibinfo {journal} {Phys. Rev.}}\ }%
  \textbf{\bibinfo {volume} {D79}},\ \bibinfo {pages} {124006} (\bibinfo {year}
  {2009}),\ \Eprint{http://arxiv.org/abs/0811.3952}{arXiv:0811.3952 [gr-qc]}%
  \bibAnnoteFile{NoStop}{GonSpeBru08}%
\end{thebibliography}%

\end{document}